\documentclass[aps,prc,10pt,amsmath,amssymb,twocolumn,floatfix,nofootinbib,
showpacs,noeprint,superscriptaddress]{revtex4-1}

\usepackage[utf8]{inputenc}

\usepackage{amssymb,amsthm,amsmath,amstext,amsbsy}
\usepackage{bbm}
\usepackage{url}
\usepackage{nicefrac}
\usepackage{graphicx}
\usepackage{hyperref}
\usepackage{leftidx}
\usepackage{xspace}
\usepackage{array}

\newcommand{\ie}{\textit{i.e.}}
\newcommand{\eg}{\textit{e.g.}}
\newcommand{\cf}{\textit{cf.}\xspace}
\newcommand{\apriori}{\textit{a priori}\xspace}

\newcommand{\etal}{\textit{et al.}\xspace}

\newcommand{\MeV}{\ensuremath{\mathrm{MeV}}}

\newcommand{\fm}{\ensuremath{\mathrm{fm}}}

\newcommand{\mathspace}{\ \ }
\newcommand{\mathtext}[1]{\mathspace\text{#1}\mathspace}

\newcommand{\vecr}{\mathbf{r}}

\newcommand{\veck}{\mathbf{k}}

\newcommand{\dd}{\mathrm{d}}

\newcommand{\ddr}{\mathrm{d}^3r}

\newcommand{\gsim}{\gtrsim}

\newcommand{\ii}{\mathrm{i}}
\newcommand{\ee}{\mathrm{e}}

\newcommand{\OO}{\mathcal{O}}

\newcommand{\YY}{Y}

\newcommand{\Rp}{\mathrm{Re}}

\newcommand{\bra}[1]{\langle #1|}

\newcommand{\ket}[1]{|#1\rangle}

\newcommand{\braket}[2]{\langle #1|#2\rangle}

\newcommand{\mbraket}[3]{\langle #1|#2|#3\rangle}

\newcommand{\kmax}{{k_{\mathrm{max}}}}

\newcommand{\Nmax}{N}

\newcommand{\PT}{\text{PT}}

\newcommand{\Vstep}{V_\text{step}}
\newcommand{\VPT}{V_\PT}

\newcommand{\SD}{\ensuremath{SD}}

\newcommand{\etaS}{\eta_0}
\newcommand{\etaD}{\eta_2}

\newcommand{\etaSk}{\ket{\etaS}}
\newcommand{\etaSb}{\bra{\etaS}}
\newcommand{\etaDk}{\ket{\etaD}}
\newcommand{\etaDb}{\bra{\etaD}}

\newcommand{\kinf}{k_{\infty}}
\newcommand{\Einf}{E_{\infty}}
\newcommand{\ELam}{E_{\Lambda}}

\newcommand{\kapinf}{\kappa_{\infty}}
\newcommand{\phiinf}{\phi_{\infty}}
\newcommand{\cinf}{c_{\infty}}

\newcommand{\flam}{f_{\lambda}}

\newcommand{\ts}{\textstyle\strut}
\newcommand{\ds}{\displaystyle}

\newcommand{\vlowk}{V_{{\rm low\,}k}}
\newcommand{\wt}{\widetilde}

\newcommand{\beq}{\begin{equation}}
\newcommand{\eeq}{\end{equation}}
\newcommand{\bea}{\begin{eqnarray}}
\newcommand{\eea}{\end{eqnarray}}

\newcommand{\hw}{\Omega}  % {\hbar\Omega}
\newcommand{\Leff}{L_{\rm eff}}
\newcommand{\Lameff}{\Lambda_{\rm eff}}

\newcommand{\fmi}{\mbox{fm}^{-1}}
\newcommand{\LambdaExpand}{\Lambda_{*}}  % {\overline\Lambda}   % {\Lambda_{*}}

\newcommand{\fiteta}{\ensuremath{\eta}}
\newcommand{\fitetagen}{$\eta$, full}
\newcommand{\fitetagenp}{\ensuremath{\eta'}}
\newcommand{\fitetadirect}{$\eta$, direct}

\begin{document}

\title{Ultraviolet extrapolations in finite oscillator bases}

\author{S.~König}
\email{koenig.389@osu.edu}
\affiliation{Department of Physics, The Ohio State University,
Columbus, Ohio 43210, USA}

\author{S.~K.~Bogner}
\email{bogner@nscl.msu.edu}
\affiliation{National Superconducting Cyclotron Laboratory
and Department of Physics and Astronomy, Michigan State University,
East Lansing, Michigan 48844, USA}

\author{R.~J.~Furnstahl}
\email{furnstahl.1@osu.edu}
\affiliation{Department of Physics, The Ohio State University,
Columbus, Ohio 43210, USA}

\author{S.~N.~More}
\email{more.13@osu.edu}
\affiliation{Department of Physics, The Ohio State University,
Columbus, Ohio 43210, USA}

\author{T.~Papenbrock}
\email{tpapenbr@utk.edu}
\affiliation{Department of Physics and Astronomy, University of Tennessee,
Knoxville, Tennessee 37996, USA}
\affiliation{Physics Division, Oak Ridge National Laboratory,
Oak Ridge, Tennessee 37831, USA}

\date{\today}

\begin{abstract}
  The use of finite harmonic oscillator spaces in many-body
  calculations introduces both infrared (IR) and ultraviolet (UV)
  errors.  The IR effects are well approximated by imposing a
  hard-wall boundary condition at a properly identified radius
  $\Leff$.  We show that duality of the oscillator implies that the UV
  effects are equally well described by imposing a sharp momentum cutoff at
  a momentum $\Lameff$ complementary to $\Leff$.  By considering
  two-body systems with separable potentials, we show that the UV
  energy corrections depend on details of the potential, in contrast
  to the IR energy corrections, which depend only on the S-matrix.
  An adaptation of the separable treatment to more general
  interactions is developed and applied to model potentials as well as to the
  deuteron with realistic potentials.  The previous success with a
  simple phenomenological form for the UV error is also explained.
  Possibilities for controlled extrapolations for $A>2$ based on
  scaling arguments are discussed.
\end{abstract}

\maketitle

\section{Introduction}
\label{sec:Intro}

When truncated harmonic oscillator (HO) model spaces are used in
wavefunction-based methods for computing atomic nuclei, both the
infrared (IR) and the ultraviolet (UV) physics is modified, leading to
systematic errors in observables~\cite{Stetcu:2006ey,
hagen:2010gd,jurgenson:2010wy,Coon:2012ab,Furnstahl:2012qg,Coon:2014nja}.
If these errors can be understood formally, then controlled
extrapolations to the results for the full model space can be made.  A
theoretical formulation for IR extrapolations was proposed in
Ref.~\cite{Furnstahl:2012qg}, and further developed in
Refs.~\cite{More:2013rma,Furnstahl:2013vda,Furnstahl:2014aaa}.  In
this paper we provide a corresponding theoretical basis for UV
extrapolations.

The IR effect of an oscillator basis truncation is practically the
same as imposing a hard-wall boundary condition (\ie, a sharp cutoff
in position space) at a radius $\Leff$.  This is a low-momentum
equivalence in the sense of an effective theory; we determine $\Leff$
by matching the smallest eigenvalue of the squared momentum operator
in the finite basis to the smallest eigenvalue in the spherical box.
The quantity $\Leff$ depends on the number of
fermions~\cite{Furnstahl:2014aaa}.  For two-body bound states,
expansions for the corrections to the energy and other observables
based on a continuation of the S-matrix have been derived in
Ref.~\cite{Furnstahl:2013vda} to next-to-leading order (NLO).  At leading
order (LO), the energy correction is proportional to $\exp{(-2k\Leff)}$,
with $k$ given by the separation energy, due to the exponential fall-off of
the wavefunction in position space.  Further tests for oxygen isotopes
show that the LO form of the corrections works very well
for $A>2$ (although the coefficients are fit rather than given as for
$A=2$)~\cite{Furnstahl:2014aaa}.  In those tests, it was possible (for
coupled-cluster calculations with moderately soft potentials) to
suppress the UV corrections by going to large values of the oscillator
frequency $\hw$, so that the IR correction could be isolated.

However, the need to understand UV corrections remains.  For many
methods the full suppression of the UV is not feasible, and in all
cases the UV effect is a systematic error that must be quantified.  In
addition, this error worsens for harder nucleon--nucleon potentials
that may still be of interest.  Finally, we seek an understanding of
the successes (and limitations) of previous phenomenological forms.
Thus we are well motivated to study the UV errors.

Here we follow the strategy of Refs.~\cite{More:2013rma} and
\cite{Furnstahl:2013vda} by focusing on the two-body problem and
exactly solvable examples to establish the true UV behavior for these
simple systems.  In doing so, the duality of the HO
tells us that part of the IR lesson carries over; namely that the
effect of the oscillator truncation in the UV is practically the same
as a hard cutoff in momentum at an appropriate $\Lameff$, with an
expression equivalent to $\Leff$ when each is expressed in
dimensionless units.  This is demonstrated in Sec.~\ref{sec:Lambda}
(and Appendix~\ref{sec:Lambda-Details}).

However, the impact of this cutoff is not dual.  While the IR result
for the bound-state energy depends only on observables (and is
therefore the same for any two interactions that predict the same S-matrix 
elements), the UV correction depends on the high-momentum
behavior of the potential, which is not an observable.  In
Sec.~\ref{sec:Separable}, we demonstrate this explicitly and derive a
correction formula by considering a rank-one separable potential with
a super-Gaussian form such as those used for effective field theory
regulators.
We then adapt the separable formulation to more general potentials by
building on the classic work by
\citeauthor{Ernst:1973zzb}~\cite{Ernst:1973zzb}.  A fitting procedure
for UV extrapolation is established, tested with model potentials, and, 
finally, applied to the deuteron calculated using realistic
nucleon--nucleon interactions.

A phenomenological scheme for UV corrections based on a Gaussian ansatz, 
applicable to interactions evolved by the similarity renormalization
group (SRG), was proposed in Ref.~\cite{Furnstahl:2012qg} without formal 
justification.  It was also used for $A > 2$ with apparent success in
Ref.~\cite{Jurgenson:2013yya}.  Other works in the literature have
also found that such an ansatz works well (although they have not
generally treated the IR and UV parts separately).  These successes
might seem puzzling in light of our more general results, but we show
in Sec.~\ref{sec:SRG} how the phenomenological ansatz arises when
fitting in a narrow window in $\Lameff$.  Some further remarks on generalizing 
the separable-approximation approach are given in Sec.~\ref{sec:Remarks}, and 
in Sec.~\ref{sec:Outlook} we summarize our results and provide an outlook on 
extensions of the UV extrapolations to $A>2$.

\section{Basis truncation and UV cutoff}
\label{sec:Lambda}

In this section we discuss the relation between the basis size $N$ and the
frequency $\Omega$ of a finite oscillator model space and the corresponding 
UV cutoff $\Lameff$ in momentum space for a two-body system.  Our 
notation and conventions are summarized in Appendix~\ref{sec:Lambda-Details}, 
where we also give a detailed derivation of the results stated in the following.

\subsection{Duality and momentum-space boxes}
\label{sec:Lambda-2}

References~\cite{More:2013rma} and~\cite{Furnstahl:2013vda} demonstrated that a
truncated oscillator basis with highest excitation energy
$N\Omega$ effectively imposes a spherical hard-wall boundary condition
at a radius depending on $N$ and $b$.  The optimal effective radius
$\Leff$ can be determined by matching the smallest eigenvalue
$\kappa^2$ of the squared momentum operator $p^2$ in the finite
basis to the corresponding eigenvalue of the spherical box, namely
$\kappa=\pi/L$ (for $\ell=0$).  The value can be established numerically,
but an accurate approximation for the two-body system is~\cite{More:2013rma}
\begin{equation}
\label{eq:L2_def}
  \Leff = L_2\equiv\sqrt{2(N+3/2+2)}b \,.
\end{equation}
Note that $L_2$ differs by $\mathcal{O}(1/N)$ from the naive estimate
$L_0\equiv\sqrt{2(N+3/2)}b$.  In localized bases that differ from the
harmonic oscillator, $L$ can also be determined from a numerical
diagonalization of the operator $p^2$. 

The dual nature of the harmonic oscillator Hamiltonian \eqref{eq:H-HO}
(\ie, under $p \leftrightarrow \mu\Omega r$) implies that the
truncation of the basis will effectively impose a sharp cutoff at a
momentum $\Lameff$ depending only on $N$ and $b$.  The analog matching
condition leads us to consider the smallest eigenvalue (denoted $\rho$) of 
the operator $r^2$ evaluated in that truncated basis.  This eigenvalue is 
identical to the smallest (squared) distance that can be realized in the
oscillator basis. Thus it corresponds to a lattice spacing on a grid
and therefore sets the highest momentum available. As we see
in Fig.~\ref{fig:deuteron_lambdas}, the square root of the largest 
eigenvalue of the squared momentum operator, which might be a natural guess for 
the effective UV cutoff, is not an accurate estimate for $\Lameff$.

The smallest eigenvalue $\rho$ is determined by Eqs.~\eqref{eq:quant-cond-ell} 
and~\eqref{eq:c-tilde-2} in Appendix~\ref{sec:Lambda-Details}.  From steps 
completely analogous (dual) to those given in 
Refs.~\cite{More:2013rma,Furnstahl:2013vda} for the IR case, we find that
the solution (in a subspace with fixed angular momentum $\ell$) is
\begin{equation}
 \rho = \frac{x_\ell b}{\sqrt2}\left(\Nmax+\frac32+\Delta\right)^{\!-1/2}
\label{eq:rho-Nmax-Delta}
\end{equation}
with $\Delta=2$ to leading order.  The constant $x_\ell$ in the prefactor is 
the first positive zero of the spherical Bessel function $j_\ell$.  Since the 
UV cutoff is given by $x_\ell/\rho$, it drops out again in our final 
result:
\begin{equation}
 \Lambda_2 \equiv \sqrt{2(\Nmax+3/2+2)}/b \,.
\label{eq:Lambda-2-simple}
\end{equation}
Hence, we have shown that the proper effective UV cutoff imposed by 
the basis truncation is given by $\Lambda_2$, which differs by a correction 
term from the naive estimate
\begin{equation}
 \Lambda_0 \equiv \sqrt{2(\Nmax+3/2)}/b
\label{eq:Lambda-0}
\end{equation}
that one obtains by simply considering the maximum single-particle energy 
level represented by the truncated basis.  We note that subleading corrections 
to $\Delta=2$, which by duality apply equally to the IR and UV cutoff, are 
derived in Appendix~\ref{sec:Subleading}.

%%%%%%%%%%%%%%%%%%%%%%%%%%%%%%%%%%%%%%%%%%%%%%%%%%%%%%%%%%%%%%%%%%%%%%%%%%%%%%%%
\begin{figure*}[thbp]
 \centering
 \includegraphics[width=0.95\columnwidth]
 {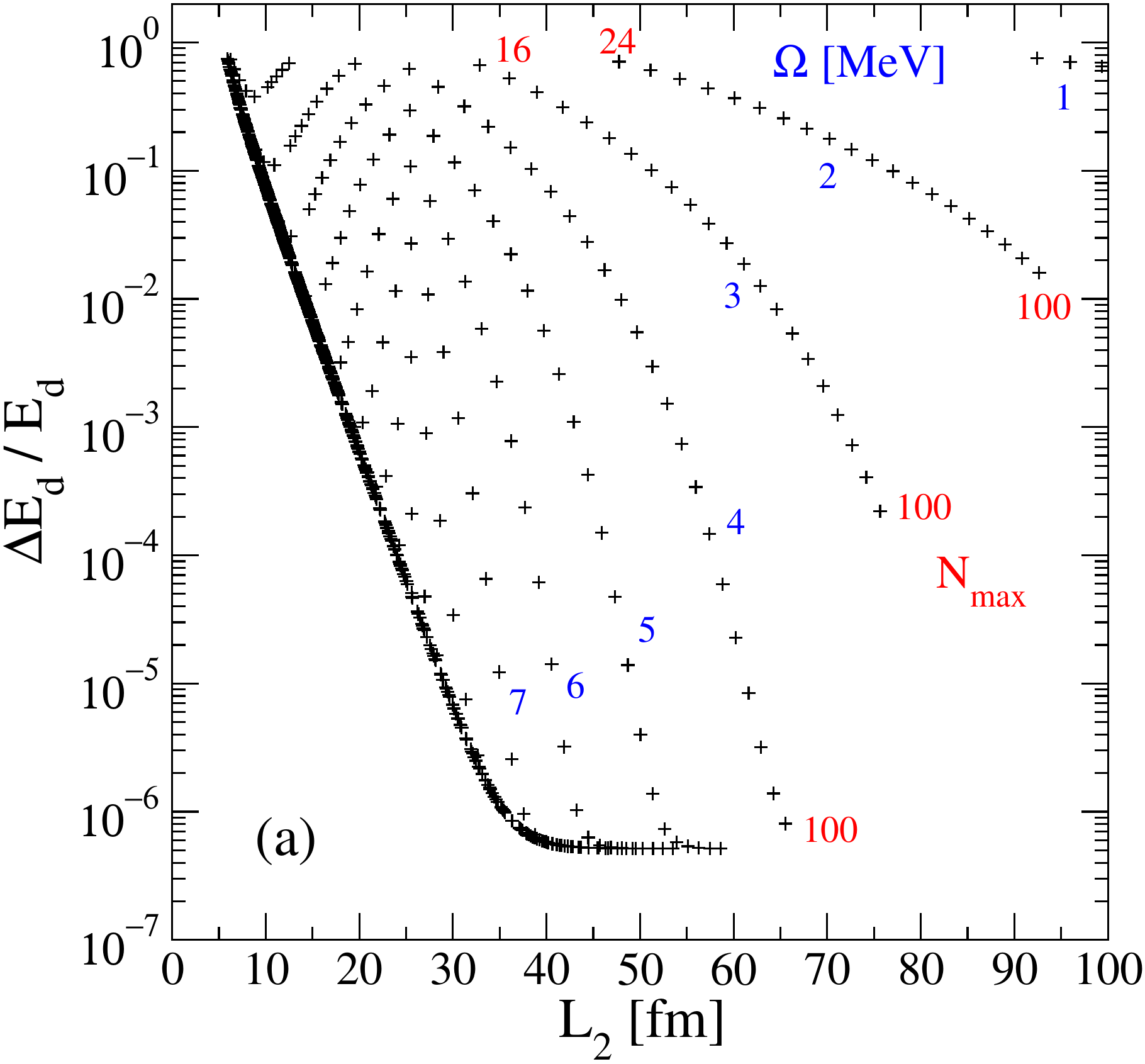}~~~%
 \includegraphics[width=0.9398\columnwidth]%
 {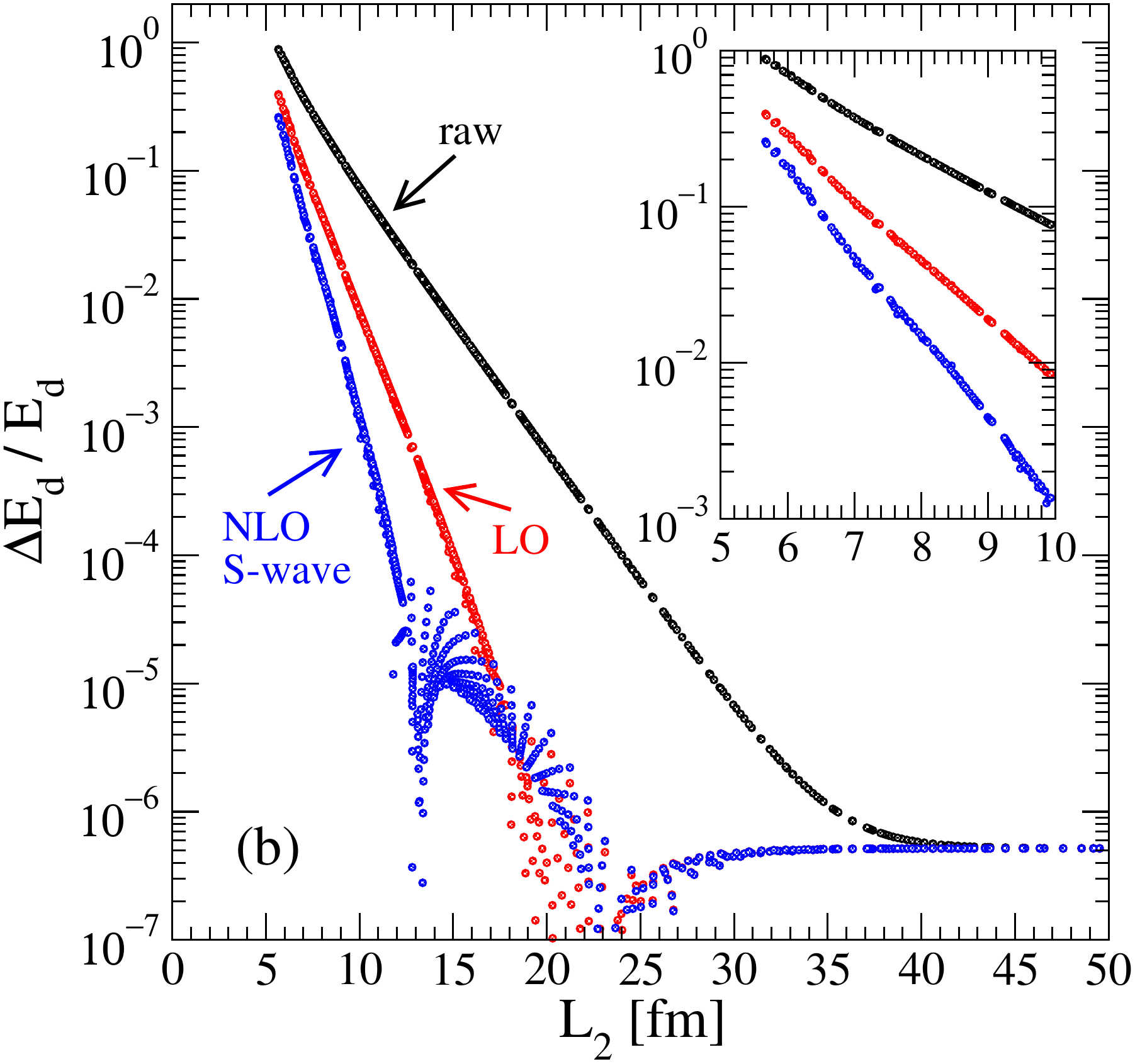}
 \caption{(Color online) (a) Relative error in the deuteron
   energy, computed in harmonic-oscillator bases, for a wide range of
   oscillator parameters $\Nmax$ and $\hw$ as a function of
   $L_2(\Nmax,\hw)$.  Red labels mark the minimum and maximum
   $\Nmax$ along sequences of constant $\Omega$ (indicated by blue
   labels along the sequence of crosses).  These calculations use the
   N$^3$LO $NN$ potential with a $500~\MeV$ regulator cutoff from
   Ref.~\cite{Entem:2003ft}, which was evolved by the similarity
   renormalization group~\cite{Bogner:2006pc} to $\lambda = 2\,\fmi$.
   (b) Subset of calculations from (a) for which the UV correction can
   be neglected compared to the IR correction (``raw''), with LO and
   NLO corrections subtracted as described in the text.  Inset:
   Curves for the lowest values of $L_2$.  }
\label{fig:IR_deuteron_HO}
\end{figure*}
%%%%%%%%%%%%%%%%%%%%%%%%%%%%%%%%%%%%%%%%%%%%%%%%%%%%%%%%%%%%%%%%%%%%%%%%%%%%%%%%

%%%%%%%%%%%%%%%%%%%%%%%%%%%%%%%%%%%%%%%%%%%%%%%%%%%%%%%%%%%%%%%%%%%%%%%%%%%%%%%%
\begin{figure*}[thbp]
 \centering
 \includegraphics[width=0.95\columnwidth]{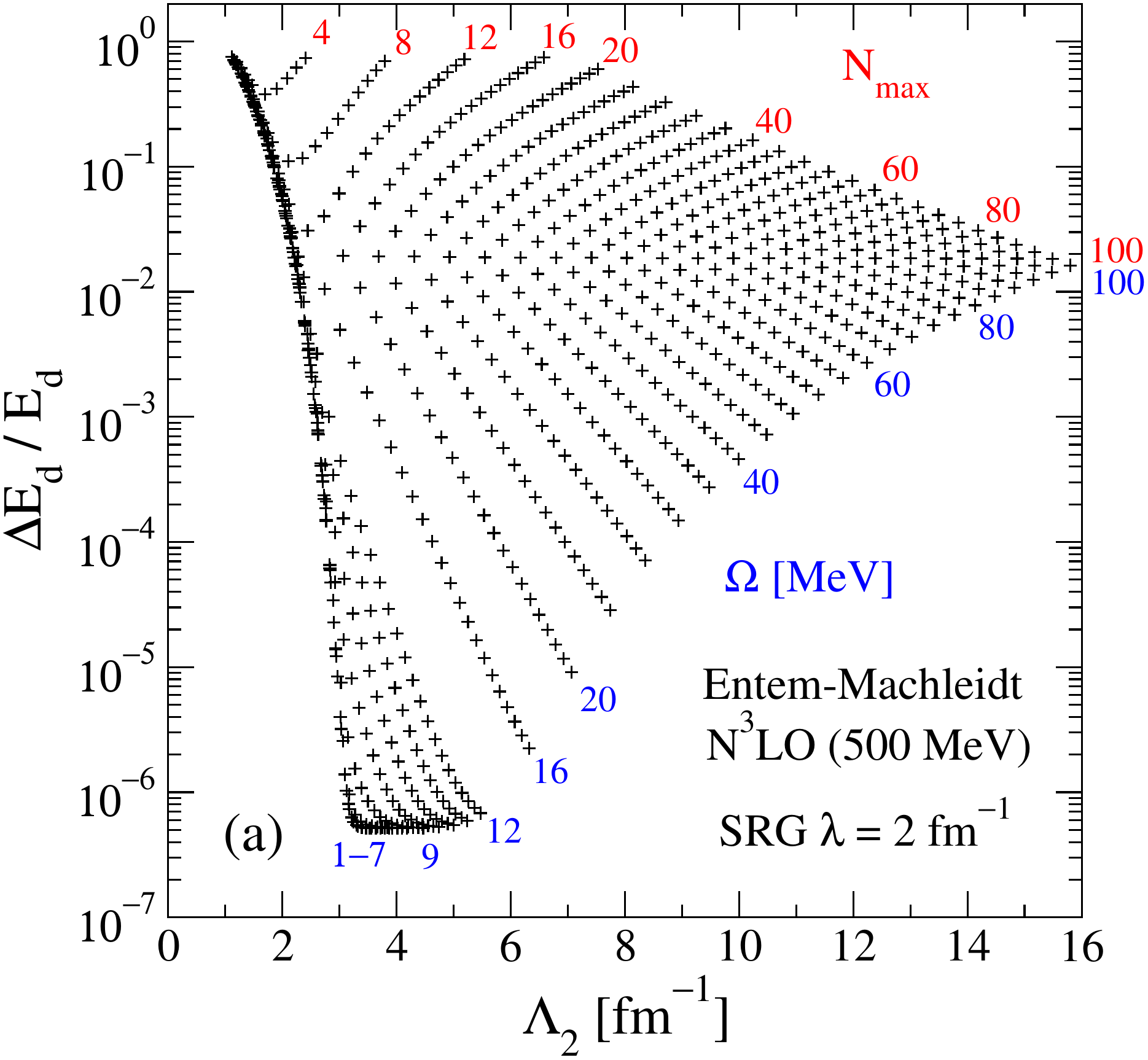}~~~%
 \includegraphics[width=0.906\columnwidth]{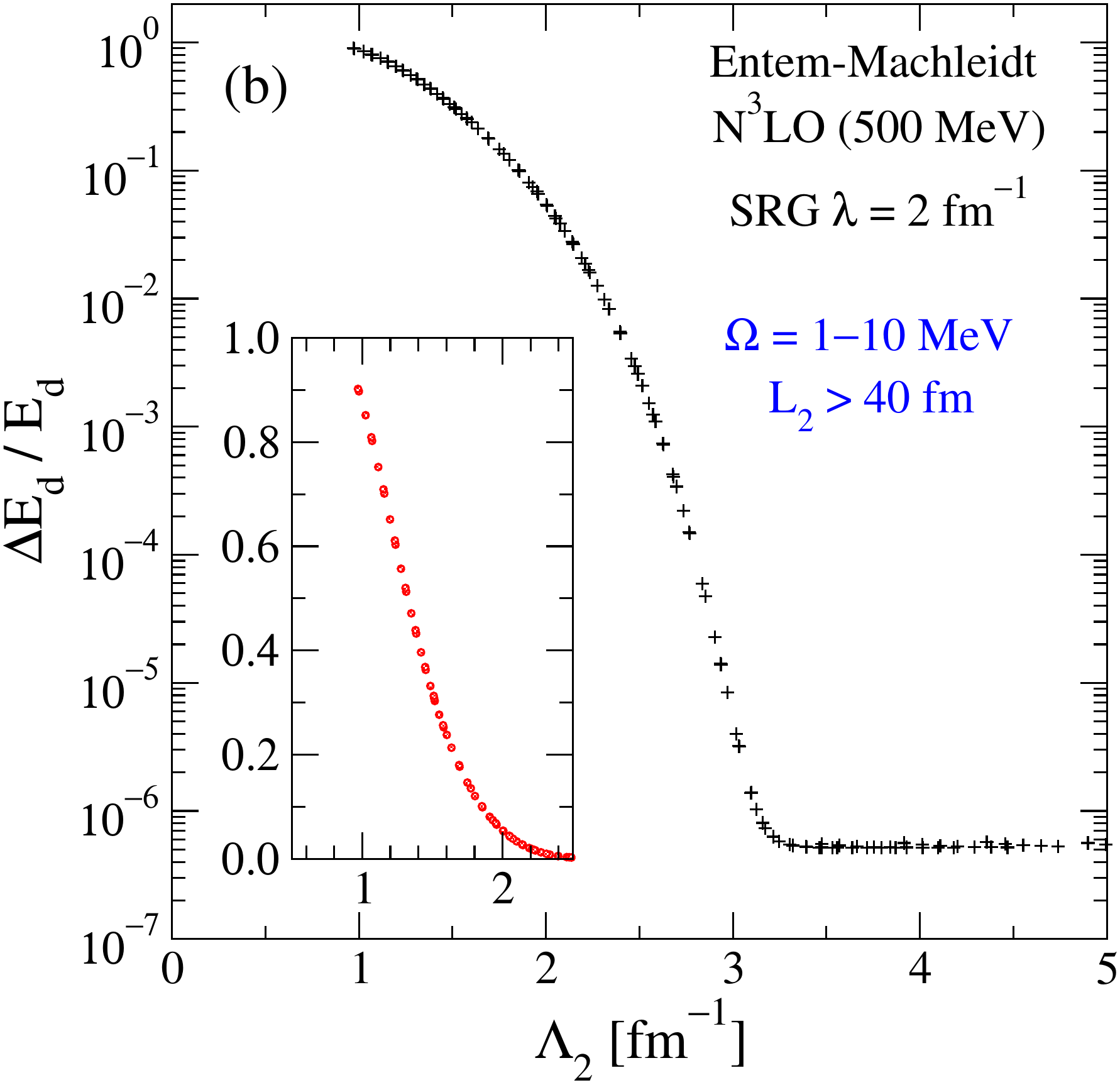}
 \caption{(Color online) (a) Oscillator calculations of the relative
   error in the deuteron energy for a wide range of oscillator
   parameters $\Nmax$ and $\hw$ as a function of
   $\Lambda_2(\Nmax,\hw)$.  These are the same calculations as in
   Fig.~\ref{fig:IR_deuteron_HO}.  (b) Subset of calculations from (a)
   for which the IR correction can be neglected compared to the UV
   correction.  Inset: linear plot.  }
\label{fig:UV_deuteron_HO}
\end{figure*}
%%%%%%%%%%%%%%%%%%%%%%%%%%%%%%%%%%%%%%%%%%%%%%%%%%%%%%%%%%%%%%%%%%%%%%%%%%%%%%%%

\subsection{Isolating UV corrections}

%%%%%%%%%%%%%%%%%%%%%%%%%%%%%%%%%%%%%%%%%%%%%%%%%%%%%%%%%%%%%%%%%%%%%%%%%%%%%%%%
\begin{figure}[thbp]
 \centering
 \includegraphics[width=0.95\columnwidth]%
 {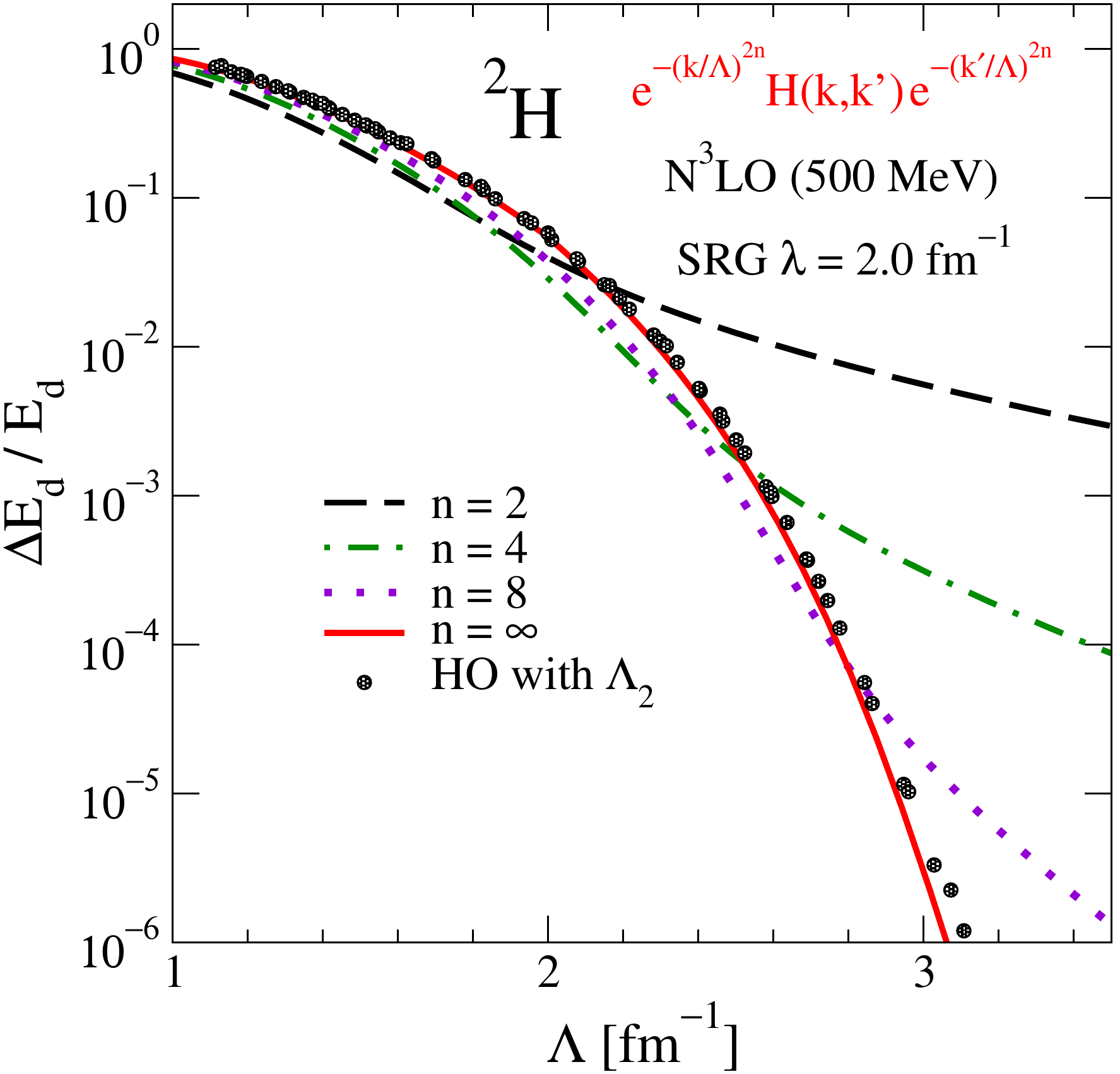}
 \caption{(Color online) Calculations of the relative error in the
   deuteron energy as a function of $\Lambda_2(\Nmax,\hw)$.  Circles 
   represent a wide range of oscillator parameters $\Nmax$ and
   $\hw$ that are IR converged.  The series of lines shows energies for
   which the Hamiltonian has been smoothly cutoff with exponent $n$.
   The solid line corresponds to a sharp cutoff.}
\label{fig:deuteron_vs_cutLam}
\end{figure}
%%%%%%%%%%%%%%%%%%%%%%%%%%%%%%%%%%%%%%%%%%%%%%%%%%%%%%%%%%%%%%%%%%%%%%%%%%%%%%%%

For an oscillator basis truncation with general $b$ and $\Nmax$, both
IR and UV errors will be significant.  However, by choosing
appropriate ranges of these parameters we can isolate one or the other
and thereby analyze them separately (with the combined effect to be
considered in future work).  In Fig.~\ref{fig:IR_deuteron_HO}(a) we
plot the relative error in the deuteron energy $\Delta E_d/E_d$ for a
large set of basis parameters with $4 < \Nmax < 100$ and
$1\,\mbox{MeV} < \Omega < 100\,\mbox{MeV}$ against the value of $L_2$
(recall $\Omega = 1/\mu b^2$) from Eq.~\eqref{eq:L2_def}.  The
calculations use the 500\,MeV N$^3$LO nucleon-nucleon $NN$ potential
of \citeauthor{Entem:2003ft}~\cite{Entem:2003ft}, evolved by the
SRG~\cite{Bogner:2006pc} to
$\lambda = 2\,\fmi$.  For sufficiently large $\Omega$, above a minimum
$\Nmax$ all points collapse to a single exponential curve that runs
over six decades (at which point numerical errors in the calculation
are reached and cause the curve to flatten)\footnote{This effect is analogous to 
what is shown in Fig.~4 in Ref.~\cite{Coon:2012ab}.  Once the calculation is 
converged in the UV regime, the curves in such error plots flatten out at a 
value determined by whatever else limits the precision of the calculation.  
In Fig.~4 in Ref.~\cite{Coon:2012ab}, the value of the plateau is different for 
each curve because the data points have not been filtered to ensure convergence 
in the IR.  In our case, the value is determined by the numerical precision of 
the calculation, which is reflected in the fact that the plateaus are the
same in Figs.~\ref{fig:IR_deuteron_HO} and \ref{fig:UV_deuteron_HO}.}). 
These are the UV-converged points; that is, those for which the UV correction is
much smaller than the IR correction. 

In Fig.~\ref{fig:IR_deuteron_HO}(b) these same UV-converged points are
plotted (labeled ``raw''). They are seen to form a smooth line
with little spread; this is a signature that $L_2$ is the correct
variable for the effective box size~\cite{More:2013rma,Furnstahl:2013vda} (if 
$L_0$ were used instead there would be a small but noticeable scatter).  It is 
also evident from the straightness of the line on a semi-log plot that the
functional form is dominantly an exponential over most of the range of
$\Delta E_d/E_d$.  This exponential is predicted by the systematic
expansion derived 
in Refs.~\cite{Furnstahl:2012qg,More:2013rma,Furnstahl:2013vda}, for
which successive orders are suppressed by powers of $\ee^{-2\kinf L_2}$,
where $\kinf$ is the deuteron binding momentum.  (There are also
pre-factors that are low-order polynomials in $L_2$.)  If we subtract
the leading correction, the result is the steeper exponential
(proportional to $\ee^{-4\kinf L_2}$) labeled ``LO.''  Finally, if we
subtract the NLO correction for only the S-wave part, we get the still steeper 
exponential (``NLO S-wave''), which is valid down to $10^{-5}$.  Thus we 
conclude that the IR corrections are well understood for the deuteron.  What is 
not evident from these plots alone, but is documented in 
Ref.~\cite{More:2013rma}, is that the same results in 
Fig.~\ref{fig:IR_deuteron_HO}(b) would be obtained with another potential as 
long as it was S-matrix equivalent at low energies (same phase shifts and 
deuteron properties, as from a unitary transformation).  In this sense, the IR 
corrections are universal.

Next we try in Fig.~\ref{fig:UV_deuteron_HO}(a) to isolate the
IR-converged points with an analogous plot of the relative error in
the deuteron energy but now as a function of $\Lambda_2$. There is a
much greater spread of points, indicating that it is more difficult to
have the IR error much smaller than the UV error, at least for a
conventional range of $\Omega$.  However, for very low $\Omega$ we do
find points collapsed to a single curve.  These points, for which $L_2
> 40\,$fm (to reach IR errors smaller than UV errors), are plotted in
Fig.~\ref{fig:UV_deuteron_HO}(b).  Just as in the case of isolated
IR corrections, we find that a signature both of IR convergence
and that $\Lambda_2$ is the appropriate variable is a smooth curve
with little scatter of points.  But the functional dependence is
manifestly \emph{not} dual: there are no straight-line segments in a
semi-log plot.  The phenomenological treatment of the UV correction
suggested in Ref.~\cite{Furnstahl:2012qg} for SRG-evolved potentials
used an ansatz for which $\Delta E_d/E_d \propto \ee^{-b_1\Lambda_2^2}$ 
(although $\Lambda_0$ instead of $\Lambda_2$ was
actually used in~\cite{Furnstahl:2012qg}, this difference is not
significant for the present discussion).  As we demonstrate in
Sec.~\ref{subsec:gaussian}, this form works for a limited range in
$\Lambda_2$ but is not generally applicable.

\begin{figure}[thbp]
 \centering
 \includegraphics[width=0.95\columnwidth]{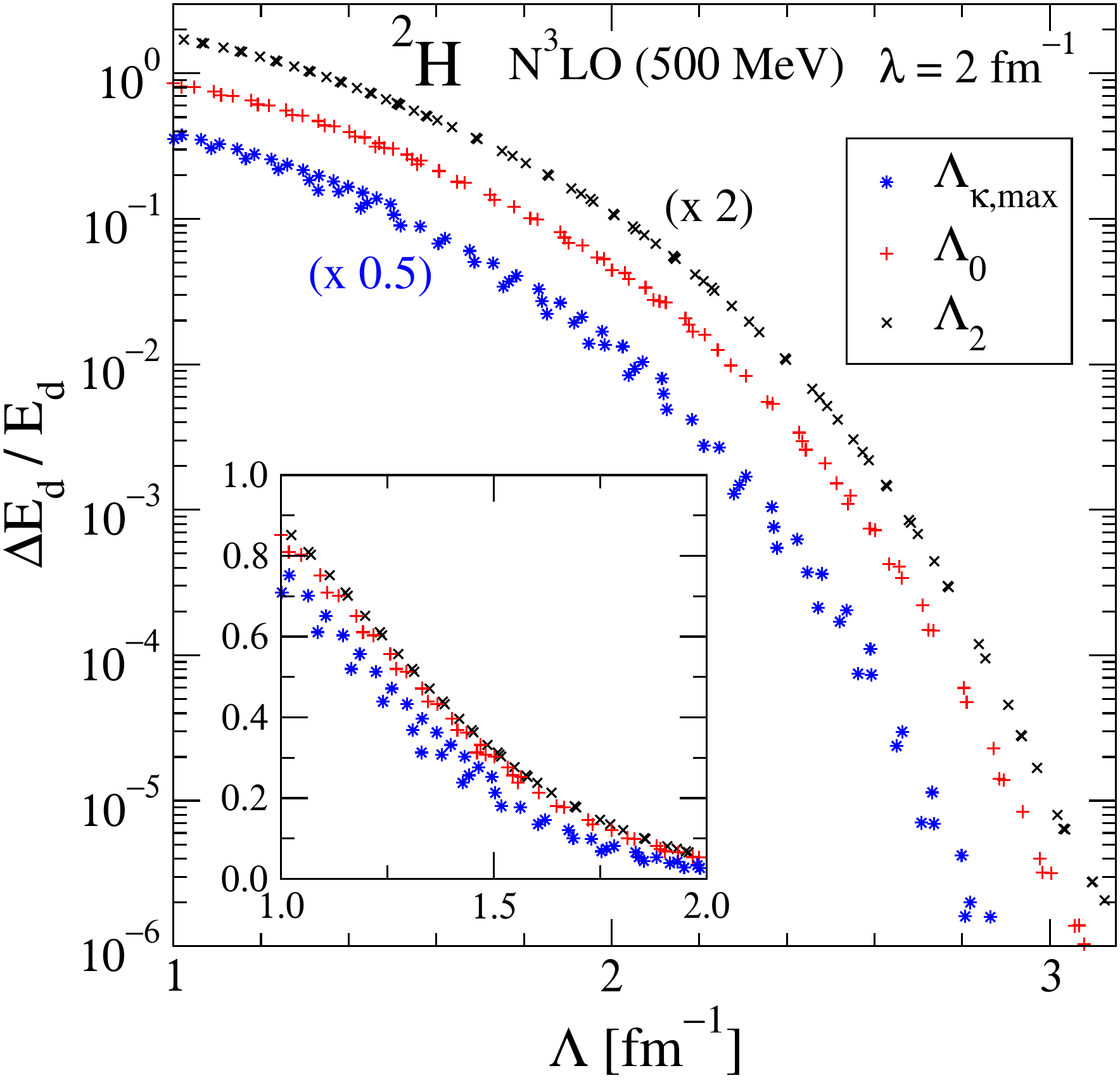}
 \caption{(Color online) Relative error of deuteron binding energy
   plotted vs. lengths $\Lambda_2$, $\Lambda_0$, and $\Lambda_{\kappa,
   {\rm max}}$ (multiplied by factors 2, 1, and $1/2$, respectively,
   to separate the curves.  Inset: The same values on a linear
   scale and without the separation factors.}
\label{fig:deuteron_lambdas}
\end{figure}

To develop a theoretical understanding of UV corrections, we first
validate the claim that the error from oscillator basis truncation is
well reproduced by applying instead a sharp cutoff in momentum at
$\Lambda_2$.  In Fig.~\ref{fig:deuteron_vs_cutLam}, the calculations
from Fig.~\ref{fig:UV_deuteron_HO}(b) are plotted as a function of
$\Lambda = \Lambda_2(\Nmax,\Omega)$ along with several other functions
of $\Lambda$ given by the relative error from the same Hamiltonian,
but now smoothly cut off as
\begin{equation}
 H_{\rm cut}(k,k') = \ee^{-(k^2/\Lambda^2)^n} H(k,k')\ee^{-(k'{}^2/\Lambda^2)^n}
 \,,
\end{equation}
for $n=2,4,8$ and $\infty$.  The latter corresponds to a sharp cutoff.
We find that the curve from a sharp cutoff tracks the
truncated-oscillator points through many orders of magnitude.  Finally,
Fig.~\ref{fig:deuteron_lambdas} shows the relative error when plotted
against three cutoff variables, $\Lambda_2$, $\Lambda_0$, and 
$\Lambda_{\kappa_{\rm max}}$.  The latter is defined as the square root
of the largest eigenvalue of the squared momentum operator in the finite
oscillator basis, which one might naively expect to be a natural choice.  
However, of the cases considered this actually gives the largest scatter
in data.  From the fact that we get an essentially
smooth curve  only for $\Lambda_2$, we conclude that this identification of the 
relevant UV cutoff is correct.

%%%%%%%%%%%%%%%%%%%%%%%%%%%%%%%%%%%%%%%%%%%%%%%%%%%%%%%%%%%%%%%%%%%%%%%%%%%%%%%%
%%%%%%%%%%%%%%%%%%%%%%%%%%%%%%%%%%%%%%%%%%%%%%%%%%%%%%%%%%%%%%%%%%%%%%%%%%%%%%%%

In the next section we take this correspondence as given and study a model
Hamiltonian for which we can analyze the UV correction exactly.

\section{Separable two-body interactions}
\label{sec:Separable}

In this section we show that the UV error from oscillator basis
truncations in the two-body problem can be determined exactly for any
rank-one separable interaction by applying the effective
sharp-momentum cutoff.  We demonstrate that, unlike the case for IR
corrections, the UV corrections depend on the high-momentum behavior
of the potential.  We derive an explicit correction formula for
separable potentials and then adapt the approach to more general
potentials, which leads to a practical fitting procedure.

\medskip

\subsection{Regularized contact interaction}
\label{sec:RegContact}

Let us consider two particles interacting via an S-wave ($\ell=0$) rank-one 
separable interaction of the form
\begin{equation}
 V(\veck,\veck') = V(k',k) = a\,\flam(k') \flam(k) \,.
\label{eq:pot-RegContact}
\end{equation}
While most of the following derivation is actually more general, we
will consider below the case of non-negative, dimensionless regulator
functions $\flam(k)$ that satisfy $\flam(0)=1$ and $\flam\to 0$ for
$k/\lambda\to\infty$.  The potential~\eqref{eq:pot-RegContact} is then
just a regularized contact interaction as it would arise, for example,
from a low-energy effective field theory, and the coupling constant
$a$ is a length scale related (up to some rescaling factors) to an
S-wave scattering length.  For convenience, in this section we work in
units with $2\mu=1$, where $\mu$ is the reduced mass of the
two-particle system.  We focus on the single bound state (assuming that $a$
is negative and large enough) with energy $\Einf \equiv -\kapinf^2$
and momentum-space wavefunction $\phi(k)$.\footnote{$\phi(k)$ is the
full three-dimensional wavefunction of the state, but it only
depends on $k=|\veck|$ due to the S-wave nature of the state, and we
have absorbed the constant factor $Y_{00}=1/\sqrt{4\pi}$ into the
definition of $\phi$.}  Thus $\kappa_\infty$ is the binding
momentum.  The Schr{\"o}dinger equation for $\phi(k)$ is
\begin{equation}
 k^2 \phi(k) + a  \flam(k) \int\! \dd^3k'\, \flam(k')\phi(k') 
 = -\kappa_\infty^2 \phi(k) 
 \,.
\label{eq:haminfty}
\end{equation}

\subsubsection{Exact extrapolation formula}

Let us assume now that we are in a limited model space with an
effective sharp momentum cutoff $\Lambda$.  In Sec.~\ref{sec:Lambda}
we have illustrated how this cutoff is related to the truncation
parameter of a finite HO basis; below we use a
model interaction to further demonstrate the result
$\Lambda=\Lambda_2$ numerically.

Given $\Lambda$ and defining $\phi_\Lambda(k) \equiv \phi(k)\Theta(\Lambda-k)$,
Eq.~\eqref{eq:haminfty} becomes
\begin{multline}
 k^2\,\phi_\Lambda(k) 
 + a
 \,\flam(k)\,\Theta(\Lambda-k) \int\!\dd^3k'\, 
 \flam(k')\,\phi_\Lambda(k') \\
 = -\kappa_\Lambda^2 \phi_\Lambda(k)
 \,.
\label{eq:hamlambda}
\end{multline}
Here, $\Theta$ denotes the unit step function, so $\Theta(\Lambda-k)$ is a 
projector, and Eq.~\eqref{eq:hamlambda} is obtained by simply introducing such 
a projector for each momentum dependence.  To indicate the cutoff dependence of 
the energy eigenvalue, we now write it as $-\kappa_\Lambda^2$.  Note that 
Eq.~\eqref{eq:hamlambda} turns into Eq.~\eqref{eq:haminfty} for 
$\Lambda\to\infty$.  We solve Eq.~\eqref{eq:hamlambda} for $\phi_\Lambda(k)$ 
and find (for $k < \Lambda$)
\begin{equation}
 \phi_\Lambda(k) = \frac{c_\Lambda \flam(k)}{\kappa^2_\Lambda +k^2} 
 \,,
\label{eq:phi}
\end{equation}
where
\begin{equation}
 c_\Lambda \equiv -a \int\! \dd^3k\, \flam(k)\,\phi_\Lambda(k) 
\label{eq:cLambda}
\end{equation}
is independent of $k$.  Thus, we know the full momentum dependence of
$\phi_\Lambda$ from Eq.~\eqref{eq:phi}.  The cutoff does not imply
that $\phi_\Lambda(k)$ goes smoothly to $0$ at $k=\Lambda$, unlike
the behavior of a coordinate-space wavefunction with a hard-wall
boundary condition, because the momentum-space potential is nonlocal.

For determination of the eigenvalue $\kappa_\Lambda$ we insert the 
solution~\eqref{eq:phi} into Eq.~\eqref{eq:hamlambda}---or just 
substitute~\eqref{eq:phi} into~\eqref{eq:cLambda} and cancel the common factor 
$c_\Lambda$---to find the quantization condition
\begin{equation}
 {-}1 = a \int\! \dd^3k\, 
 \frac{f^2_\lambda(k)\Theta(\Lambda-k)}{\kappa^2_\Lambda+k^2}
 = 4\pi a \int_0^\Lambda\!\dd k\,
 \frac{k^2\,f^2_\lambda(k)}{\kappa^2_\Lambda+k^2}
 \,,
\label{eq:quant}
\end{equation}
which is straightforward to solve numerically.  Note that
Eq.~\eqref{eq:quant} implies that there is at most one bound state, as
we have assumed.  Note also that the quantized solution
$\kappa^2_\Lambda$ \emph{increases} to $\kapinf^2$ as $\Lambda$
approaches $\infty$.

To derive an analytic formula for the dependence of $\kappa_\Lambda$
on $\Lambda$, we start by defining (recall that we set $\hbar=2\mu=1$)
\begin{equation}
 \Delta E_{\Lambda} \equiv E_\Lambda - \Einf =
 \kapinf^2 - \kappa_\Lambda^2
 \,.
\end{equation}
Inserting this into Eq.~\eqref{eq:quant} and Taylor-expanding to first order in
$\Delta E_{\Lambda}/\kappa_\infty^2$, we find
\begin{align}
 -1 &= a \int\! \dd^3k\, 
 \frac{f^2_\lambda(k)\Theta(\Lambda-k)}
 {\kappa^2_\infty+k^2-\Delta E_{\Lambda}} \nonumber \\
 &\approx 
 a\int\! \dd^3 k\, 
 \frac{f^2_\lambda(k)\Theta(\Lambda-k)}{\kappa^2_\infty+k^2}
 \left(1+{\frac{\Delta E_{\Lambda}}{\kappa_\infty^2+k^2}}\right) \nonumber \\
 &= -1 - a \int\! \dd^3k\, 
 \frac{f^2_\lambda(k)\Theta(k-\Lambda)}{\kappa^2_\infty+k^2} \nonumber \\
 & \qquad\null + \Delta E_{\Lambda} \, a\int\! \dd^3k\, 
 \frac{f^2_\lambda(k)\Theta(\Lambda-k)}{(\kappa^2_\infty+k^2)^2} 
 \,.
\end{align}
In the second step here we have employed Eq.~\eqref{eq:quant} for 
$\Lambda=\infty$, also using $\Theta(\Lambda-k) = 1 - \Theta(k-\Lambda)$.
Thus, the general result for $\Delta E_{\Lambda}$ is
\begin{equation}
 \Delta E_{\Lambda} \approx 
 \dfrac{\displaystyle\int\!\dd^3 k\, 
     \dfrac{f^2_\lambda(k)\Theta(k-\Lambda)}{\kapinf^2 + k^2}
 }
 {\displaystyle\int\!\dd^3 k\, 
     \dfrac{f^2_\lambda(k)\Theta(\Lambda-k)}{(\kappa^2_\infty+k^2)^2}}
 \,.
\label{eq:deltaE}
\end{equation}
This should be a quantitatively accurate expression in those regions
of $\Lambda$ for which $\Delta E_{\Lambda} \le \kappa_\infty^2$.

We can further approximate the result by dropping terms of 
$\mathcal{O}(\kapinf^2/\Lambda^2)$, noting that this may not be a good 
quantitative approximation when $\Lambda \approx \lambda$:
\begin{eqnarray}
 \Delta E_{\Lambda} &=&
 \dfrac{\displaystyle\int\!\dd^3 k\, 
   \dfrac{f^2_\lambda(k)\Theta(k-\Lambda)}{k^2}}
 {\displaystyle\int\! \dd^3 k\, 
   \dfrac{f^2_\lambda(k)\Theta(\Lambda-k)}{(\kappa^2_\infty+k^2)^2}}
   \Big[1+ {\mathcal O}(\kappa_\infty^2/\Lambda^2)\Big]
   \nonumber\\
 &\approx&
   \dfrac{\displaystyle\int\!\dd^3 k\, 
     \dfrac{f^2_\lambda(k)\Theta(k-\Lambda)}{k^2}}
 {\displaystyle\int\!\dd^3 k\, 
    \dfrac{f^2_\lambda(k)}{(\kappa^2_\infty+k^2)^2}} 
    \Big[1+ {\mathcal O}(\kappa_\infty^2/\Lambda^2)\Big] 
 \,.
\label{eq:deltaEapprox}
\end{eqnarray}
In the last step, we extended the integration in the denominator from
$\Lambda$ to $\infty$.  This introduces a negligible error for
reasonable regulators when $\Lambda \gg \lambda$.  However, we may
want to keep the $\Lambda$-dependence in the denominator more general.

If we make all these approximations, then the $\Lambda$-dependence of 
$\Delta E_{\Lambda}$ is simple, and we have
\begin{equation}
 \Delta E_{\Lambda} \propto \int\!\dd k\, f^2_\lambda(k)\Theta(k-\Lambda) 
 = \int\limits_\Lambda^\infty \dd k\, f^2_\lambda(k) \,.
\label{eq:final}
\end{equation}
Thus, the cutoff dependence is determined entirely by the regulator,
while the low-energy length scale $a$ has dropped out in this
approximation and only appears in weaker approximations through
$\kapinf$.  In other words, the energy correction will depend strongly
on the details of how the potential falls off at high momentum, but
only weakly on the strength of the potential.

\subsubsection{Perturbation theory}
\label{subsubsec:perturbation}

Here we show that the result~\eqref{eq:deltaEapprox} for $\Delta
E_{\Lambda}$ in the region $\Lambda > \lambda$ can also be derived
from first-order perturbation theory.  The unperturbed wavefunction is
from Eq.~\eqref{eq:phi} with $\Lambda\rightarrow\infty$:
\begin{equation}
 \phiinf(k) \equiv \langle k | \phiinf \rangle
 = \frac{\cinf \flam(k)}{\kapinf^2+k^2}
 \,,
\label{eq:phiunperturb}
\end{equation}
and the perturbation can be written  (for S-waves) as
\begin{multline}
 \delta H(k,k') =
    -\Big[
     k^2 \frac{\delta(k-k')}{4\pi kk'} \Theta(k-\Lambda) \Theta(k'-\Lambda) \\
     + a\,\flam(k')\flam(k) \big[\Theta(k-\Lambda) + \Theta(k'-\Lambda)\big]
    \Big] \,.
\label{eq:sepperturbation}
\end{multline}
In writing $\delta H(k,k')$, we have 
neglected a contribution to the second term proportional to $\Theta(k-\Lambda) 
\Theta(k'-\Lambda)$, which would be doubly suppressed by $\flam(k>\Lambda)$.

The first-order energy shift is
\begin{eqnarray}
 \Delta\ELam
 &=& \frac{\langle\phiinf|\delta H | \phiinf\rangle}
   {\langle\phiinf|\phiinf\rangle} 
 \nonumber \\
 &=& -\biggl[
  4\pi \cinf^2 \int_\Lambda^\infty\! \dd k\, k^2 \frac{k^2\flam^2(k)}
  {(\kapinf^2 + k^2)^2}
 \nonumber \\
 & &\qquad\null
  + (2a)4\pi \cinf \int_0^\infty\! \dd k'\, k'^{2} 
  \flam(k')\frac{\flam(k')}{\kapinf^2 + k'{}^2} 
  \nonumber \\
  & &\qquad\null
  \times 4\pi\cinf \int_\Lambda^\infty\! \dd k\, k^2 \flam(k)\frac{\flam(k)}
  {\kapinf^2 + k^2} 
 \biggr]
 \nonumber \\
 & &\qquad\null \times \left[4\pi \cinf^2\int_{0}^{\infty}\! \dd k\, k^2 
  \frac{\flam^2(k)}{(\kapinf^2 + k^2)^2} \right]^{-1}
 \nonumber \\  
 &=& -\left[
  \int_\Lambda^\infty\! \dd k\, \frac{k^4\flam^2(k)}{(\kapinf^2 + k^2)^2}
   - 2 \int_\Lambda^\infty\! \dd k\,  \frac{k^2\flam^2(k)}{\kapinf^2 + k^2} 
   \right] 
  \nonumber \\
  & &\qquad\null
   \times \left[\int_{0}^{\infty}\! \dd k\,
   \frac{k^2 \flam^2(k)}{(\kapinf^2 + k^2)^2} \right]^{-1}
 \nonumber \\
 &=& \left[
     \int_\Lambda^\infty\! \dd k\, \flam^2(k)
   \right]
   \times \left[\int_{0}^{\infty}\! \dd k\, \frac{k^2 \flam^2(k)}
   {(\kapinf^2 + k^2)^2} \right]^{-1}
  \nonumber \\
  & &\qquad\null
   \times \Big[1+ {\mathcal O}(\kapinf^2/\Lambda^2)\Big] 
 \,.
\end{eqnarray}
This agrees with Eq.~\eqref{eq:deltaEapprox} up to terms of order 
$\kapinf^2/\Lambda^2$.  
Note that an analogous application of first-order perturbation theory
fails if applied to the IR correction; one finds the leading
$\ee^{-2\kinf L}$ dependence, but with the wrong coefficient.

\subsubsection{Asymptotic expansion} \label{subsubsec:Asymptotic}

It is instructive to look at the large $\Lambda$ expansion of
Eq.~\eqref{eq:final} when $\flam(k)$ has the form of a Gaussian or
super-Gaussian:
\begin{equation}
  f_\lambda(k) = \ee^{-(k/\lambda)^{2n}}
  \;.
\end{equation}
We can express $\Delta\ELam$ in this case in terms of the incomplete
gamma function $\Gamma(a,z)$~\cite{Olver:2010:NHMF}:
\begin{eqnarray}
  \Delta\ELam &\propto& \int_{\Lambda}^{\infty}\! \dd 
k\,\ee^{-2(k/\lambda)^{2n}}
     \nonumber \\
  &=& \frac{\lambda}{4n}\int_{2(\Lambda/\lambda)^{2n}}^\infty \! \dd t\,
     (t/2)^{\frac{1}{2n} - 1} \ee^{-t}
     \nonumber \\
  &=& \frac{\lambda}{4n}\frac{1}{2^{\frac{1}{2n} - 1}}
  \Gamma\!\left(\frac{1}{2n},2(\Lambda/\lambda)^{2n}\right)
  \;,
\end{eqnarray}
so that for $\Lambda \gg \lambda$,
\begin{equation}
  \Delta\ELam(\Lambda) \underset{\Lambda\gg\lambda}{\longrightarrow}
   [\mbox{const.}] \times \lambda
   \left(\frac{\Lambda}{\lambda}\right)^{1-2n} \ee^{-2(\Lambda/\lambda)^{2n}}
   \;.
\end{equation}
Only for $n=1$ does this have the Gaussian form used in
phenomenological methods for extrapolation, which is further
verification of the non-universality of UV corrections.  However, the
asymptotic region where $\Lambda \gg \lambda$ is seldom reached in
practice (if it were, convergence would likely be sufficient without
extrapolation).  When $\Lambda$ is the same size as or smaller than
$\lambda$, and if the region over which a fit is made is limited, then a
Gaussian form can arise, as shown in Sec.~\ref{subsec:gaussian}.

\subsubsection{Numerical calculations}

We test the extrapolation law~\eqref{eq:final} with the specific but
arbitrary choice
\begin{equation}
 \flam(k) = \ee^{-\left({k/\lambda}\right)^4} \,,
\label{eq:regul-f4}
\end{equation}
with $\lambda=2.0$~fm$^{-1}$ and $a=-0.1$~fm.  The solution of the
quantization condition~\eqref{eq:quant} yields $\kappa_\infty\approx
0.634$~fm$^{-1}$.

%%%%%%%%%%%%%%%%%%%%%%%%%%%%%%%%%%%%%%%%%%%%%%%%%%%%%%%%%%%%%%%%%%%%%%%%%%%%%%%%
\begin{figure}[tbh-]
 \centering
 \includegraphics[width=0.95\columnwidth]{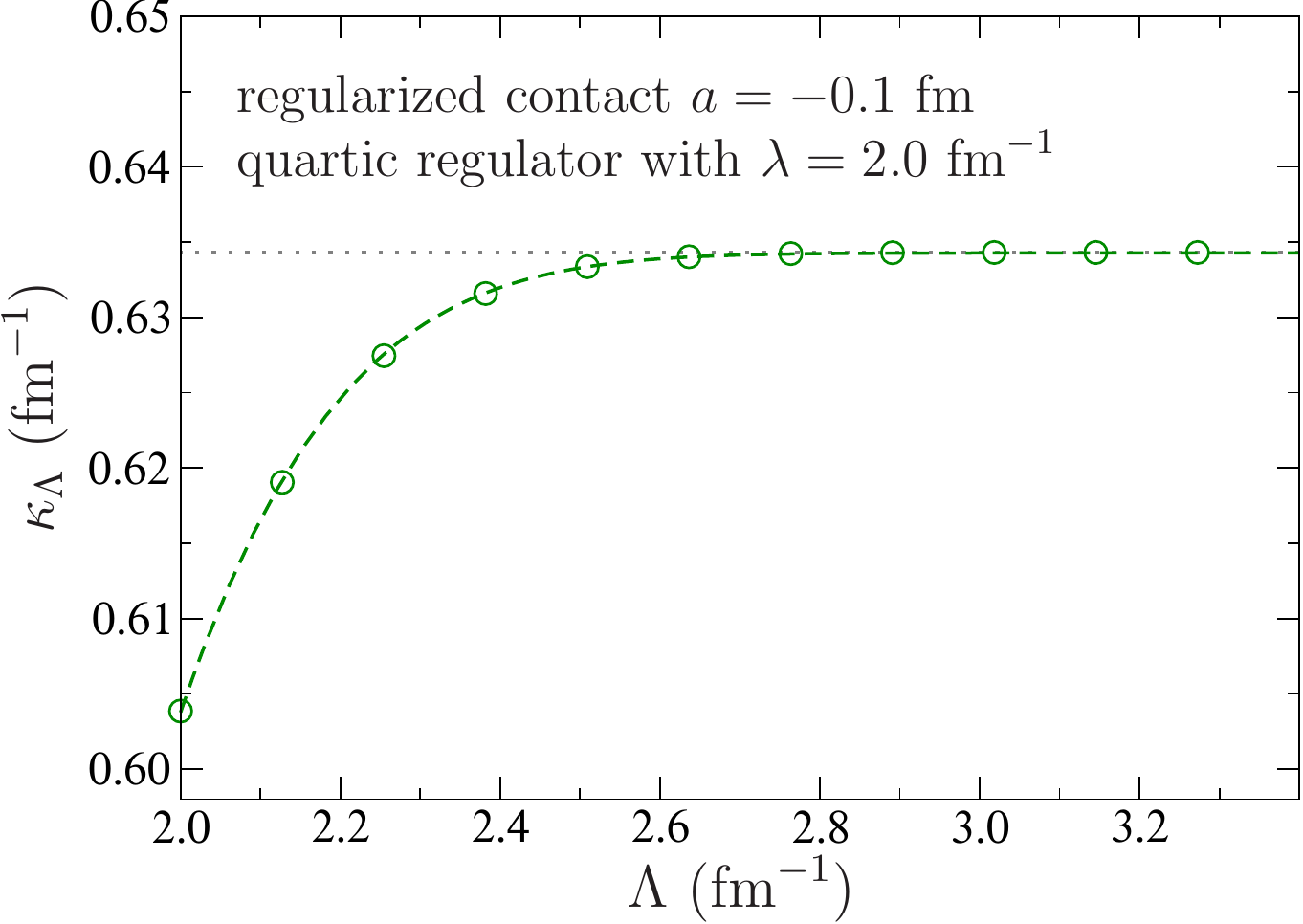}
 \caption{(Color online) Test of the extrapolation law~\eqref{eq:final} for a 
   contact $a=-0.1$~fm and the quartic regulator~\eqref{eq:regul-f4} with 
   $\lambda=2$~fm.  Points: solution of the quantization 
   condition~\eqref{eq:quant}.  Line: fit of $\kappa_\infty$ and $A$ from 
   Eq.~\eqref{eq:fit}.
 }
\label{fig:SimpleContactFit}
\end{figure}
%%%%%%%%%%%%%%%%%%%%%%%%%%%%%%%%%%%%%%%%%%%%%%%%%%%%%%%%%%%%%%%%%%%%%%%%%%%%%%%%

Figure~\ref{fig:SimpleContactFit} shows the numerical solution of the
exact quantization condition~\eqref{eq:quant} plotted as
$\kappa_\Lambda$ {\it vs.} $\Lambda$ (circles).  The line is the
extrapolation with the function~\eqref{eq:final}, \ie, we write
\begin{equation}
 \Delta\ELam = \kappa_\Lambda^2 - \kappa_\infty^2
 \approx 2\kappa_\infty \Delta\kappa_\Lambda
 \mathtext{with}
 \Delta\kappa_\Lambda = \kappa_\infty - \kappa_\Lambda
\label{eq:kappaEps}
\end{equation}
and determine $\kappa_\infty$ and the proportionality constant $A$ from a fit to
\begin{equation}
 \kappa_\Lambda
 = \kappa_\infty - \Delta\kappa
 = \kappa_\infty - A \int_\Lambda^\infty \dd k\,f^2_\lambda(k) \,.
\label{eq:fit}
\end{equation}
Note that $\kappa_\infty\ll\Lambda$, and $\Lambda>\lambda$, as
required.  Despite the approximations, the fit is very good, and in
fact the extracted value for $A$ agrees to better than 10\% with the
explicit result
\begin{equation}
 A_\infty = \left
 (2\kappa_\infty
 \times \int_0^\infty\!\dd k\,k^2 \frac{\flam(k)^2}
 {(\kappa_\infty^2+k^2)^2} \right)^{\!-1} \,,
\end{equation}
which follows directly from combining Eqs.~\eqref{eq:deltaEapprox} 
and~\eqref{eq:kappaEps}.

\noindent
This simple test already suggests that the approximations in deriving
the extrapolation law~\eqref{eq:final} are well under control.

%%%%%%%%%%%%%%%%%%%%%%%%%%%%%%%%%%%%%%%%%%%%%%%%%%%%%%%%%%%%%%%%%%%%%%%%%%%%%%%%
\begin{figure}[hbtp]
 \centering
 \includegraphics[width=0.95\columnwidth]{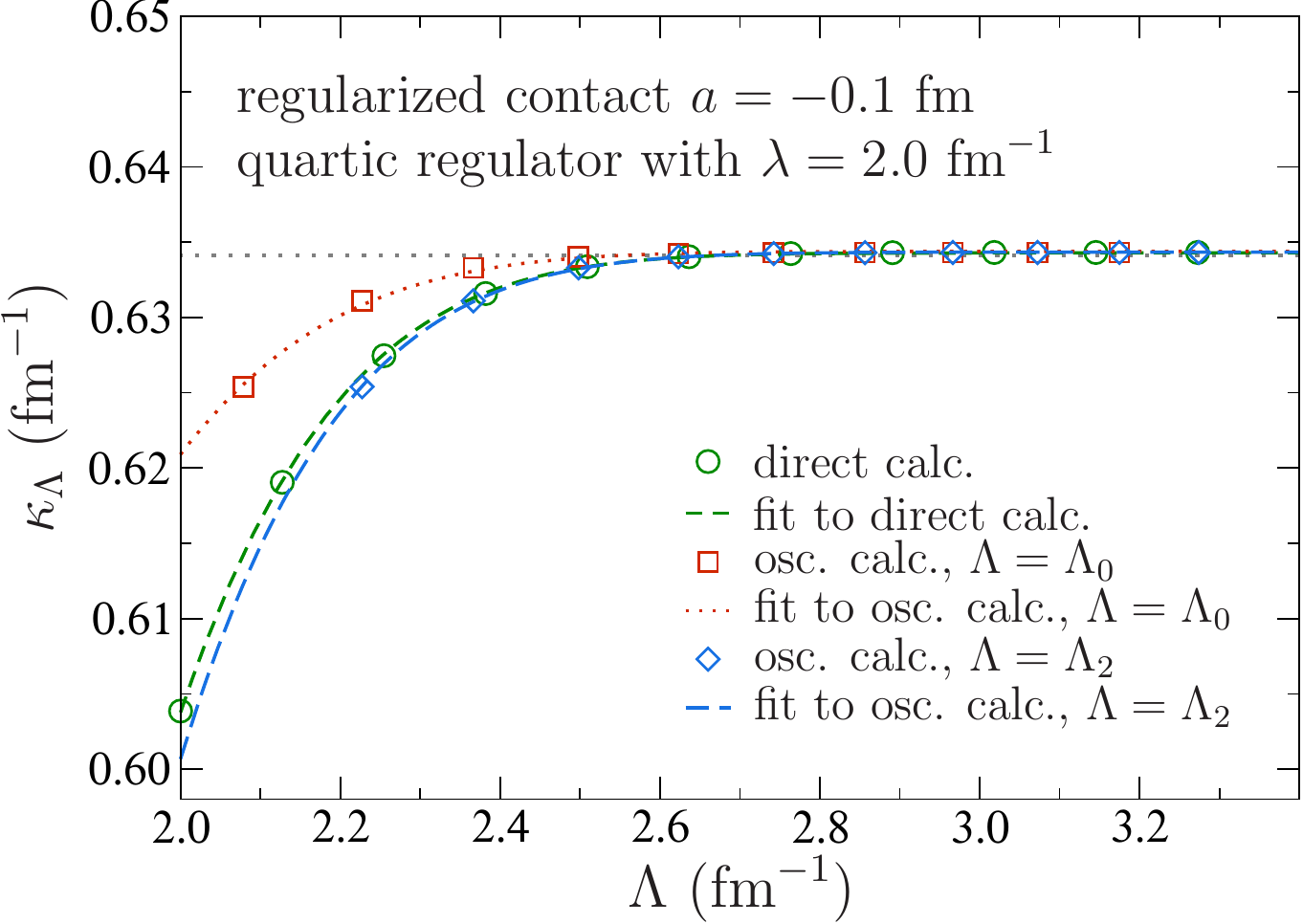}
 \caption{(Color online) Oscillator calculations (with $b=2.5$~fm and 
   $n=6,\ldots,16$) and extrapolations for a contact $a=-0.1$~fm and 
   quartic regulator~\eqref{eq:regul-f4} with $\lambda=2$~fm.
   Circles and short-dashed line: direct-quantization result and fit, as 
   in Fig.~\ref{fig:SimpleContactFit}.
   Squares: oscillator result with $\Lambda=\Lambda_0(n)$.
   Dotted line: fit of Eq.~\eqref{eq:fit} to squares.
   Diamonds: oscillator result with $\Lambda=\Lambda_2(n)$.
   Long-dashed line: fit of Eq.~\eqref{eq:fit} to diamonds.
 }
\label{fig:OscContactFit-f4}
\end{figure}
%%%%%%%%%%%%%%%%%%%%%%%%%%%%%%%%%%%%%%%%%%%%%%%%%%%%%%%%%%%%%%%%%%%%%%%%%%%%%%%%

\begin{figure}[hbtp]
 \centering
 \includegraphics[width=0.95\columnwidth]{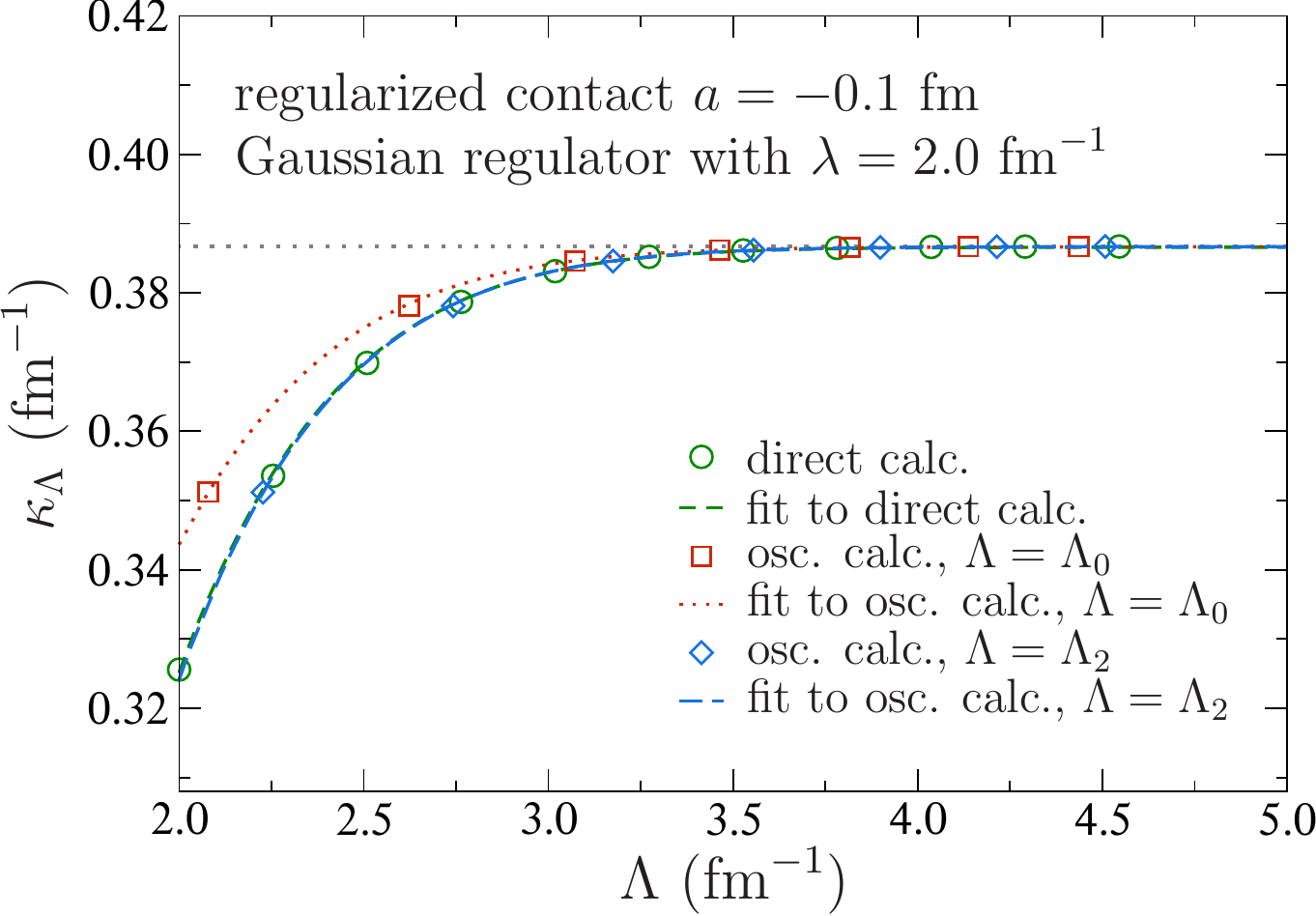}
 \caption{(Color online) Oscillator calculations (with $b=2.5$~fm and 
   $n=6,\ldots,34$) and extrapolations for a contact $a=-0.1$~fm and Gaussian
   regulator~\eqref{eq:regul-f2} with $\lambda=2$~fm.
   Symbols and curves are as in Fig.~\ref{fig:OscContactFit-f4}.
 }
\label{fig:OscContactFit-f2}
\end{figure}

Indeed, the extrapolation also works very well for calculations in
truncated oscillator bases, provided the effective UV cutoff is
calculated according to $\Lambda=\Lambda_2$ as derived in
Sec.~\ref{sec:Lambda}.  Although the overall cutoff dependence is
quite small for the simple regularized contact interaction, one can
clearly see a substantial improvement when one uses
$\Lambda=\Lambda_2$ instead of the naive estimate $\Lambda=\Lambda_0$.
As shown in Fig.~\ref{fig:OscContactFit-f4}, the difference between
the two choices is a horizontal shift of the oscillator data that
moves them almost right on top of the direct-quantization result
according to Eq.~\eqref{eq:quant}.  If instead of
Eq.~\eqref{eq:regul-f4} we use a Gaussian regulator,
\begin{equation}
 \flam(k) = \ee^{-\left({k/\lambda}\right)^2} \,,
\label{eq:regul-f2}
\end{equation}
the overall cutoff dependence is somewhat stronger, but, as shown in
Fig.~\ref{fig:OscContactFit-f2}, the qualitative behavior is exactly
the same.  In fact, the agreement is even somewhat better, at least
for the parameters chosen in the calculation.

%%%%%%%%%%%%%%%%%%%%%%%%%%%%%%%%%%%%%%%%%%%%%%%%%%%%%%%%%%%%%%%%%%%%%%%%%%%%%%%%
%%%%%%%%%%%%%%%%%%%%%%%%%%%%%%%%%%%%%%%%%%%%%%%%%%%%%%%%%%%%%%%%%%%%%%%%%%%%%%%%

\begin{figure}[hbtp]
 \centering
 \includegraphics[width=0.95\columnwidth]{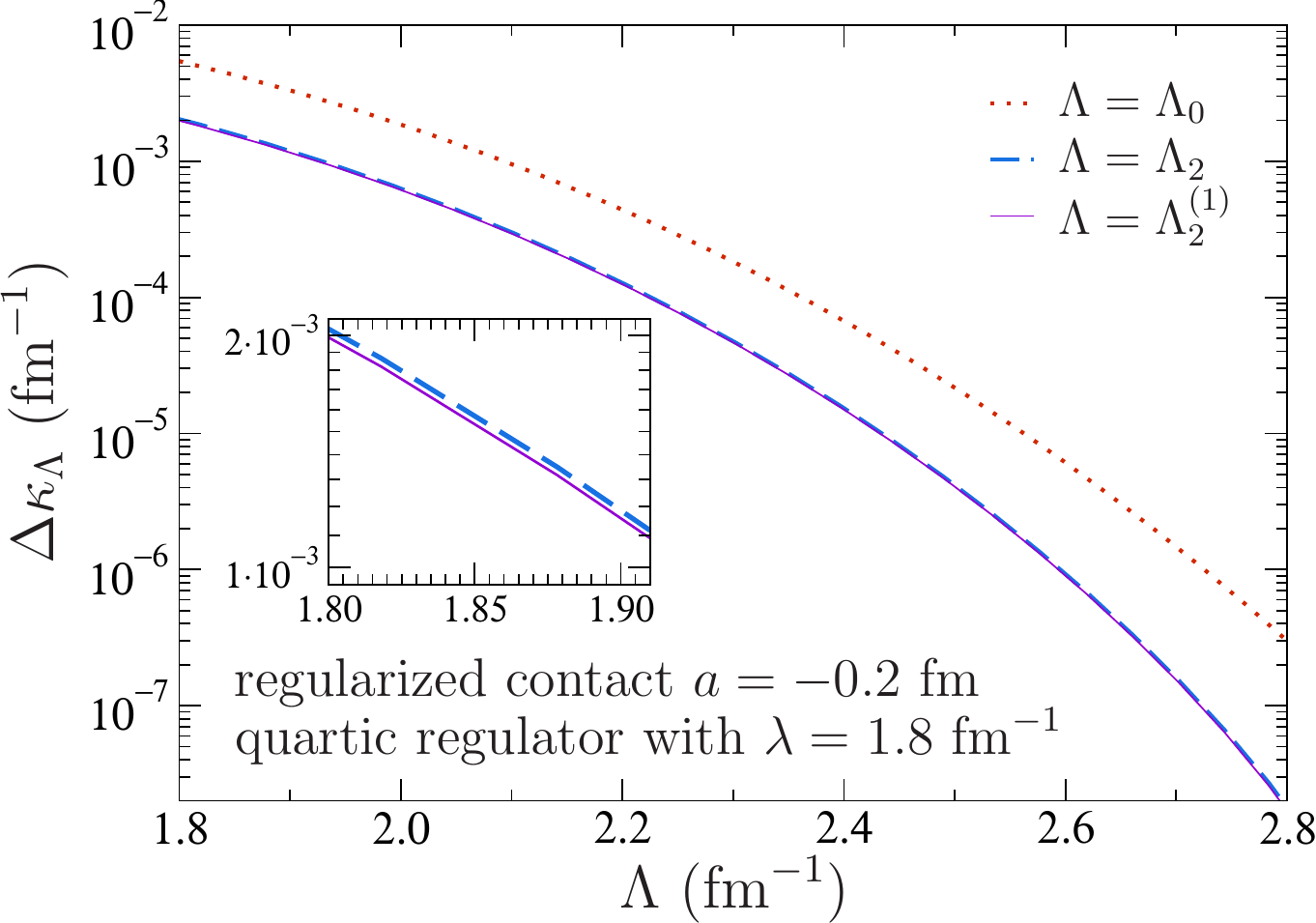}
 \caption{(Color online) Logarithmic plot of $\Delta\kappa_\Lambda$ as defined 
   in Eq.~\eqref{eq:Delta-kappa} for a contact $a=-0.2$~fm and quartic
   regulator~\eqref{eq:regul-f4} with $\lambda=1.8$~fm.  Dotted line:
   result for $\Lambda=\Lambda_0$.  Thick dashed line: result for
   $\Lambda=\Lambda_2$.  Thin dashed line: result for
   $\Lambda=\Lambda_2^{(1)}$ (including the first subleading
   correction). Inset: The small improvement from $\Lambda_2$
   to $\Lambda_2^{(1)}$.  }
\label{fig:ContactErrorPlot-f4}
\end{figure}

\begin{figure}[hbtp]
 \centering
 \includegraphics[width=0.95\columnwidth]{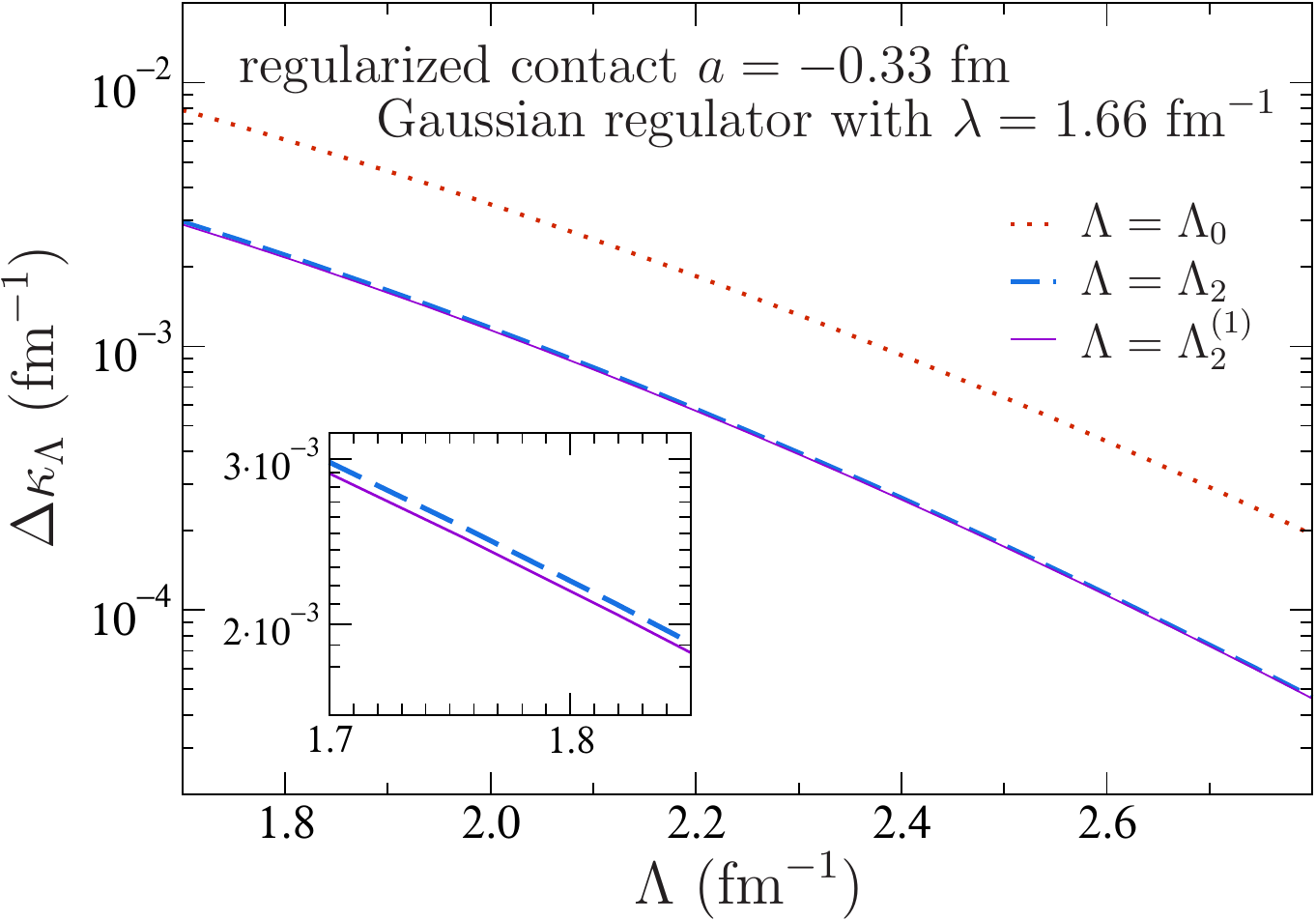}
 \caption{(Color online) Logarithmic plot of $\Delta\kappa_\Lambda$ as defined 
   in Eq.~\eqref{eq:Delta-kappa} for a contact $a=-0.33$~fm and 
   Gaussian regulator~\eqref{eq:regul-f2} with $\lambda=1.66$~fm.
   Curves and inset are as in Fig.~\ref{fig:ContactErrorPlot-f4}.
 }
\label{fig:ContactErrorPlot-f2}
\end{figure}

To get a more quantitative assessment of the agreement, in
Figs.~\ref{fig:ContactErrorPlot-f4} and~\ref{fig:ContactErrorPlot-f2}
we plot the quantity
\begin{equation}
 \Delta\kappa_\Lambda = |\kappa(m) - \kappa_{\Lambda(m)}|
\label{eq:Delta-kappa}
\end{equation}
on a logarithmic scale for different choices of $\Lambda_m$.  With this 
notation we mean that for a given truncation parameter $m$ we first calculate 
the corresponding effective cutoff $\Lambda_m$ and then apply 
Eq.~\eqref{eq:quant} to obtain the exact binding momentum for that cutoff.
In these calculations we have used a very large oscillator length $b=6.0$~fm to 
suppress IR corrections and go up to very large truncation parameters 
(up to $n=72$) to still reach sizable UV cutoffs.  For both regulators 
discussed above (quartic and Gaussian), the $\Lambda=\Lambda_2$ curve clearly 
lies below the one for $\Lambda=\Lambda_0$.

%%%%%%%%%%%%%%%%%%%%%%%%%%%%%%%%%%%%%%%%%%%%%%%%%%%%%%%%%%%%%%%%%%%%%%%%%%%%%%%%
%%%%%%%%%%%%%%%%%%%%%%%%%%%%%%%%%%%%%%%%%%%%%%%%%%%%%%%%%%%%%%%%%%%%%%%%%%%%%%%%
%%%%%%%%%%%%%%%%%%%%%%%%%%%%%%%%%%%%%%%%%%%%%%%%%%%%%%%%%%%%%%%%%%%%%%%%%%%%%%%%

In these plots we have also included the result with the first
subleading correction to $\Lambda=\Lambda_2$ (see 
Appendix~\ref{sec:Lambda-Details}).  It is reassuring to see that there is some 
small improvement (the curves for $\Lambda=\Lambda_2^{(1)}$ lie consistently 
below those for $\Lambda=\Lambda_2$), but we conclude that these corrections can
safely be neglected for all present practical purposes.

\subsection{Separable approximations}
\label{sec:Sep-Approx}

For a general rank-1 separable potential
\begin{equation}
 \hat{V}_\text{sep} = g \, \ket{\eta}\bra{\eta} \,,
\end{equation}
which in momentum space simply becomes (with $\eta(k)\equiv\braket{k}{\eta}$)
\begin{equation}
 V_\text{sep}(k,k') = g \, \eta(k)\eta(k') \,,
\end{equation}
the quantization condition~\eqref{eq:quant} can be written as
\begin{equation}
 -1 = 4\pi g 
 \int_0^\Lambda\!\dd k\, k^2 \frac{\eta(k)^2}{\kappa^2_\Lambda+k^2} \,.
\label{eq:quant-sep}
\end{equation}
This is, of course, merely a change of notation, $a \rightarrow g$ and
$\flam(k)\rightarrow\eta(k)$ compared to Eq.~\eqref{eq:quant}.  Most
interactions used in practical calculations, however, do not have this
convenient simple form (at least not in nuclear physics).  Still,
Eq.~\eqref{eq:quant-sep} can be put to some use.

Methods to obtain separable approximations for a given potential have
been known and used for quite a while (see, \eg,
Refs.~\cite{Harms:1970hd,Ernst:1973zzb,elgaroy1998} and further references
therein).  We use the technique here in its simplest form, also called
the \emph{unitary pole approximation 
(UPA)}~\cite{Ernst:1973zzb,Lovelace:1964aa}.  Assuming that for an
arbitrary potential $\hat{V}$ we know a (bound) eigenstate $\ket{\psi}$, we can 
construct a rank-1 separable approximation in momentum space by setting
\begin{equation}
 \hat{V}_\text{sep} = \frac{\hat{V}\ket{\psi}\bra{\psi}\hat{V}}
 {\mbraket{\psi}{\hat{V}}{\psi}} \,.
\label{eq:UPA-op}
\end{equation}
In other words, we have
\begin{equation}
 \eta(k) = \mbraket{k}{\hat{V}}{\psi}
\end{equation}
for the momentum-space ``form factor,'' and the coupling strength
$g=\mbraket{\psi}{\hat{V}}{\psi}$ is, of course, independent of any 
particular representation.  From Eq.~\eqref{eq:UPA-op} one immediately sees that
\begin{equation}
 \hat{V}_\text{sep}\ket{\psi} = \hat{V}\ket{\psi} \,.
\label{eq:Vsep-psi-V-psi}
\end{equation}
This means that the separable approximation is constructed in such a
way that it exactly reproduces the state $\ket{\psi}$ used for its
construction.  The potential~\eqref{eq:UPA-op} reproduces the exact
half off-shell T-matrix at the energy corresponding to the state
$\psi$, and more sophisticated approximations (separable potential of
rank $>1$) can be constructed by using more than a single
state~\cite{Ernst:1973zzb}.  Since we are only interested in
performing the UV extrapolation for a single state here, however, the
rank-1 approximation should be sufficient.  In fact, based on our
expectation that the UV extrapolation we seek should depend on
short-range/high-momentum modes of the potential and the state under
consideration, Eq.~\eqref{eq:UPA-op} looks very promising.  To assess
to what extent it actually reflects the UV behavior of a calculation
based on the \emph{original} potential, we first consider some
examples where the separable approximation can be constructed
analytically.

\begin{figure}[thbp]
 \centering
 \includegraphics[width=0.95\columnwidth]{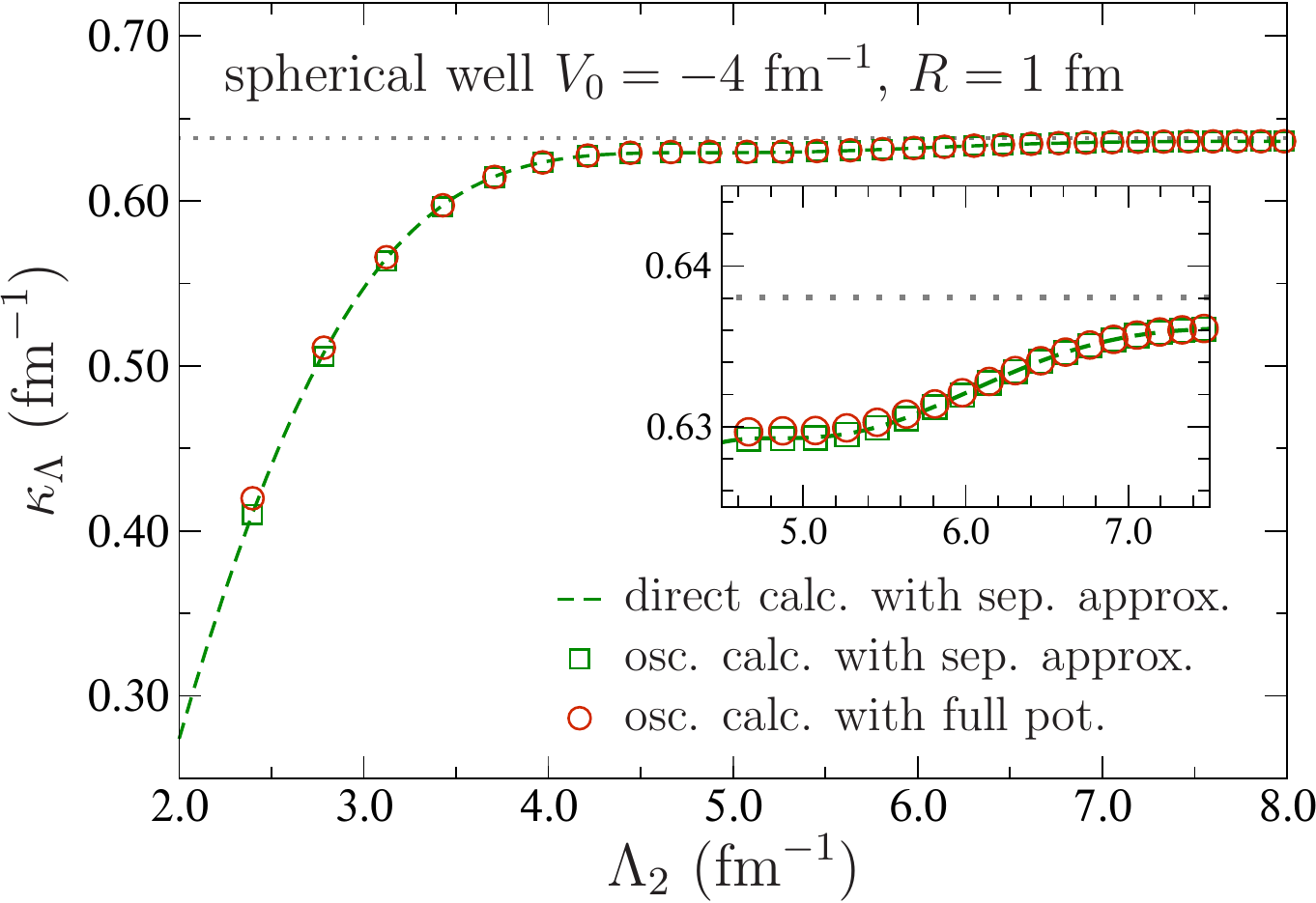}
 \caption{(Color online) Oscillator calculations (with $b=2.0$~fm and 
   $n=4,\ldots,64$) for a  spherical step potential with $V_0=-4~\fm^{-1}$ and 
   $R=1~\fm$ and its separable approximation.  Dashed line: 
   direct-quantization result according to Eqs.~\ref{eq:quant-sep} 
   and~\eqref{eq:eta-step}.  Squares: oscillator result with separable 
   approximation~\eqref{eq:eta-step}.  Circles: oscillator result with the 
   original (full) potential~\eqref{eq:V-step}.  The horizontal dotted line 
   indicates the exact result for the binding momentum.
 }
\label{fig:SepCompare-SW}
\end{figure}

\paragraph{Spherical well.} One of the simplest potentials for which the 
bound-state wavefunctions can be written down explicitly in closed form is the 
spherical well (step),
\begin{equation}
 \Vstep(r) = V_0\,\Theta(R - r) \mathtext{,} V_0 < 0 \,.
\label{eq:V-step}
\end{equation}
The eigenfunctions for this standard textbook example are spherical Bessel 
functions.  Separable approximations for these potential have been discussed in 
Ref.~\cite{Bund:1980aa}.  If $\Vstep$ supports an S-wave bound-state at energy 
$E=-\kappa^2$, we find from the results presented there that
\begin{subequations}%
\begin{align}
 \eta_{\text{step}}(k) &= \frac{2}{\pi} V_0 R^2
 \frac{Z(E;V_0,kR)}{K^2 - k^2} \,, \\
 g_{\text{step}} &= \left(\frac{2}{\pi}V_0 R^2\,F(E;V_0,R)\right)^{\!\!-1} \,,
\end{align}
\label{eq:eta-step}
\end{subequations}%
with $K \equiv K(E;V_0) = \sqrt{E-V_0}$ and
\begin{align}
 L(E;V_0,R) &= K\,\frac{j_0'(KR)}{j_0(KR)} \,, \\
 Z(E;V_0,kR) &= k\,j_0'(KR) - L(E;V_0,R)\,j_0(KR) \,, \\
 F(E;V_0,R) &= \frac{1}{2KR}
 \Bigl[R^2 L(E;V_0,R) \nonumber \\
  & \quad\null + R L(E;V_0,R) + R^2 K\Bigr] \,.
\end{align}
In Fig.~\ref{fig:SepCompare-SW} we show how the separable 
approximation~\eqref{eq:eta-step} (squares) performs compared to the original 
potential~\eqref{eq:V-step} (circles) in an oscillator calculation.  We use 
$V_0=-4~\fm^{-1}$ and $R=1~\fm$, which produces a bound state at 
$\kappa_\infty\approx0.638~\fm^{-1}$ (determined numerically from the 
quantization condition for attractive step potentials and shown as the 
dotted line in Fig.~\ref{fig:SepCompare-SW}).  The dashed line furthermore 
shows the result of a direct calculation based on Eqs.~\eqref{eq:quant-sep} 
and~\eqref{eq:eta-step} (see inset).

%%%%%%%%%%%%%%%%%%%%%%%%%%%%%%%%%%%%%%%%%%%%%%%%%%%%%%%%%%%%%%%%%%%%%%%%%%%%%%%%
%%%%%%%%%%%%%%%%%%%%%%%%%%%%%%%%%%%%%%%%%%%%%%%%%%%%%%%%%%%%%%%%%%%%%%%%%%%%%%%%

The results of all three calculations agree remarkably well.  The fact that the 
separable approximation used in the oscillator calculations follows the result 
from the direct quantization according to Eq.~\eqref{eq:quant-sep} is primarily 
reassuring.  More interestingly, the latter also traces the result of a ``full'' 
oscillator calculation based on the original step potential---including the 
rather slow convergence towards the exact result and even the peculiar step 
around $\Lambda\approx6~\fm^{-1}$ in Fig.~\ref{fig:SepCompare-SW}.

These features are due to the rather pathological (oscillatory) 
behavior of the step potential in momentum space.  In the next subsection, we 
avoid this complication by studying another class of exactly solvable 
interactions, which are smooth.

\medskip
\paragraph{Pöschl--Teller potential.}  It is convenient for us to consider a 
so-called Pöschl--Teller potential of the form
\begin{equation}
 \VPT(r) = -\frac{\alpha^2 \beta(\beta-1)}{\cosh^2(\alpha r)}
\label{eq:V-PT}
\end{equation}
Originally, this potential describes a one-dimensional problem on the interval
$(-\infty,\infty)$.  However, restricting ourselves to S-waves (and to 
states with odd wavefunctions), we can use it as a solvable problem in 
three dimensions.  For given values of $\alpha$ and $\beta$, this potential has 
an analytically known bound-state spectrum.  Labeling different states by an 
index $\nu$, we have, for example, a single bound state ($\nu=0$) with binding 
momentum $\kappa=\alpha$ for $\beta=3$.  For $\beta=5$, there are two bound 
states at $\kappa=3\alpha$ ($\nu=0$) and $\kappa=\alpha$ ($\nu=1$).  The 
wavefunctions for this potentials are known analytically as well, which allows 
us to derive explicit expressions for the form factors $\eta(k)$ as well.  
These details are given in Appendix~\ref{sec:PT-Details}.

%%%%%%%%%%%%%%%%%%%%%%%%%%%%%%%%%%%%%%%%%%%%%%%%%%%%%%%%%%%%%%%%%%%%%%%%%%%%%%%%
\begin{figure}[hbtp]
 \centering
 \includegraphics[width=0.95\columnwidth]{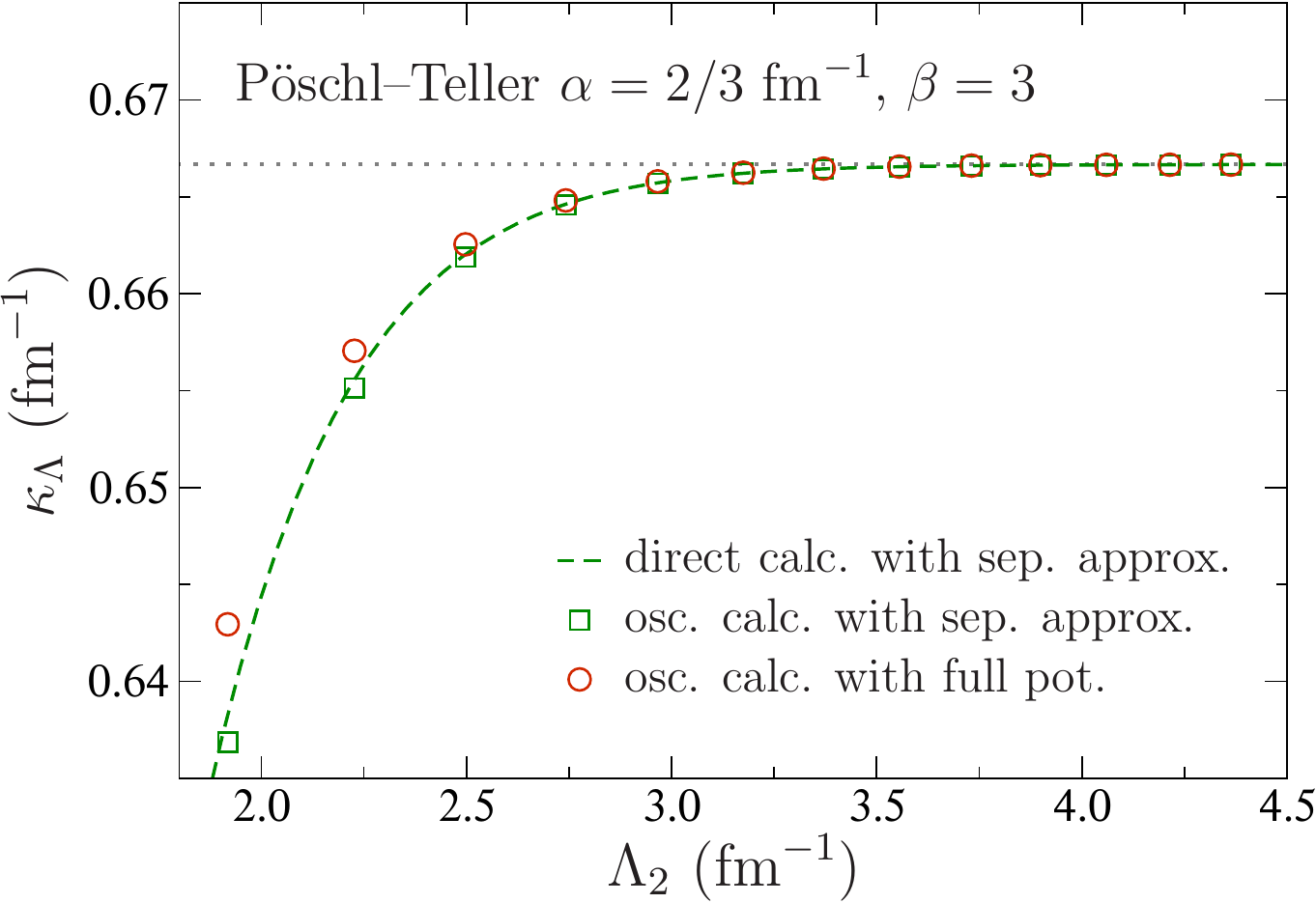}
 \caption{(Color online) Oscillator calculations (with $b=2.5$~fm and 
   $n=4,\ldots,32$) for a Pöschl--Teller potential with $\beta=3$ and 
   $\alpha=2/3~\fm^{-1}$ and its separable approximation.  Dashed line: 
   direct-quantization result according to Eqs.~\ref{eq:quant-sep} 
   and~\eqref{eq:eta-PT-3}.  Squares: oscillator result with separable 
   approximation~\eqref{eq:eta-PT-3}.  Circles: oscillator result with the 
   original (full) potential~\eqref{eq:V-PT}.  The horizontal dotted line 
   indicates the exact result for the binding momentum.
 }
\label{fig:SepCompare-PT}
\end{figure}
%%%%%%%%%%%%%%%%%%%%%%%%%%%%%%%%%%%%%%%%%%%%%%%%%%%%%%%%%%%%%%%%%%%%%%%%%%%%%%%%

%%%%%%%%%%%%%%%%%%%%%%%%%%%%%%%%%%%%%%%%%%%%%%%%%%%%%%%%%%%%%%%%%%%%%%%%%%%%%%%%
\begin{figure}[thbp]
 \centering
 \includegraphics[width=0.95\columnwidth]{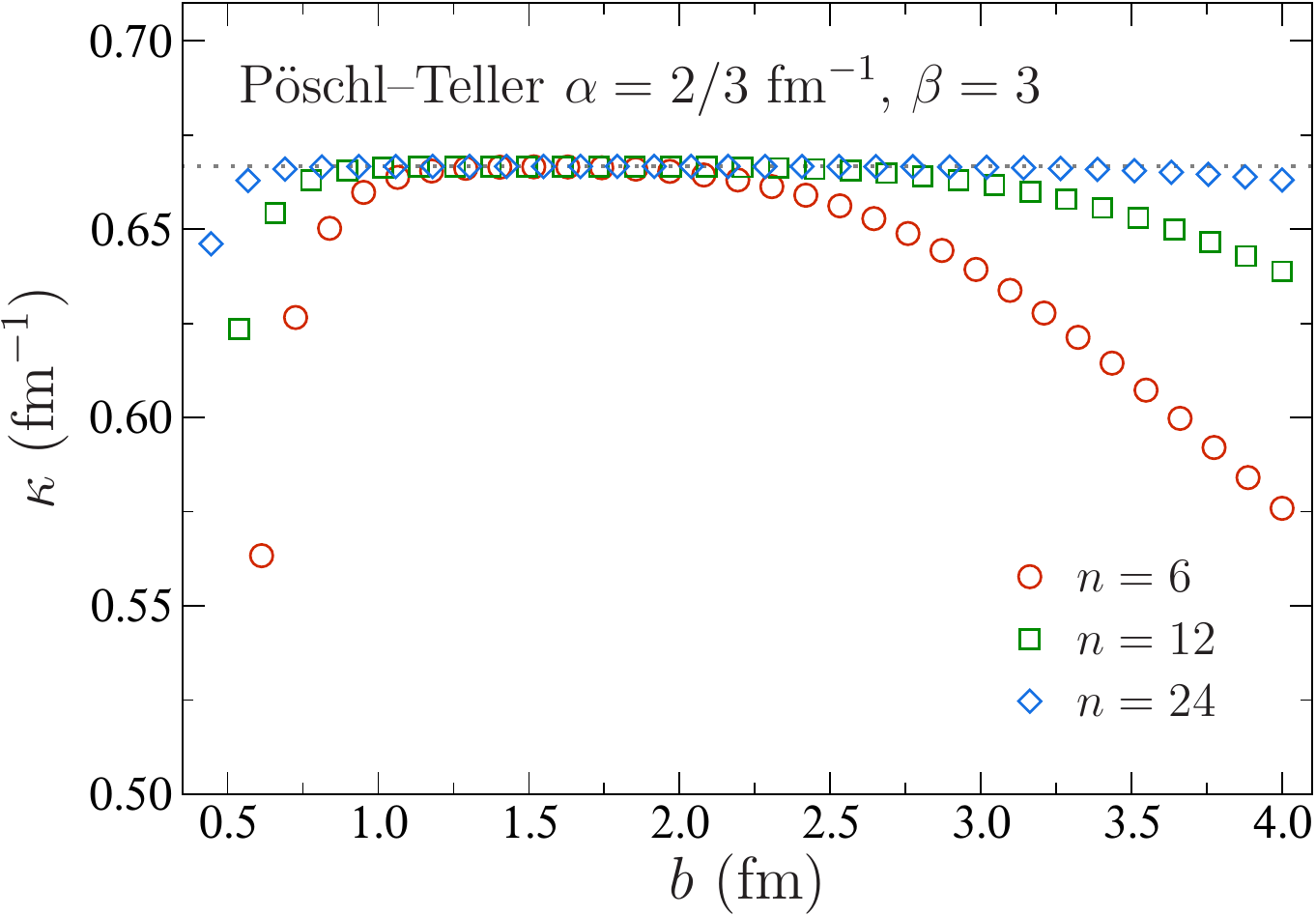}
 \caption{(Color online) Binding momentum as a function of $b$ obtained from 
   oscillator with different basis sizes.  Circles: $n=6$.  Boxes: $n=12$.  
   Diamonds: $n=24$.  The horizontal dotted line 
   indicates the exact result for the binding momentum.
 }
\label{fig:bDep-PT3-23}
\end{figure}
%%%%%%%%%%%%%%%%%%%%%%%%%%%%%%%%%%%%%%%%%%%%%%%%%%%%%%%%%%%%%%%%%%%%%%%%%%%%%%%%

In Fig.~\ref{fig:SepCompare-PT} we show results for a Pöschl--Teller potential 
with $\beta=3$ and a bound-state at $\kappa=\alpha=2/3~\fmi$.  The curves are 
analogous to those shown in Fig.~\ref{fig:SepCompare-SW} for the step 
potential.  While the agreement of the calculations with the original 
potential and with the separable approximation is not as impressive as for the 
step potential, it is still very good for cutoffs $\Lambda_2\gsim2.5~\fm^{-1}$ 
($n>8$).  

In general, the regime where UV cutoff effects dominate the energy
correction can be found from plots like the one shown in
Fig.~\ref{fig:bDep-PT3-23}, where we plot the $b$-dependence of
$\kappa$ for our P{\"o}schl--Teller potential with $\beta=3$ and
$\alpha=2/3~\fm$.  Recalling that large $b$ correspond to large
configuration-space boxes $L_2 = \sqrt{2(2n+3/2+2)}b$ and thus
negligible IR correction, we identify the UV-dominated region as the
one with $b\gsim2$.

\subsection{UV extrapolation for Pöschl-Teller potential}
\label{sec:Bootstrap}

Based on these encouraging results, we now turn to actual
extrapolations.  The simplest fit formula one can write down for that
purpose is
\begin{equation}
 \mathtext{(fit ``\fiteta'')}\mathspace
 \kappa_\Lambda = \kappa_\infty
  - A \int\nolimits_\Lambda^\infty \dd k\,\eta(k)^2 \,,
\label{eq:fit-eta-simple}
\end{equation}
which is just Eq.~\eqref{eq:fit} in a more general notation 
($\flam\rightarrow\eta$).  In the absence of an explicit scale $\lambda$ 
associated with the separable ``form factor,'' however, it is not \apriori 
clear that the various approximations made in Sec.~\ref{sec:RegContact} are 
rigorously justified.  Without any of those approximations, the most general 
fit formula---based directly on Eq.~\eqref{eq:deltaE}---is
\begin{equation}
 \mathtext{(fit ``\fitetagen'')}\mathspace
 \kappa_\Lambda = \kappa_\infty
  - A 
  \dfrac{\displaystyle\int\nolimits_\Lambda^\infty \dd k\,   
   \dfrac{k^2\,\eta(k)^2}{\kappa_\infty^2+k^2}}
  {\displaystyle\int\nolimits_0^\Lambda \dd k\,
   \dfrac{k^2\,\eta(k)^2}{(\kappa_\infty^2+k^2)^2}} \,.
\label{eq:fit-eta-gen}
\end{equation}
This is actually quite restrictive since for an exact calculation one would 
expect $A\approx1/(2\kappa_\infty)$ here, and in a fit to the binding energies 
instead of the binding momenta one should expect a prefactor $\sim 1$.  As one
more alternative, one can choose a middle ground and write
\begin{equation}
 \mathtext{(fit ``\fitetagenp'')}\mathspace
 \kappa_\Lambda = \kappa_\infty
  - A 
  \displaystyle\int\nolimits_\Lambda^\infty \dd k\,
   \dfrac{k^2\,\eta(k)^2}{\kappa_\infty^2+k^2} \,,
\label{eq:fit-eta-gen-prime}
\end{equation}
which is obtained from Eq.~\eqref{eq:fit-eta-gen} by extending the integral in 
the denominator up to infinity---rendering it independent of $\Lambda$---and 
then absorbing the whole term into the fit constant $A$.  In the next section, 
we compare the three approaches to one another and to phenomenological 
fits (Gaussian, exponential).

Of course, we are ultimately interested in performing these fits for potentials
for which we have no analytical knowledge of the wavefunctions.  Fortunately, 
the diagonalization calculation in the truncated oscillator basis does provide 
us at least with approximate wavefunctions, so it is natural to simply use the 
``best'' solution available\footnote{Typically, ``best'' would refer to the 
result from the largest available oscillator space and the most UV-converged 
(small $b$) calculation.  In practice, one could also make several choices for 
the extrapolation and take the spread in the result as a lower bound for the fit 
uncertainty.} from a set of calculations, \ie, set
\begin{equation}
 \eta(k) = \mbraket{k}{\hat{V}}{\psi}_{\text{HO, best}}
\label{eq:eta-HO}
\end{equation}
in what can be called a ``bootstrap extrapolation'' because---aside from the 
original potential---it only uses information that comes out of the numerical 
calculation.  If the interaction $\hat{V}$ is already given on a 
momentum-space mesh for the numerical calculation, Eq.~\eqref{eq:eta-HO} is 
very simple (and fast) to implement.  Using that wavefunction, one can 
perform a direct extrapolation to $\Lambda\to\infty$ by simply using the 
corresponding $\eta(k)$ in the separable quantization 
condition~\eqref{eq:quant-sep} without fitting a range of data points.  Below, 
we refer to this approach as ``\fitetadirect.''

Possible phenomenological approaches for extrapolation fits include a simple 
exponential,
\begin{equation}
 \mathtext{(fit ``$E$'')}\mathspace
 \kappa_\Lambda = \kappa_\infty
  - A \, \ee^{-B \Lambda} \,,
\label{eq:fit-exp}
\end{equation}
or a Gaussian
\begin{equation}
 \mathtext{(fit ``$G$'')}\mathspace
 \kappa_\Lambda = \kappa_\infty
  - A \, \ee^{-B \Lambda^2} \,.
\label{eq:fit-Gauss}
\end{equation}
We now investigate how well our separable extrapolations perform in comparison 
to Eqs.~\eqref{eq:fit-exp} and~\eqref{eq:fit-Gauss}.  We point out that they 
are quite a bit more constrained because they use only two fit parameters 
($\kappa_\infty$ and $A$) instead of three ($\kappa_\infty$, $A$, and $B$).  As 
described above, we follow the bootstrap procedure and take the wavefunction 
from the ``best'' numerical calculation available to construct the $\eta(k)$ 
used for the extrapolation.  Since we have analytical expressions for the 
wavefunctions, we additionally show the extrapolation results obtained with 
those.  This allows us to get at least an idea of how much influence it has on 
the extrapolation if the wavefunction is only given in a truncated basis.

%%%%%%%%%%%%%%%%%%%%%%%%%%%%%%%%%%%%%%%%%%%%%%%%%%%%%%%%%%%%%%%%%%%%%%%%%%%%%%%%
\begin{table*}[hbtp]
\caption{Comparison of different extrapolations for a Pöschl--Teller 
  potential with $\alpha=2/3$ and $\beta=3$.  For calculations where $n$ 
  is varied, it is increased in steps of $2$, and $b$ is held fixed at 
  $4.0~\fm$.  For calculations with variable $b$ (increasing in steps of 
  $0.5~\fm$), $n$ is held fixed at $12$.  The dimension of $\kappa_{\infty}$ is 
  always $\fm^{-1}$ and has been omitted in the table.  Percentage 
  values in parentheses give the relative difference, defined here as
  $100\times\left(1-|\kappa_{\infty} / \kappa_{\infty,\text{exact}}|\right)$,
  of the extrapolated values to the exact answer
  $\kappa_{\infty,\text{exact}} \approx 0.6667~\fm^{-1}$.
}
\label{tab:Compare-PT-beta3}
\centering
\begingroup
\newcommand{\skncol}{6}
\renewcommand\arraystretch{1.25}
\begin{tabular}{c|
 >{\centering\arraybackslash}p{6.2em}|
 >{\centering\arraybackslash}p{6.2em}|
 >{\centering\arraybackslash}p{7.0em}||
 >{\centering\arraybackslash}p{6.7em}|
 >{\centering\arraybackslash}p{7.0em}
}
 \hline\hline
 \multicolumn{\skncol}{l}{
   $V_\PT$ with $\alpha=2/3~\fm^{-1}$, $\beta=3$
   \ $\rightarrow$\ \;$\kappa_{\infty,\text{exact}} \approx 0.6667~\fm^{-1}$
 } \\
 \hline\hline
 Calculation
 & $n=2$--$8$ & $n=4$--$12$ & $n=6$--$16$
 & $b=4.5$--$6.5~\fm$ & $b=3.5$--$5.5~\fm$ \\
 \hline
 $\kappa_{\Lambda_{\text{max}}}$ 
 & $0.607~(8.9\%)$ & $0.639~(4.2\%)$ & $0.6530~(2.04\%)$
 & $0.619~(7.2\%)$ & $0.6535~(1.97\%)$ \\
 \hline
 \multicolumn{\skncol}{l}{Phenomenological fits} \\
 \hline
 $\kappa_{\infty}$, ``$E$''
 & $0.694~(4.1\%)$ & $0.678~(1.7\%)$ & $0.6719~(0.79\%)$
 & $0.685~(2.8\%)$ & $0.6726~(0.89\%)$ \\
 $\kappa_{\infty}$, ``$G$''
 & $0.650~(2.5\%)$ & $0.659~(1.2\%)$ & $0.6633~(0.51\%)$
 & $0.656~(1.6\%)$ & $0.6633~(0.51\%)$ \\
 \hline
 \multicolumn{\skncol}{l}{Separable fits with exact $\eta$} \\
 \hline
 $\kappa_{\infty}$, ``\fiteta''
 & $0.651~(2.4\%)$ & $0.659~(1.2\%)$ & $0.6628~(0.59\%)$
 & $0.655~(1.8\%)$ & $0.6620~(0.70\%)$ \\
 $\kappa_{\infty}$, ``\fitetagenp''
 & $0.661~(0.9\%)$ & $0.662~(0.7\%)$ & $0.6642~(0.38\%)$
 & $0.660~(1.0\%)$ & $0.6635~(0.47\%)$ \\
 $\kappa_{\infty}$, ``\fitetagen''
 & $0.644~(3.4\%)$ & $0.658~(1.3\%)$ & $0.6631~(0.54\%)$
 & $0.653~(2.1\%)$ & $0.6622~(0.67\%)$ \\
 \hline
 \multicolumn{\skncol}{l}{Separable fits with $\eta$
 from best oscillator calculation} \\
 \hline
 $\kappa_{\infty}$, ``\fiteta''
 & $0.633~(5.1\%)$ & $0.651~(2.4\%)$ & $0.6593~(1.11\%)$ 
 & $0.642~(3.7\%)$ & $0.6585~(1.23\%)$ \\
 $\kappa_{\infty}$, ``\fitetagenp''
 & $0.639~(4.2\%)$ & $0.654~(1.9\%)$ & $0.6604~(0.94\%)$
 & $0.646~(3.1\%)$ & $0.6598~(1.03\%)$ \\
 $\kappa_{\infty}$, ``\fitetagen''
 & $0.630~(5.5\%)$ & $0.651~(2.4\%)$ & $0.6596~(1.06\%)$
 & $0.641~(3.9\%)$ & $0.6587~(1.19\%)$ \\
 \hline
 \multicolumn{\skncol}{l}{Direct quantization with $\eta$
 from best oscillator calculation} \\
 \hline
 $\kappa_{\infty}$, ``\fitetadirect''
 & $0.652~(2.2\%)$ & $0.661~(0.9\%)$ & $0.6643~(0.36\%)$
 & $0.655~(1.8\%)$ & $0.6644~(0.35\%)$ \\
 \hline\hline
\end{tabular}
\endgroup
\end{table*}
%%%%%%%%%%%%%%%%%%%%%%%%%%%%%%%%%%%%%%%%%%%%%%%%%%%%%%%%%%%%%%%%%%%%%%%%%%%%%%%%

%%%%%%%%%%%%%%%%%%%%%%%%%%%%%%%%%%%%%%%%%%%%%%%%%%%%%%%%%%%%%%%%%%%%%%%%%%%%%%%%
\begin{figure}[thbp]
 \centering
 \includegraphics[width=0.93\columnwidth]{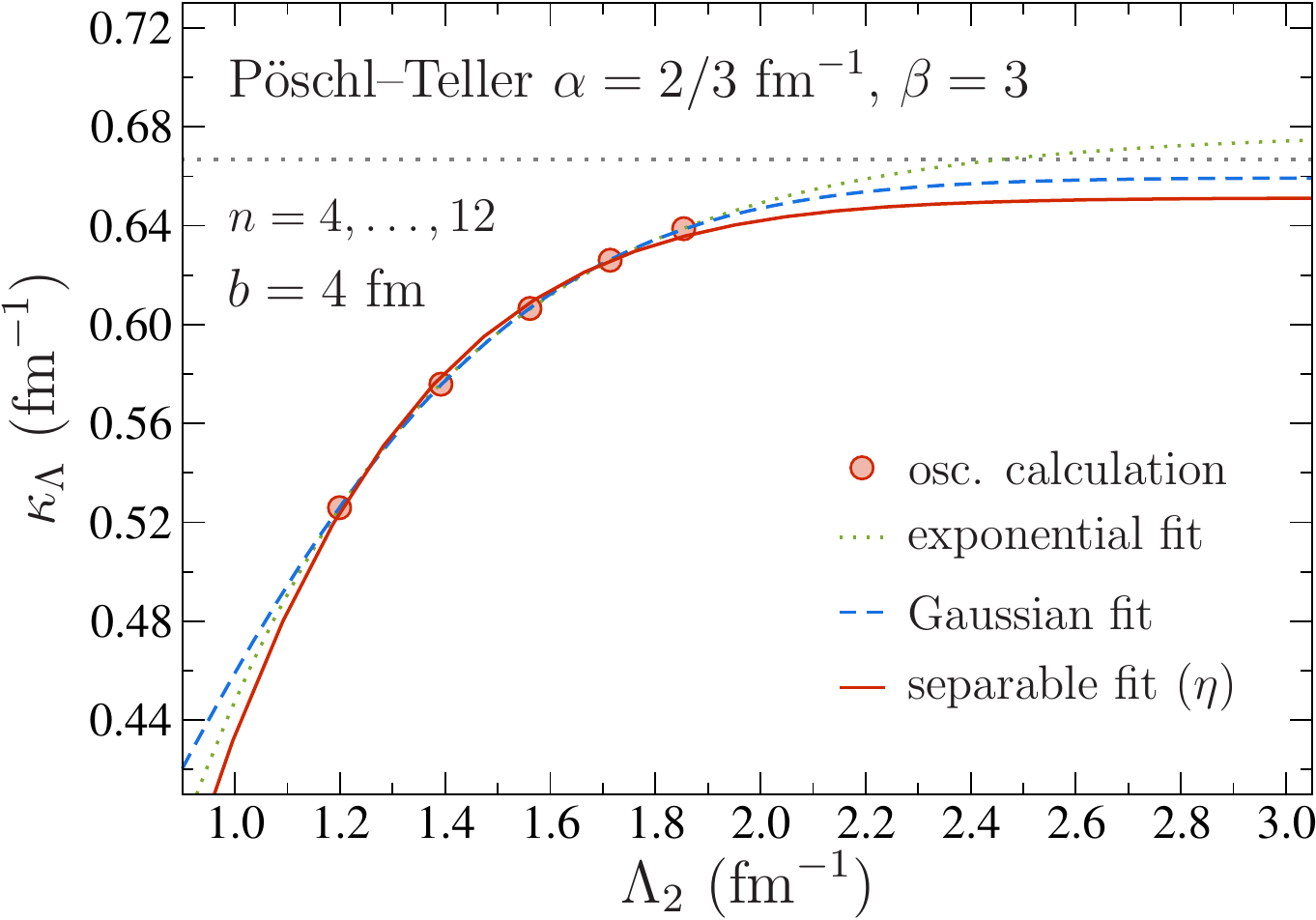}
 \caption{(Color online) Comparison of UV extrapolations for an oscillator 
   calculation (fixed $b=4.0~\fm$, running $n=4,\ldots,12$) with a $\beta=3$,
   $\alpha=2/3~\fm^{-1}$  P\"oschl--Teller potential.   Circles: oscillator 
   results.  Dotted line: exponential extrapolation (fit ``$E$'').
   Dashed line: Gaussian extrapolation (fit ``$G$'').
   Solid line: simplest separable extrapolation (fit ``$\eta$'').
   The horizontal dotted line indicates the exact result for the binding 
   momentum.
 }
\label{fig:ExtCompare-PT3-23-n12}
\end{figure}
%%%%%%%%%%%%%%%%%%%%%%%%%%%%%%%%%%%%%%%%%%%%%%%%%%%%%%%%%%%%%%%%%%%%%%%%%%%%%%%%

%%%%%%%%%%%%%%%%%%%%%%%%%%%%%%%%%%%%%%%%%%%%%%%%%%%%%%%%%%%%%%%%%%%%%%%%%%%%%%%%
\begin{figure}[hbtp]
 \centering
 \includegraphics[width=0.95\columnwidth]{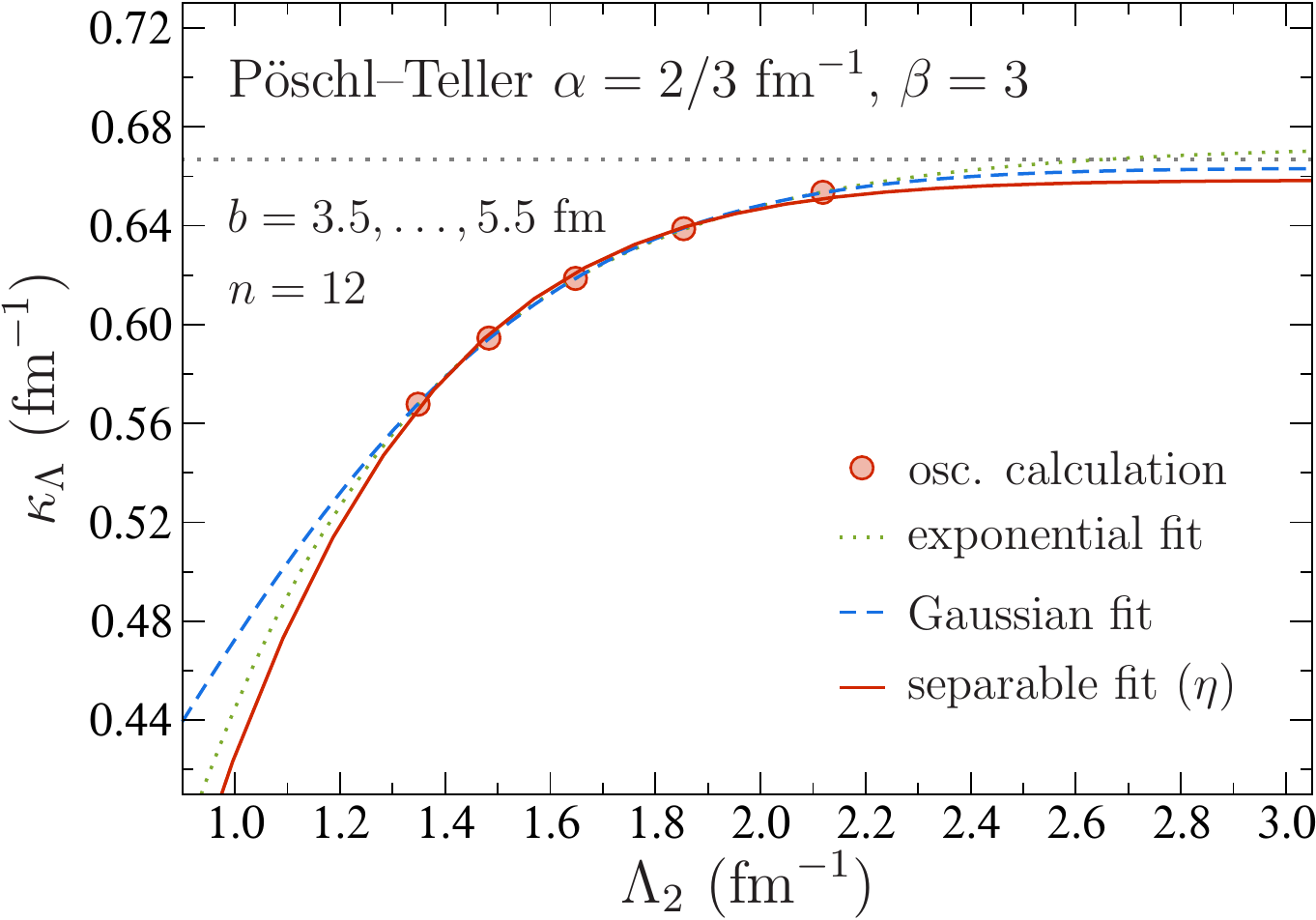}
 \caption{(Color online) Same as Fig.~\ref{fig:ExtCompare-PT3-23-n12}, but now 
   with fixed basis size $n=12$ and running oscillator length 
   $b=3.5,\dotsc,5.5~\fm$.
 }
\label{fig:ExtCompare-PT3-23-b55}
\end{figure}
%%%%%%%%%%%%%%%%%%%%%%%%%%%%%%%%%%%%%%%%%%%%%%%%%%%%%%%%%%%%%%%%%%%%%%%%%%%%%%%%

In Table~\ref{tab:Compare-PT-beta3} we give a detailed account of the results 
for a P\"oschl--Teller potential with $\alpha=2/3~\fm^{-1}$ and $\beta=3$,
which supports a single bound state with binding momentum 
$\kappa_{\infty,\text{exact}} = \alpha$.  Comparison plots 
for the $n=4,\dotsc,12$ and $b=3.5,\dotsc,5.5~\fm$ datasets are shown in 
Figs.~\ref{fig:ExtCompare-PT3-23-n12} and~\ref{fig:ExtCompare-PT3-23-b55} 
respectively.  To avoid cluttering, only the two phenomenological fits 
according to Eqs.~\eqref{eq:fit-exp} and~\eqref{eq:fit-Gauss} and the simplest 
separable one---Eq.~\eqref{eq:fit-eta-simple} with $\eta(k)$ constructed from 
the numerical data---are shown in the figures.  
Table~\ref{tab:Compare-PT-beta5} furthermore shows a detailed comparison for 
the excited state (at $\kappa_{\infty,\text{exact}} = 1/3~\fm^{-1}$) of a 
P\"oschl--Teller potential with $\beta=5$ and $\alpha = 
1/3~\fm^{-1}$.\footnote{This potential has a deeper ground state with binding 
momentum $\kappa=3\alpha=1~\fm^{-1}$.}

%%%%%%%%%%%%%%%%%%%%%%%%%%%%%%%%%%%%%%%%%%%%%%%%%%%%%%%%%%%%%%%%%%%%%%%%%%%%%%%%
\begin{table*}[hbtp]
\caption{Comparison of different extrapolations for a Pöschl--Teller 
  potential with $\alpha=1/3$ and $\beta=5$.  For calculations where $n$ 
  is varied, it is increased in steps of $2$, and $b$ is held fixed at 
  $4.5~\fm$.  For calculations with variable $b$ (increasing in steps of 
  $0.5~\fm$), $n$ is held fixed at $16$.  See Table~\ref{tab:Compare-PT-beta3} 
  and text for further explanation.
}
\label{tab:Compare-PT-beta5}
\centering
\begingroup
\newcommand{\skncol}{6}
\renewcommand\arraystretch{1.25}
\begin{tabular}{c|
 >{\centering\arraybackslash}p{6.2em}|
 >{\centering\arraybackslash}p{6.2em}|
 >{\centering\arraybackslash}p{7.0em}||
 >{\centering\arraybackslash}p{6.7em}|
 >{\centering\arraybackslash}p{7.0em}
}
 \hline\hline
 \multicolumn{\skncol}{l}{
   Excited state of $V_\PT$ with $\alpha=2/3~\fm^{-1}$, $\beta=5$
   \ $\rightarrow$\ \;$\kappa_{\infty,\text{exact}} \approx 0.3333~\fm^{-1}$
 } \\
 \hline\hline
 Calculation
 & $n=4$--$12$ & $n=6$--$16$ & $n=8$--$18$
 & $b=5.5$--$7.5~\fm$ & $b=4.5$--$6.5~\fm$ \\
 \hline
 $\kappa_{\Lambda_{\text{max}}}$ 
 & $0.313~(6.1\%)$ & $0.326~(2.2\%)$ & $0.3223~(3.31\%)$
 & $0.301~(9.7)\%)$ & $0.3264~(2.08\%)$ \\
 \hline
 \multicolumn{\skncol}{l}{Phenomenological fits} \\
 \hline
 $\kappa_{\infty}$, ``$E$''
 & $0.348~(4.4\%)$ & $0.340~(2.0\%)$ & $0.3421~(2.63\%)$
 & $0.357~(7.1\%)$ & $0.3394~(1.82\%)$ \\
 $\kappa_{\infty}$, ``$G$''
 & $0.332~(0.4\%)$ & $0.334~(0.2\%)$ & $0.3341~(0.23\%)$
 & $0.335~(0.5\%)$ & $0.3339~(0.17\%)$ \\
 \hline
 \multicolumn{\skncol}{l}{Separable fits with exact $\eta$} \\
 \hline
 $\kappa_{\infty}$, ``\fiteta''
 & $0.328~(1.6\%)$ & $0.330~(1.0\%)$ & $0.3287~(1.39\%)$
 & $0.324~(2.8\%)$ & $0.3297~(1.09\%)$ \\
 $\kappa_{\infty}$, ``\fitetagenp''
 & $0.328~(1.6\%)$ & $0.330~(1.0\%)$ & $0.3291~(1.27\%)$
 & $0.325~(2.5\%)$ & $0.3300~(1.00\%)$ \\
 $\kappa_{\infty}$, ``\fitetagen''
 & $0.325~(2.5\%)$ & $0.329~(1.3\%)$ & $0.3282~(1.54\%)$
 & $0.322~(3.4\%)$ & $0.3294~(1.18\%)$ \\
 \hline
 \multicolumn{\skncol}{l}{Separable fits with $\eta$
 from best oscillator calculation} \\
 \hline
 $\kappa_{\infty}$, ``\fiteta''
 & $0.317~(4.9\%)$ & $0.326~(2.2\%)$ & $0.3240~(2.80\%)$ 
 & $0.310~(7.0\%)$ & $0.3267~(1.99\%)$ \\
 $\kappa_{\infty}$, ``\fitetagenp''
 & $0.318~(4.6\%)$ & $0.327~(1.9\%)$ & $0.3243~(2.71\%)$
 & $0.310~(7.0\%)$ & $0.3269~(1.93\%)$ \\
 $\kappa_{\infty}$, ``\fitetagen''
 & $0.315~(5.5\%)$ & $0.326~(2.2\%)$ & $0.3237~(2.89\%)$
 & $0.309~(7.3\%)$ & $0.3265~(2.05\%)$ \\
 \hline
 \multicolumn{\skncol}{l}{Direct quantization with $\eta$
 from best oscillator calculation} \\
 \hline
 $\kappa_{\infty}$, ``\fitetadirect''
 & $0.327~(1.9\%)$ & $0.332~(0.4\%)$ & $0.3303~(0.91\%)$
 & $0.322~(0.0\%)$ & $0.3316~(0.52\%)$ \\
 \hline\hline
\end{tabular}
\endgroup
\end{table*}
%%%%%%%%%%%%%%%%%%%%%%%%%%%%%%%%%%%%%%%%%%%%%%%%%%%%%%%%%%%%%%%%%%%%%%%%%%%%%%%%

From the results presented in the tables and figures, we draw the following 
conclusions at this point:

\begin{itemize}
\item None of the fits produces the correct (exact) binding momentum right 
  away, not even if the calculation is already converged to within only 2\%.  
  It should be noted, however, that we made no effort (\eg, weighting) here to 
  improve the fits.  The only exception to this is the excited state of the 
  $\beta=5$ P\"oschl--Teller potential (see Table~\ref{tab:Compare-PT-beta5}),
  where the Gaussian fit works remarkably well.  This may be an 
  accidental property of that potential and just supports our previous
  statement that in general it is desirable to have an extrapolation approach 
  that really takes into account information from the particular potential and 
  state under consideration.
\item On average, the Gaussian fit mostly produces the best (closest to the 
   exact answer) results.  Except for the least-converged oscillator 
   calculations, however, the separable fits based on our analytical 
   knowledge of the exact wavefunctions work as well as the corresponding 
   Gaussian ones.  This indicates that the main limitation of the separable 
   approach is the incomplete knowledge of the wavefunction that one gets 
   from the numerical calculations.
 \item For the more converged calculations, however, the completely numerical 
   separable fits come close to the Gaussian results---although, as we
   have pointed out, the latter approach uses one more fit parameter.
 \item Reassuringly, there is little scatter in the different separable fits, 
   Eqs.~\eqref{eq:fit-eta-simple}, \eqref{eq:fit-eta-gen}, 
   and~\eqref{eq:fit-eta-gen-prime}.  Except for the most converged 
   calculations, the fit based on Eq.~\eqref{eq:fit-eta-gen-prime} produces 
   significantly better (in the above sense) results than the other two.
   Returning to the discussion in Sec.~\ref{sec:Bootstrap}, this might 
   indicate that the Eq.~\eqref{eq:fit-eta-simple} is not rigorously   
   justified, whereas Eq.~\eqref{eq:fit-eta-gen} is too constraining to fit 
   the whole range of data.  While it may be tempting to thus suggest 
   Eq.~\eqref{eq:fit-eta-gen-prime} as the optimal fit strategy, it is not clear
   that our speculation here is correct in general or even for the specific 
   potentials considered here.  Since the overhead of the analysis is small 
   compared to the original diagonalization, in practice it should be useful 
   to perform all three fits and take the scatter as an indicator for the 
   stability and/or uncertainty of the method.
\item Finally, it is interesting to see that the ``direct'' extrapolation based 
   on the separable quantization condition~\eqref{eq:quant-sep} is able to 
   produce results quite close to the exact answer based on just a single 
   oscillator calculation with fixed $n$ and $b$.
\end{itemize}

\subsection{Separable deuteron extrapolation}

At this point, we finally turn to extrapolations for the deuteron bound state 
as it comes out from oscillator calculations with realistic nucleon--nucleon 
interactions.  While for this simple system one can actually choose oscillator 
spaces which yield results converged so well that no extrapolation is actually 
necessary, it is still the most interesting two-body system we can look at here 
and provides a starting point for extrapolations of many-body 
calculations to be looked at in the future.

\subsubsection{Separable deuteron potential}

The deuteron is the bound state in the $^3S_1$--$^3D_1$ coupled-channel system 
of the $n$--$p$ interaction.  We write this potential as
\begin{equation}
 \hat{V}_{\SD} =
 \left(\!\begin{array}{cc}
  \hat{V}_{00} & \hat{V}_{02} \\
  \hat{V}_{20} & \hat{V}_{22}
 \end{array}\!\right) \,,
\label{eq:V-deut}
\end{equation}
where $\hat{V}_{\ell\ell'}$ are the angular momentum components of a given 
realistic nucleon--nucleon potential (naturally, 
$\hat{V}_{20}=\hat{V}_{02}^\dagger$).  For simplicity, we have omitted here 
the remaining quantum numbers and just note that for the deuteron one has 
$S=J=1$ and $T=0$, for the spin, total angular momentum, and isospin, 
respectively.  If we now write the deuteron wavefunction found from the 
potential~\eqref{eq:V-deut} as
\begin{equation}
 \ket{\psi_d} = \left(\!\begin{array}{c} \ket{\psi_0} \\
 \ket{\psi_2} \end{array}\!\right) \,,
\label{eq:psi-deut}
\end{equation}
we can construct a separable potential of the form
\begin{equation}
 \hat{V}_{\SD,\text{sep}} = g 
 \left(\!\begin{array}{cc}
  \etaSk\etaSb & \etaSk\etaDb \\
  \etaDk\etaSb & \etaDk\etaDb
 \end{array}\!\right)
 = g 
 \left(\!\begin{array}{c} \etaSk \\ \etaDk \end{array}\!\right)
 \left(\!\begin{array}{c} \etaSb \\ \etaDb \end{array}\!\right)^{\!\!T}
\label{eq:V-deut-sep}
\end{equation}
if we set
\begin{subequations}%
\begin{align}
 \etaSk &= \hat{V}_{00}\ket{\psi_0} + \hat{V}_{02}\ket{\psi_2} \,, \\
 \etaDk &= \hat{V}_{20}\ket{\psi_0} + \hat{V}_{22}\ket{\psi_2} \,,
\end{align}
\end{subequations}%
and
\begin{equation}
 g = \left(\mbraket{\psi_0}{\hat{V}_{00}}{\psi_0}
  + \mbraket{\psi_2}{\hat{V}_{22}}{\psi_2}
  + 2\,\Rp\,\mbraket{\psi_0}{\hat{V}_{02}}{\psi_2} \right)^{-1} \,.
\end{equation}

\subsubsection{Coupled-channel separable extrapolation}

To derive the extrapolation formula for this coupled-channel separable 
potential, we start by writing the Schrödinger equation as
\begin{multline}
 \left(\!\begin{array}{cc} \hat{k}^2 & 0 \\ 0 &\hat{k}^2 \end{array}\!\right)
 \left(\!\begin{array}{c} \ket{\psi_0} \\ \ket{\psi_2} \end{array}\!\right)
 + \left[g \left(\!\begin{array}{c} \etaSk \\ \etaDk \end{array}\!\right)
 \left(\!\begin{array}{c} \etaSb \\ \etaDb \end{array}\!\right)^{\!\!T}\,\right]
 \left(\!\begin{array}{c} \ket{\psi_0} \\ \ket{\psi_2} \end{array}\!\right) \\
 = -\kappa^2
 \left(\!\begin{array}{c} \ket{\psi_0} \\ \ket{\psi_2} \end{array}\!\right) \,.
\label{eq:SG-deut-sep}
\end{multline}
Just as in the single-channel case (\cf~Sec.~\ref{sec:RegContact}), it is 
straightforward to solve as
\begin{equation}
 \left(\!\begin{array}{c} \ket{\psi_0} \\ \ket{\psi_2} \end{array}\!\right)
 = c_\infty
  \left(\!\begin{array}{cc} \hat{k}^2+\kappa^2 & 0 \\
 0 & \hat{k}^2+\kappa^2 \end{array}\!\right)^{\!\!-1}
 \left(\!\begin{array}{c} \etaSk \\ \etaDk \end{array}\!\right)
\label{eq:psi-deut-sep}
\end{equation}
with a constant
\begin{equation}
 c_\infty
 = g \, \big[\braket{\etaS}{\psi_0} + \braket{\etaD}{\psi_2}\big] \,.
\end{equation}
Noting that the operator inversion in Eq.~\eqref{eq:psi-deut-sep} can be 
carried out for the two diagonal terms individually and inserting the result 
back into Eq.~\eqref{eq:SG-deut-sep}, we get
\begin{widetext}\begin{equation}
 \left[
 \left(\!\begin{array}{cc} 1 & 0 \\ 0 & 1 \end{array}\!\right)
 + g \left(\!\begin{array}{cc}
  \etaSb\big(\hat{k}^2+\kappa^2\big)^{-1}\etaSk & 0 \\
  0 & \etaDb\big(\hat{k}^2+\kappa^2\big)^{-1}\etaDk
 \end{array}\!\right)
 \right]
 \times \left(\!\begin{array}{c} \etaSk \\ \etaDk \end{array}\!\right) = 0 \,.
\end{equation}\end{widetext}
Finally, by multiplying from the left with $\left(\etaSb,\etaDb\right)$, we 
arrive at a simple quantization condition for the binding momentum $\kappa$, 
which in momentum space reads
\begin{equation}
 -1 = 4\pi g  \int_0^\infty\!\dd k\, k^2 \,
 \frac{\etaS(k)^2 + \etaD(k)^2}{\kappa_\infty^2+k^2} \,.
\label{eq:quant-sep-SD-inf}
\end{equation}
Repeating the whole procedure with appropriate projection operators to enforce 
a momentum cutoff $\Lambda$, we find
\begin{equation}
 -1 = 4\pi g  \int_0^\Lambda\!\dd k\, k^2 \,
 \frac{\etaS(k)^2 + \etaD(k)^2}{\kappa_\Lambda^2+k^2} \,.
\label{eq:quant-sep-SD}
\end{equation}
This is just Eq.~\eqref{eq:quant-sep} with the replacement
\begin{equation}
 \eta(k)^2 \longrightarrow \etaS(k)^2 + \etaD(k)^2 \,,
\end{equation}
so it is simple to read off the coupled-channel extrapolation formulas from 
Eqs.~\eqref{eq:fit-eta-simple} to~\eqref{eq:fit-eta-gen-prime}.  For example, 
the analog of Eq.~\eqref{eq:fit-eta-simple} is just
\begin{equation}
 \kappa_\Lambda = \kappa_\infty
  - A \int\nolimits_\Lambda^\infty \dd k\,
  \left[\etaS(k)^2 + \etaD(k)^2\right] \,.
\label{eq:fit-eta-simple-deut}
\end{equation}

%%%%%%%%%%%%%%%%%%%%%%%%%%%%%%%%%%%%%%%%%%%%%%%%%%%%%%%%%%%%%%%%%%%%%%%%%%%%%%%%
\begin{figure*}[ptbh]
\centering
\begin{minipage}{0.5\textwidth}
 \includegraphics[width=0.95\textwidth]{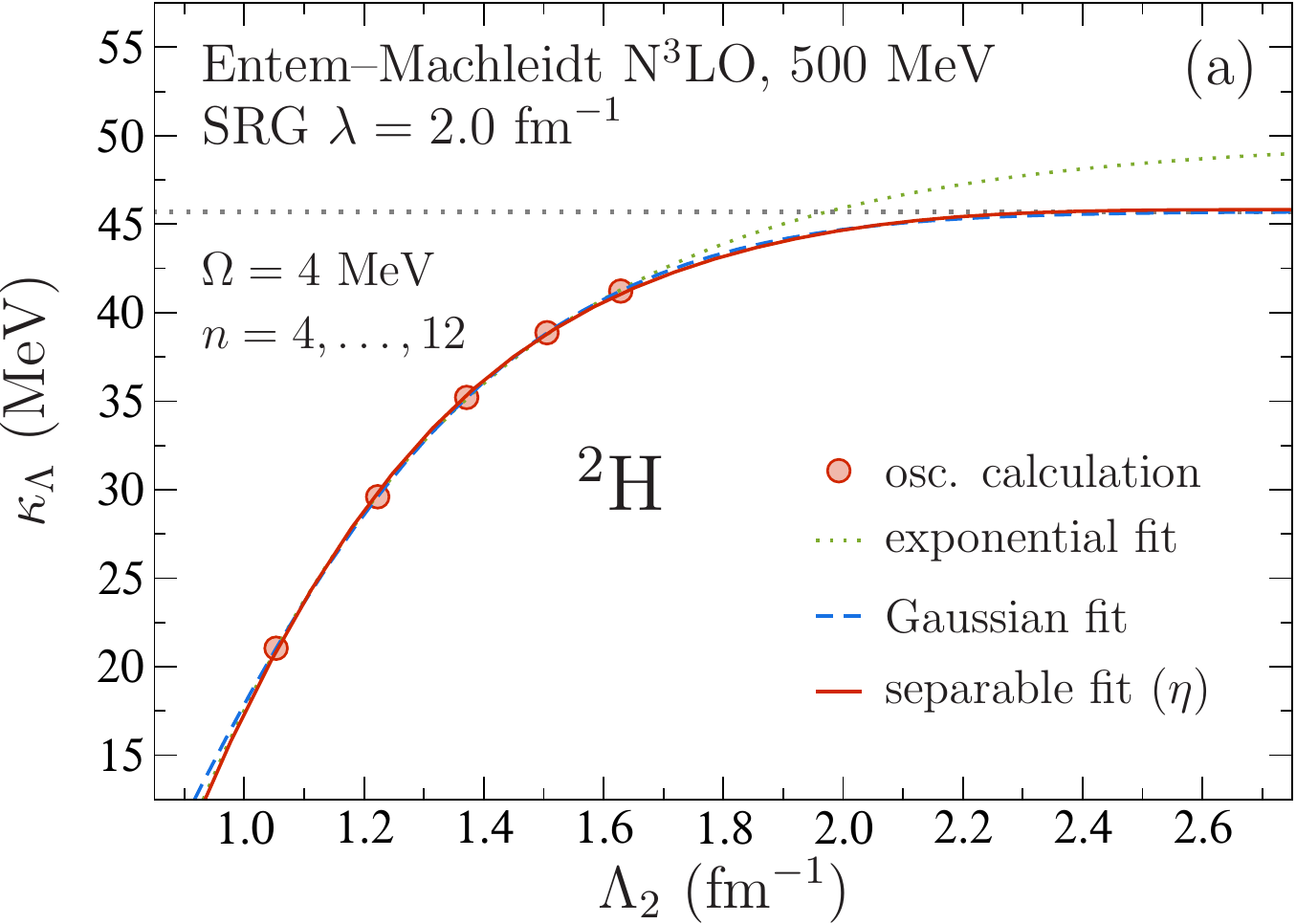}
\end{minipage}\begin{minipage}{0.5\textwidth}
 \includegraphics[width=0.962\textwidth]{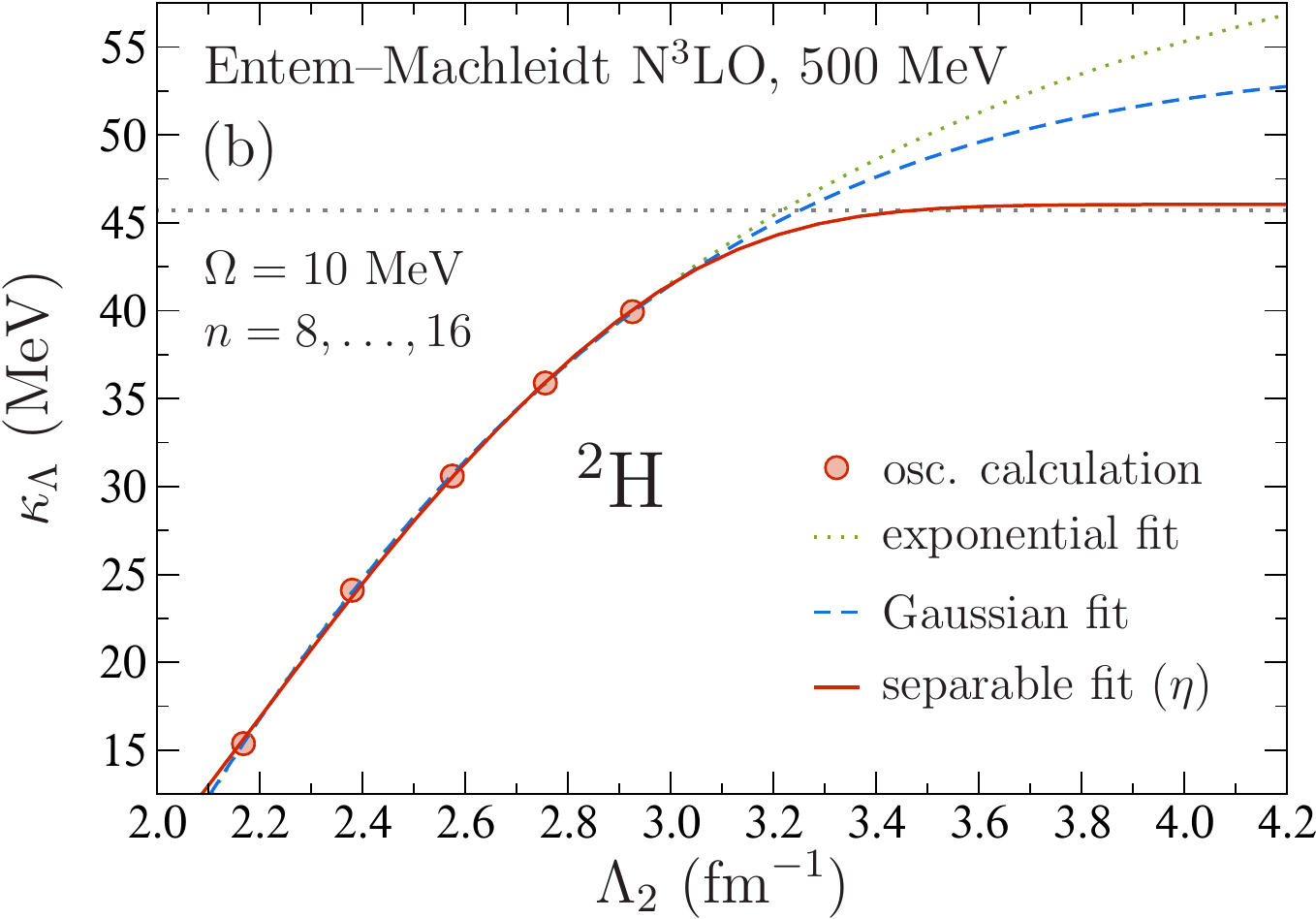}
\end{minipage}
 \caption{(Color online) Comparison of UV extrapolations for a deuteron state 
   calculated with the Entem--Machleidt N3LO ($500~\MeV$ cutoff) potential, 
   (a) SRG-evolved down to a resolution scale $\lambda=2.0~\fm^{-1}$ and
   (b) with the ``bare'' (unevolved) interaction.
   Circles: oscillator results.
   Dotted line: exponential extrapolation (fit ``$E$'').
   Dashed line: Gaussian extrapolation (fit ``$G$'').
   Solid line: simplest separable extrapolation (fit ``\fiteta'').
   Dotted horizontal lines indicate the exact result for the binding 
   momentum.
 }
\label{fig:Deut-EM500}
\end{figure*}
%%%%%%%%%%%%%%%%%%%%%%%%%%%%%%%%%%%%%%%%%%%%%%%%%%%%%%%%%%%%%%%%%%%%%%%%%%%%%%%%
\begin{figure*}[ptbh]
\centering
\begin{minipage}{0.5\textwidth}
 \includegraphics[width=0.95\textwidth]{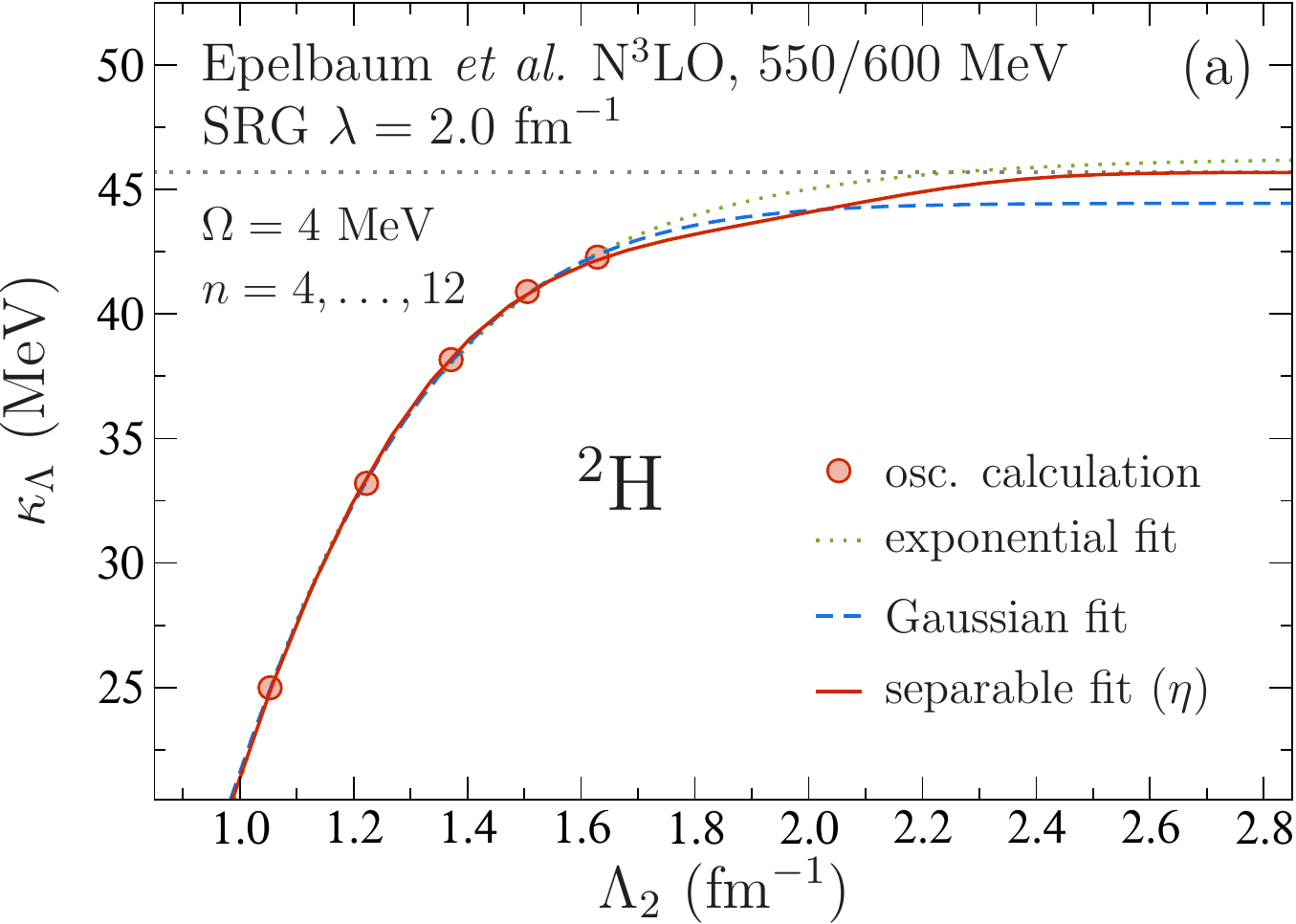}
\end{minipage}\begin{minipage}{0.5\textwidth}
 \includegraphics[width=0.95\textwidth]{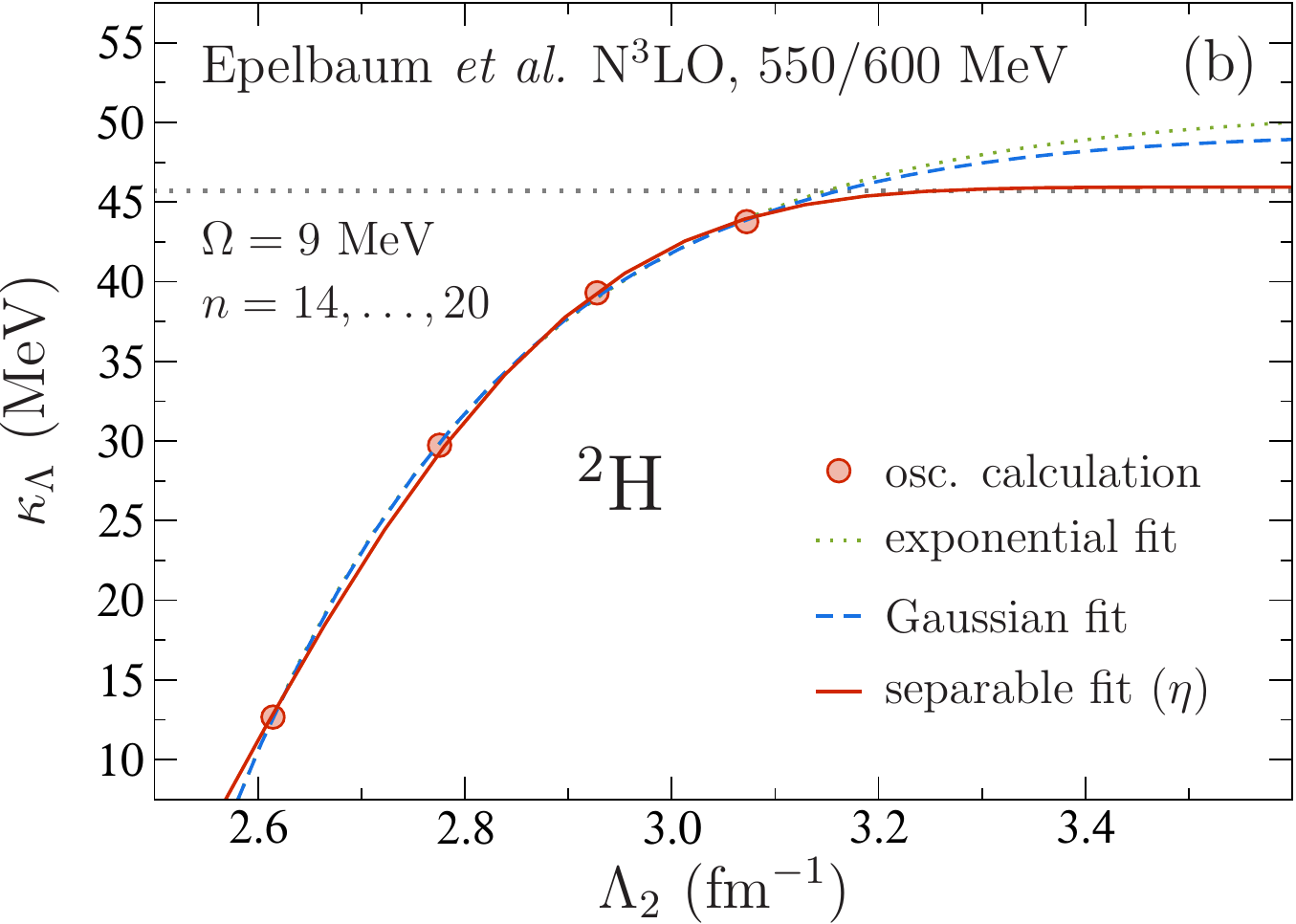}
\end{minipage}
\caption{(Color online) Comparison of UV extrapolations for a deuteron state
  calculated with the Epelbaum~\etal N3LO ($550$/$600~\MeV$ cutoff)
  potential, (a) SRG-evolved down to a resolution scale
  $\lambda=2.0~\fm^{-1}$ and (b) with the ``bare''
  (unevolved) interaction.  Symbols and curves are
  as in Fig.~\ref{fig:Deut-EGM550}.  }
\label{fig:Deut-EGM550}
\end{figure*}
%%%%%%%%%%%%%%%%%%%%%%%%%%%%%%%%%%%%%%%%%%%%%%%%%%%%%%%%%%%%%%%%%%%%%%%%%%%%%%%%

%%%%%%%%%%%%%%%%%%%%%%%%%%%%%%%%%%%%%%%%%%%%%%%%%%%%%%%%%%%%%%%%%%%%%%%%%%%%%%%%
\begin{figure}[phbt]
 \centering
 \includegraphics[width=0.95\columnwidth]{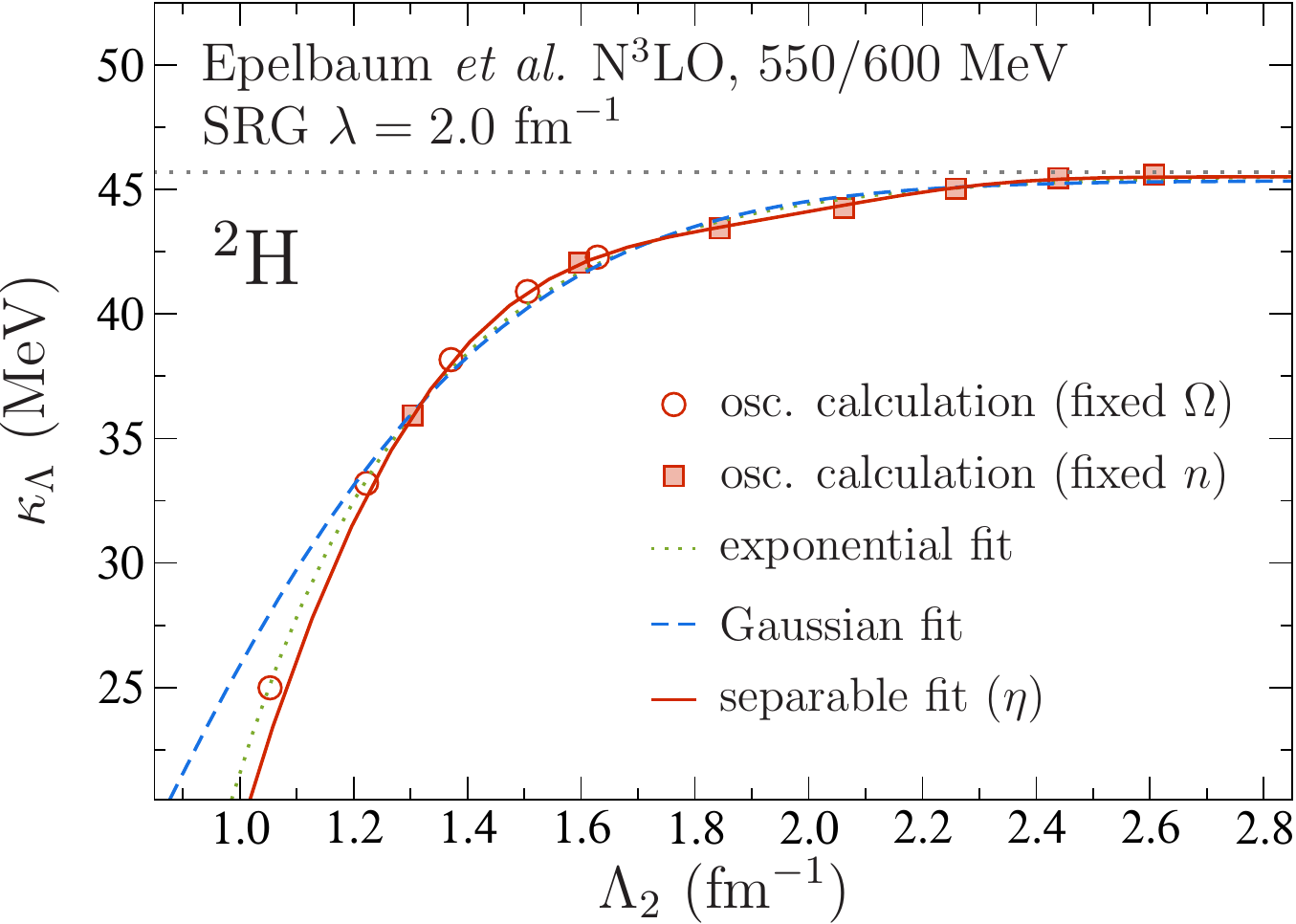}
 \caption{(Color online) Comparison of UV extrapolations for a deuteron state
   calculated with the Epelbaum~\etal N3LO ($550$/$600~\MeV$ cutoff)
   potential.  Circles: oscillator results with fixed $\hw=4~\MeV$ and
   $n=4,\ldots,12$.  Squares: oscillator results with fixed $n=10$ and
   $\hw=3,\ldots,12~\MeV$ (in steps of $1.5~\MeV$.  Curves are as
   in Fig.~\ref{fig:Deut-EGM550} and show fits to the squares only.  }
\label{fig:Deut-EGM550_srg-combined}
\end{figure}
%%%%%%%%%%%%%%%%%%%%%%%%%%%%%%%%%%%%%%%%%%%%%%%%%%%%%%%%%%%%%%%%%%%%%%%%%%%%%%%%

%%%%%%%%%%%%%%%%%%%%%%%%%%%%%%%%%%%%%%%%%%%%%%%%%%%%%%%%%%%%%%%%%%%%%%%%%%%%%%%%
\begin{figure}[phbt]
 \centering
 \includegraphics[width=0.95\columnwidth]{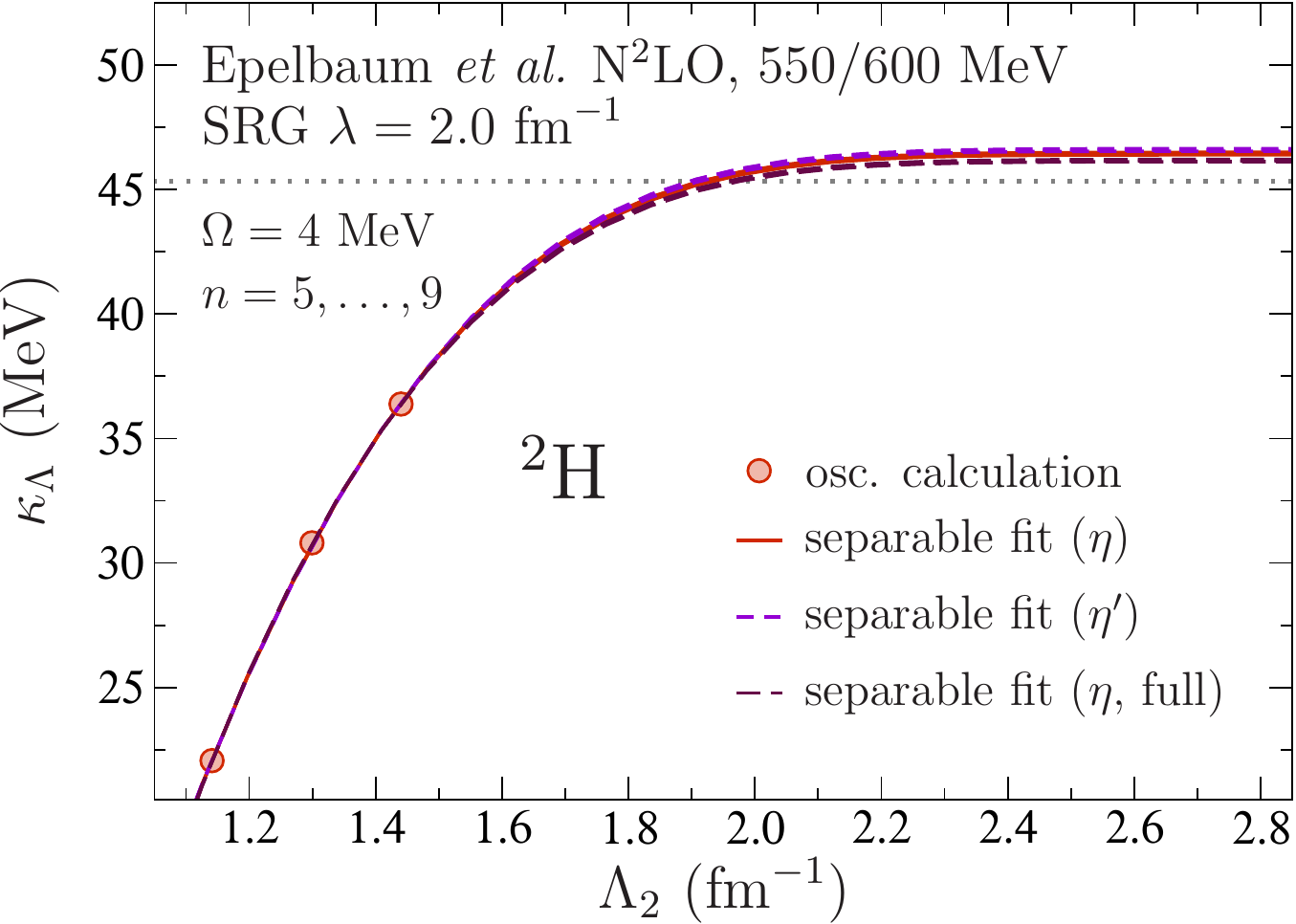}
 \caption{(Color online) Comparison of UV extrapolations for a deuteron state 
   calculated with the Epelbaum~\etal N2LO ($550$/$600~\MeV$ cutoff) 
   potential.  Circles: oscillator results.
   Solid line: simplest separable extrapolation (fit ``$\eta$'').
   Long-dashed line: general separable extrapolation
   (fit ``$\eta, \mathrm{gen.}$'').
   Short-dashed line: modified general separable extrapolation
   (fit ``$\eta, \mathrm{gen.}\!\!'\,$'').  The dotted horizontal line 
   indicates the exact result for the binding momentum.
 }
\label{fig:Deut-EGM550_N2LO_srg-mult}
\end{figure}
%%%%%%%%%%%%%%%%%%%%%%%%%%%%%%%%%%%%%%%%%%%%%%%%%%%%%%%%%%%%%%%%%%%%%%%%%%%%%%%%

\subsubsection{Extrapolation results}

In Fig.~\ref{fig:Deut-EM500} we show results obtained with the
Entem--Machleidt N3LO ($500~\MeV$ cutoff) potential.  For the
interaction SRG-evolved down to a resolution scale
$\lambda=2.0~\fm^{-1}$ (left panel), the picture is similar to what we
found for the P{\"o}schl--Teller potential in
Sec.~\ref{sec:Bootstrap}.  With the oscillator calculation
(performed at fixed $\Omega=4~\MeV \rightarrow b\approx 4.55~\fm$),
the Gaussian fit actually yields the exact answer $\kappa_d=45.702$
(for the given interaction) to within $0.01\%$.  The separable fits,
however, also work very well and give the right answer to within
$0.15$ to $0.55$ percent.  The simple exponential fit does not perform
well at all in this case.

Seeing how \emph{all three} curves actually fit the data points 
very well with negligible residuals, however, the ``danger'' of purely 
phenomenological extrapolations becomes quite evident.  If one does not use a 
known answer as guideline---as clearly one should not in a completely rigorous 
approach---it would be hard to judge which fit should be trusted.

For the results based on the ``bare'' (unevolved) interaction
Fig.~\ref{fig:Deut-EM500}, right panel), the situation is even more
dramatic: in this case, both phenomenological approaches fail badly
(based on comparing their results to the known answer), whereas the
separable approximation still works remarkably well (better than $1\%$
agreement with exact answer).  This should finally serve to exhibit
the true value of this physically motivated extrapolation approach.

We find the same situation also for other nucleon--nucleon
interactions.  As a further example, we show in
Fig.~\ref{fig:Deut-EGM550} results for the Epelbaum~\etal N3LO
potential ($550$/$600~\MeV$ cutoff).  For the SRG-evolved interaction
we see an interesting feature at $\Lambda_2\sim2.0~\fm^{-1}$.  The
curve for the separable fit has a ``bump'' structure around this
cutoff, but it ends up almost exactly at the converged value.  To
prove that this is not a peculiar artifact of the separable fit, we
show in Fig.~\ref{fig:Deut-EGM550_srg-combined} results for the same
potential but up to larger cutoffs.  To also demonstrate once more the
validity of identifying $\Lambda_2 = \Lambda_2(\Nmax,\hw)$ as the
relevant UV cutoff, we use in this case data points obtained at fixed
$\Nmax$ and varying $\hw$ for the fits.  The data points from
Fig.~\ref{fig:Deut-EGM550} are shown at the same time for comparison.
The plot shows that the bump structure really is a feature that is in
the oscillator data.  We point out that the simple exponential and
Gaussian fits shown for comparison cannot possibly capture this kind
of behavior.  We hence claim that the separable fit approach is
superior to the phenomenological ones also for SRG-softened
interactions (at least for fits over a large cutoff range, \cf the
following section).

Finally, to look at one more $NN$ potential, we show in
Fig.~\ref{fig:Deut-EGM550_N2LO_srg-mult} results for the
Epelbaum~\etal interaction at N2LO.  Since the overall situation is
the same, we focus in this case on assessing the stability of the
separable fits alone.  To this end, we show now fit curves for the
three versions---Eqs.~\eqref{eq:fit-eta-simple}
to~\eqref{eq:fit-eta-gen-prime} with $\eta(k)^2 \to \etaS(k)^2 +
\etaD(k)^2$---obtained from oscillator calculations with $n=5,7,9$ 
(instead of using just the one with the largest $n$).  Although the
overall spread is remarkably small, we suggest this procedure in order
to assess the stability of the fit.  Since the band generated this way
unfortunately does not cover the exact answer for this potential
($\kappa_\infty\approx45.3~\MeV$), it is clear that this obviously
gives a lower bound on the overall theoretical uncertainty of the
calculation.  Note, however, that the best oscillator result shown in
the plot is only converged to within about $20\%$.  We also point out
that the separable fits still perform better than the phenomenological
ones (not shown in the plot).

\section{Re-examining SRG-based extrapolations}
\label{sec:SRG}

In the preceding section we showed that the separable extrapolation applied
to the deuteron worked very well for bare potentials and SRG-evolved
potentials.  In this section, we re-examine prior results in the
literature for SRG interactions.  These include the phenomenological
result that a Gaussian ansatz for the UV correction,
\begin{equation}
  \Delta \Einf \propto \ee^{-b_1 \Lambda_2^2} \,,
  \label{eq:gaussansatz}
\end{equation}
gives good fits with $b_1 \approx 4/\lambda^2$ at resolution scale
$\lambda$~\cite{Furnstahl:2012qg,Jurgenson:2013yya}.%
\footnote{Note that these earlier works used an oscillator parameter $b$
  defined with the nucleon mass rather than the reduced mass as used
  here.  Thus the numerical values of the effective IR and UV cutoffs
  differ by a factor of $\sqrt{2}$ compared to the SRG results given
  here.}

\subsection{Perturbation theory for SRG potentials}
\label{sec:SRGperturbation}

Here we reconsider evaluating the UV correction in perturbation theory
as in Sec.~\ref{subsubsec:perturbation}, but instead of a separable
potential we only assume that we have a potential $V_\lambda(k,k')$
with a UV scale $\lambda$, the (dominant) behavior of which is known when
one argument is small ($<\lambda$) and one argument is large
($>\lambda$).  In particular, we expect the dominant dependence for
SRG evolved potentials to be roughly~\cite{Anderson:2010aq,Bogner:2012zm}
\begin{equation}
 V_\lambda(k,k') \overset{k>\lambda}{\underset{k'\ll\lambda}{\longrightarrow}}
 V_\infty(k,k')\ee^{-k^4/\lambda^4}
 \approx V_\infty(k,0)\ee^{-k^4/\lambda^4}
 \,,
\label{eq:Vasymptotic}
\end{equation}
where $V_\infty(k,0)$ varies relatively slowly compared to
$\ee^{-k^4/\lambda^4}$ in the relevant range of $k$.  
Equation~\eqref{eq:Vasymptotic} 
follows from the SRG flow equations because of
the dominance of the kinetic energy for far off-diagonal matrix elements
(see Eq.~(12) in Ref.~\cite{Anderson:2010aq}),
together with an expansion about $k' = 0$.
Another class of potentials with analogous behavior is the smooth $\vlowk$ 
potential with super-Gaussian regulators with a cutoff
$\lambda$~\cite{Bogner:2006vp}, for which the regulator dependence is
strongly imposed on the potential.

The momentum space Schrödinger equation with $\Lambda=\infty$ is
\begin{equation}
k^2\phiinf(k) + \int\!\dd^3k'\, 
 V_\lambda(k,k') \phiinf(k')
 = -\kapinf^2\phiinf(k)
 \,.
\end{equation}
So the analog equation 
to~\eqref{eq:phiunperturb} for the unperturbed wavefunction is
\begin{equation}
 \phiinf(k) = \frac{{-\!\displaystyle\int}\dd^3k'\, V_\lambda(k,k') 
  \phiinf(k')}{k^2+\kapinf^2}
 \,.
\end{equation}
If we look at this wavefunction where $k > \lambda$, 
then we can take advantage of the integral being dominated by where 
$\phi_\infty(k')$ is large, which is at low $k'$, to expand $\wt V(k,k')$
about $k'=0$ (here keeping only the leading term):
\begin{equation}
 \phiinf(k) \overset{k > \lambda}{\longrightarrow}
  {-}\frac{V_\lambda(k,0) }{k^2+\kapinf^2}\int\!\dd^3k'\,\phiinf(k')
 \,. 
\label{eq:phiinf_asymptotic}
\end{equation}
Given that the integration over the wavefunction is now a constant, and given 
Eq.~\eqref{eq:Vasymptotic} for $V_\lambda(k,0)$, which looks like $f_\lambda(k)$
(with some weaker $k$ dependence), we see a close correspondence to the
expression for the wavefunction in a pure separable potential given by 
$\phi(k)$ in Eq.~\eqref{eq:phi}.

The cutoff Hamiltonian is
\begin{equation}
 H_\Lambda = \left[
   k^2\frac{\delta(k-k')}{kk'} + V_\lambda(k,k')
    \right]\Theta(\Lambda-k)\Theta(\Lambda-k')
 \,,
\end{equation}
so the perturbation is $\delta H(k,k') = H_\Lambda - H_\infty$.
Using $\Theta(\Lambda-k) = 1 - \Theta(k-\Lambda)$, we find 
(\cf~Eq.~\eqref{eq:sepperturbation})
\begin{multline}
 \delta H(k,k') = -
    \Biggl[
     k^2 \frac{\delta(k-k')}{kk'} \Theta(k-\Lambda) \Theta(k'-\Lambda)
 \\
     +  V_\lambda(k,k') [\Theta(k-\Lambda) + \Theta(k'-\Lambda)]
    \Biggr]
 \,.
\end{multline}
The $\delta$-function makes the second $\Theta$ function multiplying
the kinetic energy redundant, while again we have dropped the
$-\Theta(k-\Lambda)\Theta(k'-\Lambda)$ term.
The first-order energy shift is
\begin{align}
 \Delta\Einf
  &= \frac{\langle\phiinf|\delta H| \phiinf\rangle}
                                          {\langle\phiinf|\phiinf\rangle} 
    \nonumber \\
  &= -\biggl[
   \int_\Lambda^\infty\! \dd k\, k^2\, k^2 \phiinf^2(k)
  \nonumber \\
  & \qquad\;\; \null
  + 8\pi  \int_0^\infty\! \dd k'\, k'^{2} \int_\Lambda^\infty\! \dd k\, k^2 
  \nonumber \\
  & \qquad \qquad \;\; \times \phiinf(k')  V_\lambda(k',k) \phiinf(k) 
 \biggr]
   \nonumber \\ & \;\;\;\; \null
 \times \left[ \int_{0}^{\infty}\! \dd k\, k^2 \phiinf^2(k) \right]^{-1}
 \,.
\end{align}
Now $\Lambda > \lambda$ in the present discussion, so we can
apply Eq.~\eqref{eq:phiinf_asymptotic} twice in the first (kinetic energy)
integral and once in the second of the double integrals, also taking
$ V(k',k)\rightarrow  V(0,k)$ at the same level of approximation:
\begin{widetext}
\begin{align}
 \Delta\Einf &\approx
   -4\pi\left[\int\!\dd^3k'\,\phiinf(k')\right]^2
   \left[
     \int_\Lambda^\infty\! \dd k\, \frac{k^4\, V_\lambda(0,k) V_\lambda(k,0)}
     {(\kapinf^2 + k^2)^2}
     - 2 \int_\Lambda^\infty\! \dd k\,  \frac{k^2\, V_\lambda(0,k)  
          V_\lambda(k,0)}
     {\kapinf^2 + k^2} 
   \right]
 \times \left[\int_{0}^{\infty}\! \dd^3k\, \phiinf^2(k) \right]^{-1}
   \nonumber \\ &\approx
   4\pi
   \frac{\left[\displaystyle\int\!\dd^3k'\,\phiinf(k')\right]^2}
   {\left[\displaystyle\int\! \dd^3k\, \phiinf^2(k) \right]}
   \times \left[
     \displaystyle\int_\Lambda^\infty\! \dd k\,   V_\lambda(0,k)  V_\lambda(k,0)
   \right]
    \times \left[1+ {\mathcal O}(\kapinf^2/\Lambda^2)\right] 
 \,.
 \label{eq:SRGpertE}
\end{align}
\end{widetext}
In the second line we have again just kept the leading term in 
$\kapinf^2/\Lambda^2$, which lets us combine the integrals.

%%%%%%%%%%%%%%%%%%%%%%%%%%%%%%%%%%%%%%%%%%%%%%%%%%%%%%%%%%%%%%%%%%%%%%%%%%%%%%%%
\begin{figure}[thbp]
 \centering
 \includegraphics[width=0.95\columnwidth]%
 {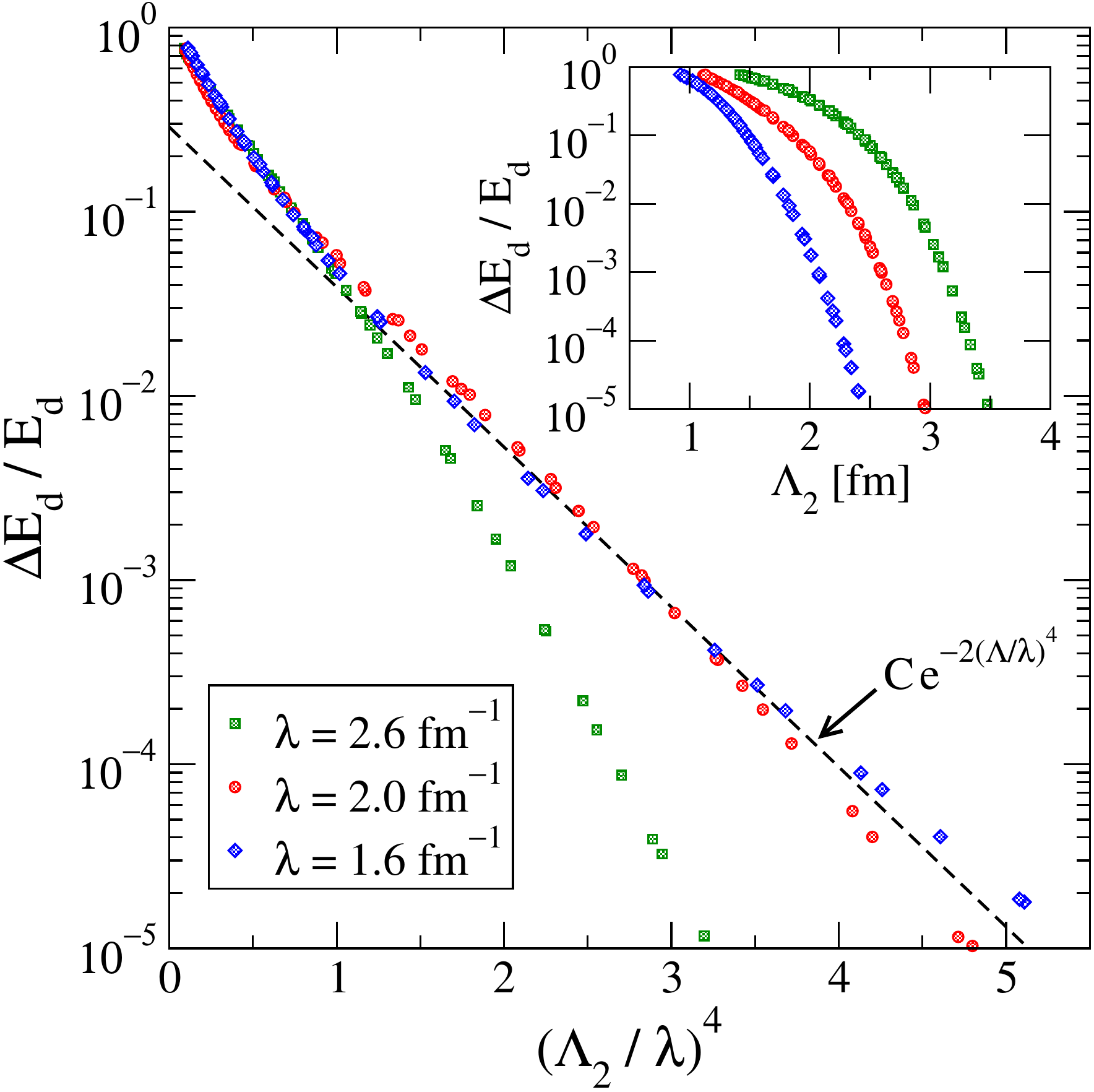}
 \caption{(Color online) Relative error for the deuteron energy from HO basis
 truncation as a function of $(\Lambda_2/\lambda)^4$ for $(N,\hw)$ values for
 which the IR correction can be neglected. 
 Several SRG-evolved potentials are used, all with the same initial
 potential as in Fig.~\ref{fig:Deut-EM500}.
 The dashed line shows the expected slope (up to prefactors) for
 $\Lambda_2/\lambda \gg 1$ according to the analysis in 
 Sec.~\ref{subsubsec:Asymptotic}.
 Inset: the relative error is plotted against the unscaled $\Lambda_2$.
 }
\label{fig:deuteron_fourth}
\end{figure}
%%%%%%%%%%%%%%%%%%%%%%%%%%%%%%%%%%%%%%%%%%%%%%%%%%%%%%%%%%%%%%%%%%%%%%%%%%%%%%%%

Several observations can be made based on Eqs.~\eqref{eq:SRGpertE} and
\eqref{eq:Vasymptotic}.  First, we have additional confirmation that
the UV energy correction is not universal in the sense that unitarily
equivalent potentials (such as SRG potentials at different $\lambda$
values) will give different corrections, unlike the case for the IR
correction (\eg, see Fig.~21 in Ref.~\cite{More:2013rma}).  We see the
same $\Lambda$ dependence at this level as in the separable case.
Therefore, the same analysis should apply when looking at the
dependence of the energy correction in the asymptotic regime where
$\Lambda > \lambda$.  Note also that at LO (at least) we
should find the correction is a function of $\Lambda/\lambda$.  Both
of these are consistent with numerical studies of the SRG-evolved
deuteron energy in this regime with the SRG Hamiltonian cut off at
$\Lambda > \lambda$.  For example, in Fig.~\ref{fig:deuteron_fourth},
the relative error in the deuteron is plotted as a function of
$(\Lambda_2/\lambda)^4$ for SRG-evolved potentials ranging from
$\lambda = 2.6\,\fmi$ to $\lambda = 1.6\,\fmi$.  The inset shows how
different the corrections are as a function of the \emph{unscaled}
$\Lambda_2$.  When scaled, the errors largely coincide for $\Lambda_2
\leq \lambda$ for all three potentials, but up to much higher cutoffs
for the two lower values of $\lambda$ (and for any $\lambda$ below about
$2.2\,\fmi$).  Apparently a sufficient degree of evolution is needed
to modify the high-momentum tail of the potential so that it follows
the universal SRG asymptotic form for the correction ($\propto
\ee^{-2(\Lambda_2/\lambda)^4}$).

\subsection{Gaussian ansatz for UV extrapolations}
\label{subsec:gaussian}

Based on the results in the last section, if we are in the asymptotic
region where $\Lambda\gg\lambda$, we would not expect to find that the
energy behaves like Eq.~\eqref{eq:gaussansatz}, but for SRG-evolved
potentials roughly like $\ee^{-2(\Lambda/\lambda)^4}$ times some
slower-varying function of $\Lambda$.  This is verified in
Fig.~\ref{fig:deuteron_fourth}.  More generally the separable
extrapolation has the form of an integral and not a simple functional
form; so how might an approximate Gaussian dependence on $\Lambda_2$
arise?

The key is that in practice UV extrapolations have typically been
applied in a \emph{limited}, non-asymptotic region {$\Lambda_{\rm min}
  < \Lambda < \Lambda_{\rm max}$} for which $\Lambda/\lambda$ is about
unity (\eg, past NCSM fits were in the range $0.7 < \Lambda/\lambda <
1.1$ and the fit was primarily determined by the points at the lower
end~\cite{Furnstahl:2012qg,Jurgenson:2013yya}).  While we expect
$\Delta E_\Lambda$ to decrease rapidly with increasing $\Lambda$,
$\log\Delta E_\Lambda$ should be well approximated by a low-order
Taylor expansion in a small region.  If $\Delta E_\Lambda$ is a
function only of $\Lambda^2$ rather than $\Lambda$, then by keeping
only through the linear term in the $\Lambda^2$ expansion we will have the 
phenomenological Gaussian ansatz for $\Delta E_\Lambda$, with a prediction for 
$b_1$ possible from our separable expansion formalism.

%%%%%%%%%%%%%%%%%%%%%%%%%%%%%%%%%%%%%%%%%%%%%%%%%%%%%%%%%%%%%%%%%%%%%%%%%%%%%%%%
\begin{figure}[thbp]
 \centering
 \includegraphics[width=0.95\columnwidth]%
 {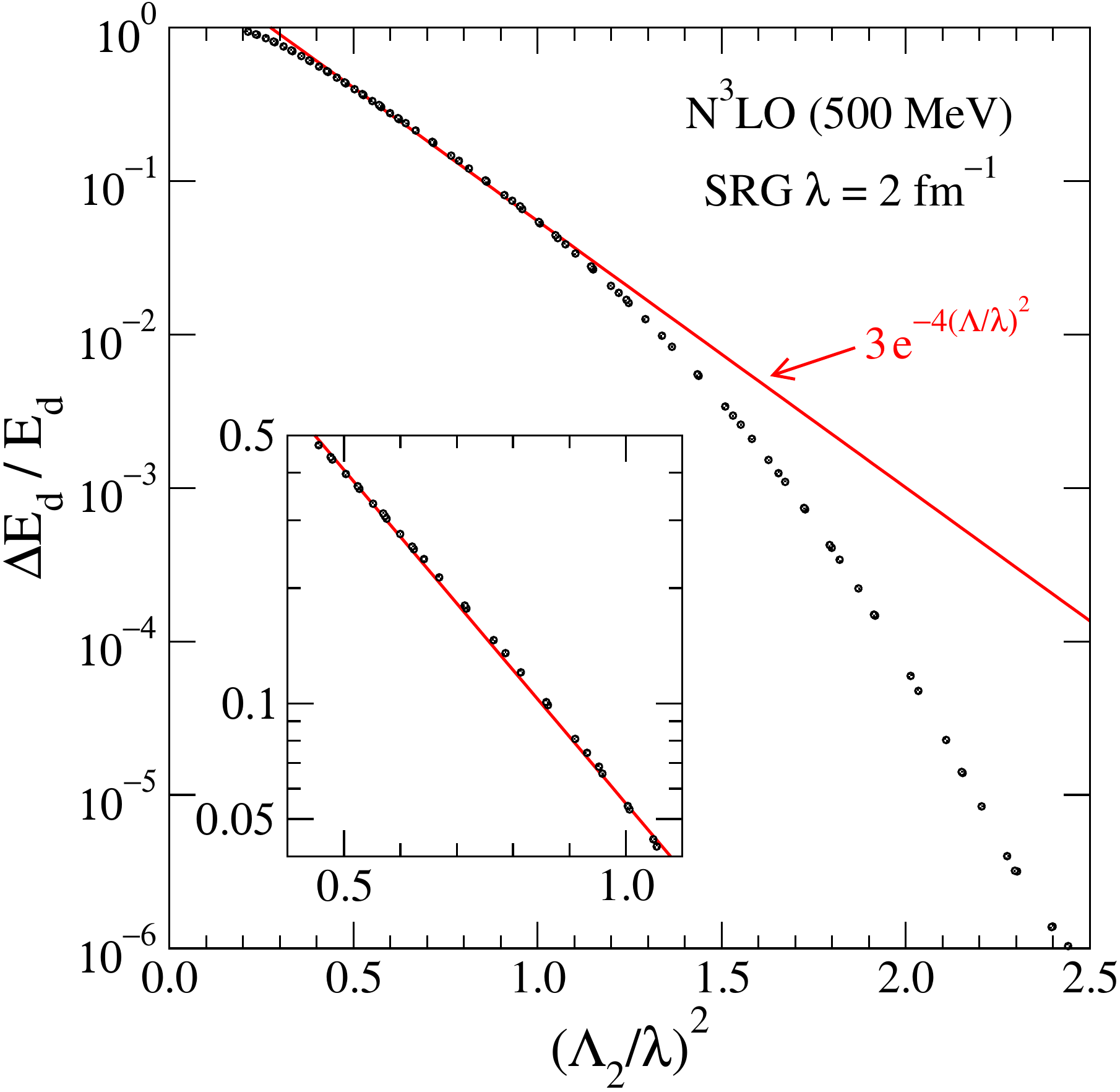}
 \caption{(Color online) Relative error for the deuteron energy from HO basis
 truncation as a function of $(\Lambda_2/\lambda)^2$ for $(\Nmax,\hw)$ values 
 for which the IR correction can be neglected.  The potential is the same as in 
 Fig.~\ref{fig:Deut-EM500}. The solid line is an approximate fit to a region 
 near $\Lambda_2/\lambda = 1$.
 }
\label{fig:deuteron_gaussian_test}
\end{figure}
%%%%%%%%%%%%%%%%%%%%%%%%%%%%%%%%%%%%%%%%%%%%%%%%%%%%%%%%%%%%%%%%%%%%%%%%%%%%%%%%

We first consider separable potentials and demonstrate that $\Delta E_\Lambda = 
\Delta E_\Lambda(\Lambda^2)$ for any $f_\lambda(k)$ that is a function of 
$k^2$.  (For $\ell>0$, we expect that $f_\lambda(k)$ will be of the form 
$k^\ell$ times a function of $k^2$, and the demonstration is trivially 
generalized.)  We start with Eq.~\eqref{eq:deltaE}, which we expect to be 
quantitatively accurate in the region of interest.  We first make the $\Lambda$ 
dependence explicit in the limits and then change variables to $u = k/\Lambda$:
\begin{eqnarray}
  \Delta E_\Lambda &=& 
   \frac{\ds\int\! \dd^3 k\, 
     \frac{\ts f^2_\lambda(k)\Theta(k-\Lambda)}{\ts \kapinf^2 + k^2}
   }
   {\ds\int\! \dd^3 k\, 
     \frac{\ts f^2_\lambda(k)\Theta(\Lambda-k)}{\ts (\kappa^2_\infty+k^2)^2}}
   =  
      \frac{\ds\int_{\Lambda}^{\infty}\! \dd k\, 
     \frac{\ts k^2  f^2_\lambda(k)}{\ts \kapinf^2 + k^2}
   }
   {\ds\int_0^{\Lambda}\! \dd k\, 
     \frac{\ts k^2 f^2_\lambda(k)}{\ts (\kappa^2_\infty+k^2)^2}}
     \nonumber \\ 
   &=&
      \frac{\ds\int_{1}^{\infty}\! \dd u\, 
     \frac{\ts u^2 f^2_\lambda(u\Lambda)}{\ts \kapinf^2/\Lambda^2 + u^2}
   }
   {\ds\frac{1}{\Lambda^2}\int_0^{1}\! \dd u\, 
     \frac{\ts u^2 f^2_\lambda(u\Lambda)}{\ts (\kapinf^2/\Lambda^2 + u^2)^2}}
     \,.
    \label{eq:deltaE_explicit} 
\end{eqnarray}
But by assumption $f_\lambda(u\Lambda)$ depends only on the argument
squared and therefore only on $\Lambda^2$, so we have shown $\Delta
E_\Lambda = \Delta E_\Lambda(\Lambda^2)$.  Next we write:
\begin{multline}
  \!\!\!\!\! g(\Lambda^2) \equiv \log\Delta E_\Lambda(\Lambda^2) \\
  \,= \log \int_{\Lambda}^{\infty}\! \dd k\, 
     \frac{\ts k^2  f^2_\lambda(k)}{\ts \kapinf^2 + k^2}
     - \log \int_0^{\Lambda}\! \dd k\, 
     \frac{\ts k^2 f^2_\lambda(k)}{\ts (\kappa^2_\infty+k^2)^2}
     \,,
\end{multline}
and expand about $\Lambda^2 = \LambdaExpand^2$,
\begin{equation}
  g(\Lambda^2) = g_0 + g_1 (\Lambda^2 - \LambdaExpand^2)
      + \frac12 g_2 (\Lambda^2 - \LambdaExpand^2)^2 + \cdots
      \,.
\end{equation}
Truncating at the linear term, we obtain (with $b_1 = -g_1$)
\begin{equation}
  \Delta E_\Lambda = [\ee^{(g_0 - g_1\LambdaExpand^2)}]\, \ee^{g_1 \Lambda^2}
    = (\mbox{const.})\times\ee^{-b_1 \Lambda^2} \;,
    \label{eq:gaussianform}
\end{equation}
which is the Gaussian form we are looking for.  We can directly
evaluate the $g_i$ for $i>0$ using
\begin{equation}
  \frac{\dd}{\dd\Lambda^2} = \frac{1}{2\Lambda} \frac{\dd}{\dd\Lambda}
  \;.
\end{equation}
Thus,
\begin{eqnarray}
 g_1 &=& \left.\frac{\dd g}{\dd\Lambda^2}\right|_{\LambdaExpand^2}
  \nonumber \\
   &=& \frac{1}{2\LambdaExpand}
   \left\{
    \frac{ -f_\lambda^2(\LambdaExpand) \frac{\ts\LambdaExpand^2}
    {\ts\kapinf^2 + \LambdaExpand^2}}
         {\ds \int_{\LambdaExpand}^{\infty}\! d k\, 
     \frac{\ts k^2  f^2_\lambda(k)}{\ts \kapinf^2 + k^2}} \right.
  \nonumber \\
    & & \qquad\ \null - \left.
     \frac{f_\lambda^2(\LambdaExpand) \frac{\ts\LambdaExpand^2}
     {\ts(\kapinf^2 + \LambdaExpand^2)^2}}
      {\ds\int_0^{\LambdaExpand}\! \dd k\, 
     \frac{\ts k^2 f^2_\lambda(k)}{\ts (\kappa^2_\infty+k^2)^2}}
   \right\} < 0 \,.
\end{eqnarray}
Note that this is a negative-definite function of $\LambdaExpand^2$
(\eg, change variables again to $u = k/\LambdaExpand$), so $b_1 > 0$.
Finally, let us consider $g_2$.  We need the second derivative
of $g$:
\begin{multline}
 \frac{\dd}{\dd\Lambda^2}
  \left(\frac{1}{\Delta E_\Lambda}
  \frac{\dd\Delta E_\Lambda}{\dd\Lambda^2}\right)
 = \frac{\dd}{\dd\Lambda^2}
  \left(\frac{\dd\log\Delta E_\Lambda}{\dd\Lambda^2}\right) \\
 = -\frac{1}{\Delta E_\Lambda^2}
  \left(\frac{\dd\Delta E_\Lambda}{\dd\Lambda^2}\right)^{\!2}
   + \frac{1}{\Delta E_\Lambda}\frac{\dd^2\Delta E_\Lambda}{\dd(\Lambda^2)^2}
      \,.
   \label{eq:gderiv}   
\end{multline}
Now the first term on the right-hand side of the last equality is negative
definite.  In the other term, $\Delta E_\Lambda(\Lambda^2)$ is positive
definite and the curvature with respect to $\Lambda^2$ is positive.
So we expect cancellation here for $\LambdaExpand \approx \lambda$,
which is verified numerically.  With $g_2$ small, the linear
approximation and therefore the Gaussian ansatz are valid.  An example
showing the Gaussian region for an SRG potential is given in
Fig.~\ref{fig:deuteron_gaussian_test}, for which $b_1 = 4/\lambda^2$
is found to be a good fit, with $g_2 \approx 0$.  This same value
works with other light nuclei.  Note that when fitting to the
functional form $E(\Lambda) = \Einf + B_0 \ee^{-b_1 \Lambda^2}$, the
choice of $\LambdaExpand$ is made implicitly by the fit to $B_0$ and
$b_1$.

\begin{figure}[thbp]
 \centering
 \includegraphics[width=0.95\columnwidth]{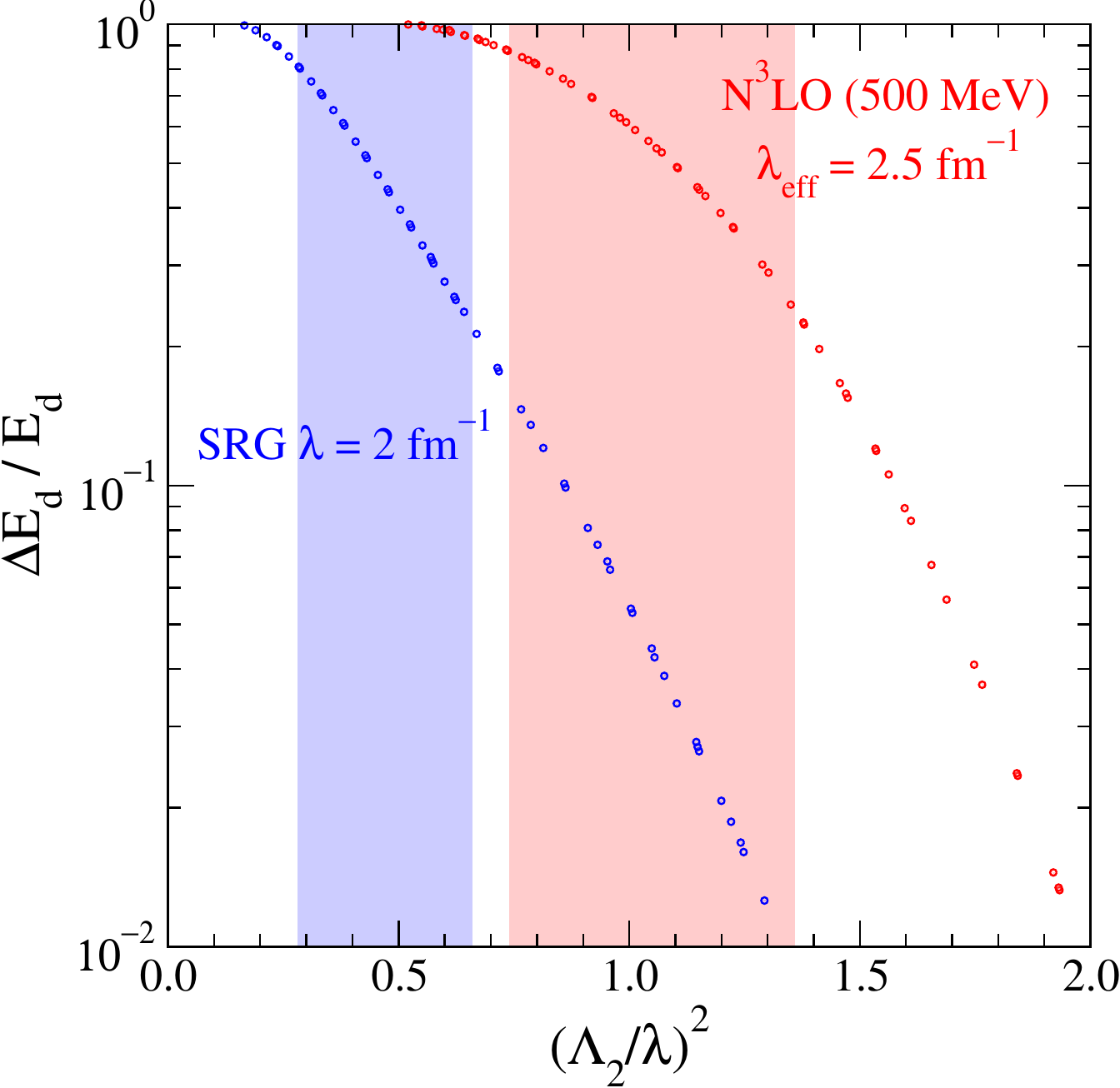}
 \caption{(Color online) Relative error for the deuteron energy from HO basis
   truncation as a function of $(\Lambda_2/\lambda)^2$ for
   $(\Nmax,\hw)$ values for which IR corrections can be neglected for
   the SRG-evolved N3LO potential by Entem and Machleidt (left) and
   the bare potential (right). Also indicated are the regions where the
   fit to a Gaussian was attempted.}
\label{fig:gaussian_fit_comparison}
\end{figure}

Let us briefly speculate why the Gaussian fit does not work well in
general (see, \eg, the right panel in Fig.~\ref{fig:Deut-EM500} and
both panels in \ref{fig:Deut-EGM550}). For the Gaussian fit to be
applicable, $g_2$ needs to be sufficiently small that $g_1$
dominates for an accessible range of $\Lambda$. This condition is not
met in general: Figure~\ref{fig:gaussian_fit_comparison} shows the
relative error of the deuteron binding energy as a function of
$(\Lambda_2/\lambda)^2$, and the shaded regions indicate where the
Gaussian fit was attempted (compare to Fig.~\ref{fig:Deut-EM500}).
The condition that $g_1$ dominates means that $\Delta E_d/E_d$ is well 
approximated by a straight line, which is satisfied for the SRG-evolved 
potential (left shaded region) but not the unevolved potential (right
shaded region).

%%%%%%%%%%%%%%%%%%%%%%%%%%%%%%%%%%%%%%%%%%%%%%%%%%%%%%%%%%%%%%%%%%%%%%%%%%%%%%%%
%%%%%%%%%%%%%%%%%%%%%%%%%%%%%%%%%%%%%%%%%%%%%%%%%%%%%%%%%%%%%%%%%%%%%%%%%%%%%%%%

\section{Further remarks}
\label{sec:Remarks}

Before we summarize our results and conclude in the next section, we return
here to some general remarks about the separable-approximation approach 
introduced in Sec.~\ref{sec:Separable}.

\subsection{More general derivation}

While it was instructive to derive our general extrapolation formulas based on 
writing down a separable approximation for the original potential and then 
taking over the results obtained for explicitly separable interactions from 
Sec.~\ref{sec:RegContact}, we can actually also take a more direct approach.  
If we consider a Hamiltonian $H = H_0 + V$ giving rise to a bound state 
$\ket{\psi}$ with binding energy $-E_B = -\kappa_\infty^2$, we can write the 
Schrödinger equation as
\begin{equation}
 \ket{\psi} = G_0(-\kappa_\infty^2)V \ket{\psi} \,,
\label{eq:SG-G0}
\end{equation}
where $G_0$ is the Green's function (free resolvent)
\begin{equation}
 G_0(z) = (z - H_0)^{-1} \,.
\end{equation}
Acting with $V$ on both sides and taking the matrix element with $\bra{\psi}$, 
we get
\begin{equation}
 1 = \frac{\bra{\psi}V G_0(-\kappa_\infty^2) V\ket{\psi}}
 {\mbraket{\psi}{V}{\psi}} \,.
\label{eq:quant-eta-gen}
\end{equation}
This already looks similar to our unitary-pole-approximation 
potential~\eqref{eq:UPA-op}.  Indeed, if we define $g \equiv 
\mbraket{\psi}{V}{\psi}^{-1}$ and $\ket{\eta} \equiv V\ket{\psi}$, we get
\begin{equation}
 1 = g \times \mbraket{\eta}{G_0(-\kappa_\infty^2)}{\eta} \,,
\label{eq:quant-eta-op}
\end{equation}
or, explicitly in momentum space,
\begin{equation}
 {-}1 = 4\pi g \int\dd k\,\frac{k^2\,\eta(k)^2}{\kappa_\infty^2 + k^2} \,.
\label{eq:quant-eta}
\end{equation}
Our extrapolation formulas follow from this if we assume that cutting off the 
integral at a cutoff $\Lambda$ can be compensated by shifting $\kappa_\infty^2 
\to \kappa_\Lambda^2 = \kappa_\infty^2 - \Delta E_\Lambda$.

Equation~\eqref{eq:quant-eta-op} is furthermore interesting because it might be 
possible to use it for deriving extrapolation relations for bound states 
of more than two particles by considering appropriate many-body Green's 
functions.
Note also that as an alternative to Eq.~\eqref{eq:quant-eta-gen} we can obtain 
from Eq.~\eqref{eq:SG-G0} a quantization condition of the form
\begin{equation}
 1 = \frac{\bra{\psi}G_0(-\kappa_\infty^2) V\ket{\psi}}
 {\braket{\psi}{\psi}} \,.
\label{eq:quant-alt}
\end{equation}
This could be used to derive alternative extrapolation relations that involve 
$\psi(k)\eta(k)$ instead of $\eta(k)^2$.  From the discussion in the following 
subsection, however, it will become clear that Eq.~\eqref{eq:quant-eta} is the 
better choice.

\subsection{The form factors}

%%%%%%%%%%%%%%%%%%%%%%%%%%%%%%%%%%%%%%%%%%%%%%%%%%%%%%%%%%%%%%%%%%%%%%%%%%%%%%%%
\begin{figure*}[thbp]
 \centering
 \includegraphics[width=0.95\columnwidth]{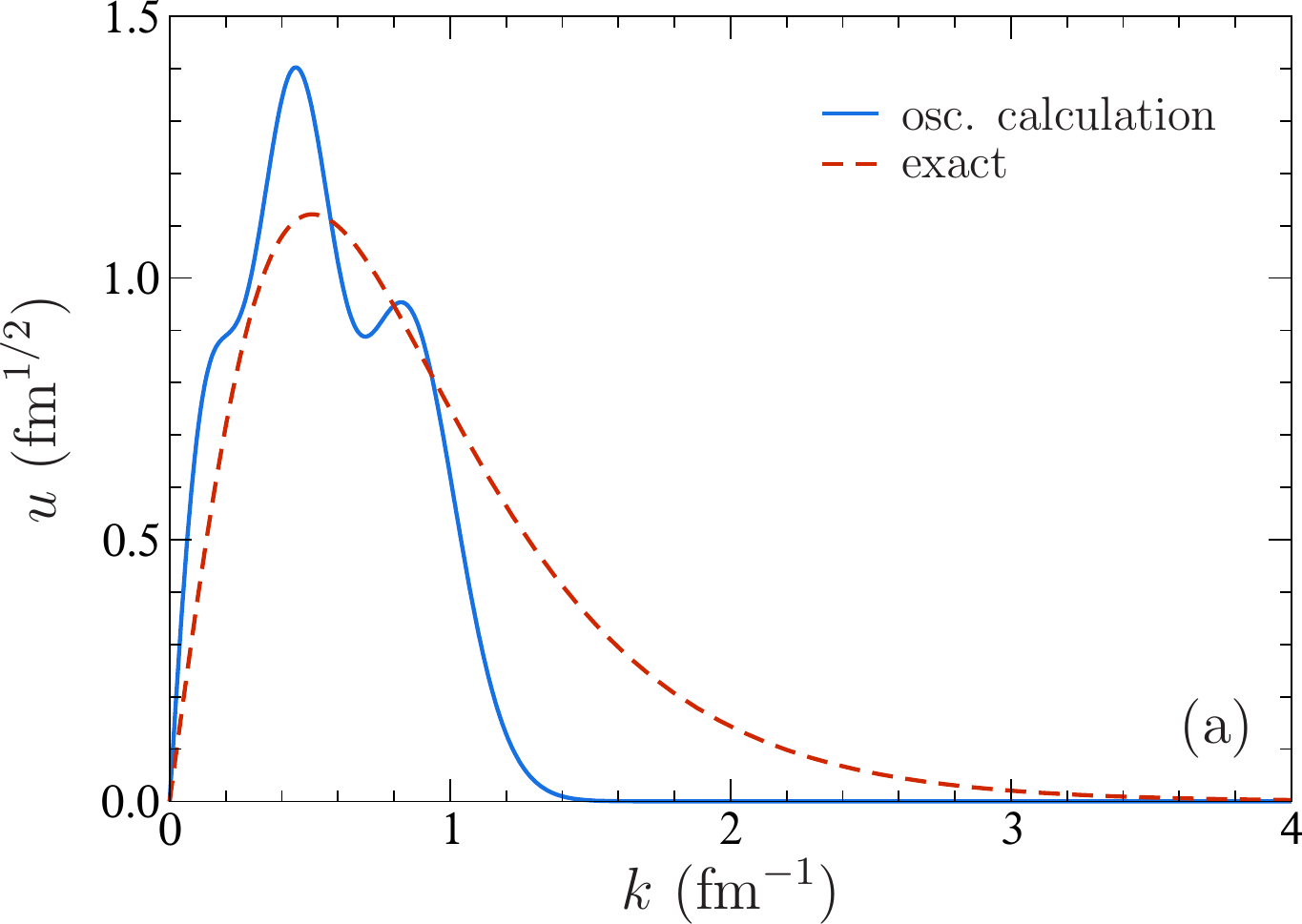}~~~
 \includegraphics[width=0.968\columnwidth]{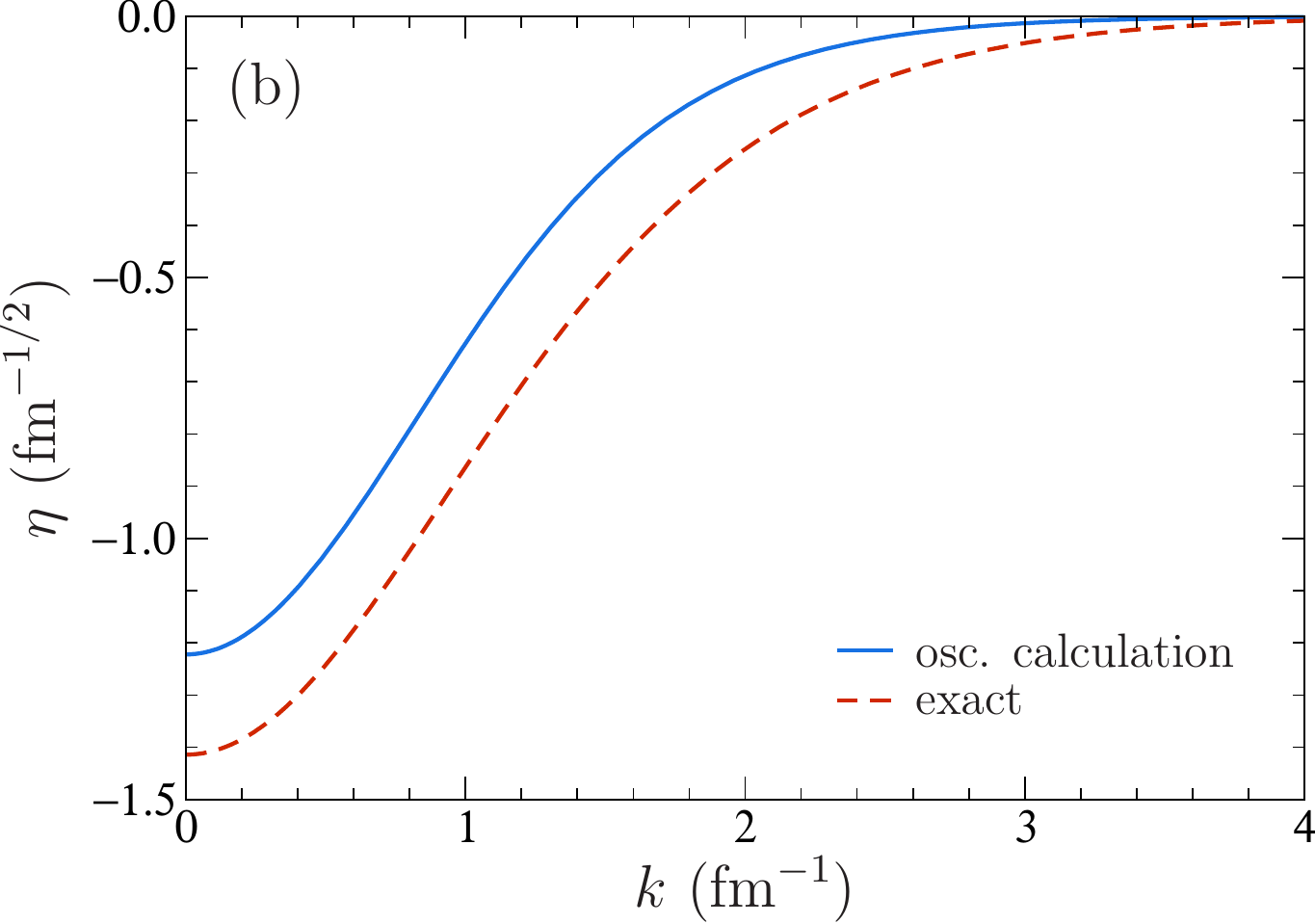}
 \caption{(Color online) (a) Wave functions and (b) corresponding separable 
   form factor for a Pöschl--Teller potential with $\alpha=2/3$ 
   and $\beta=3$.  Solid lines: results from oscillator calculation with 
   $b=4.0~\fm$ and $n=4$.  Dashed lines: exact (analytically known) results for 
   comparison.
 }
\label{fig:u-eta-PT3-23-n4}
\end{figure*}
%%%%%%%%%%%%%%%%%%%%%%%%%%%%%%%%%%%%%%%%%%%%%%%%%%%%%%%%%%%%%%%%%%%%%%%%%%%%%%%%
\begin{figure*}[thbp]
 \centering
 \includegraphics[width=0.95\columnwidth]{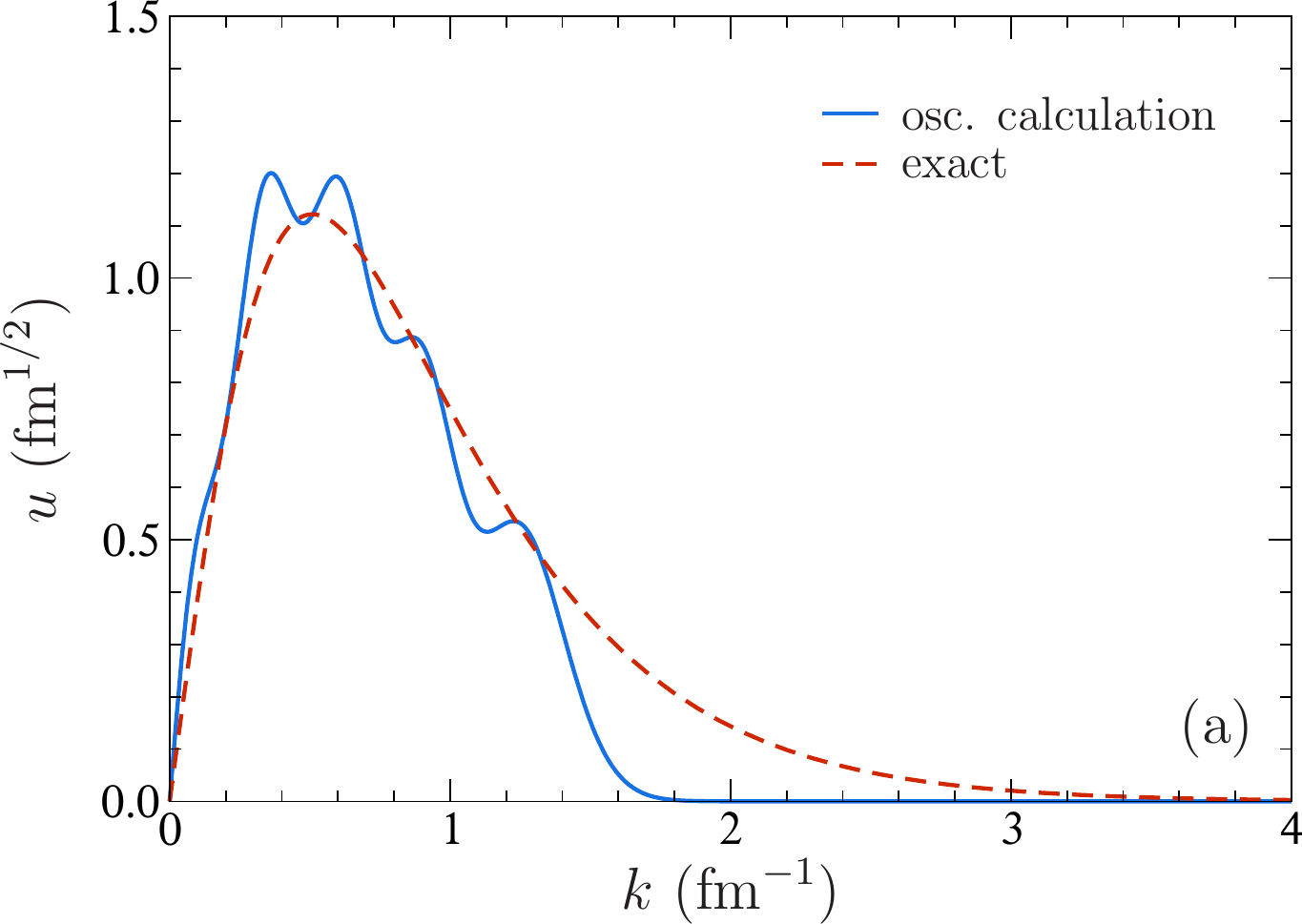}~~~
 \includegraphics[width=0.968\columnwidth]{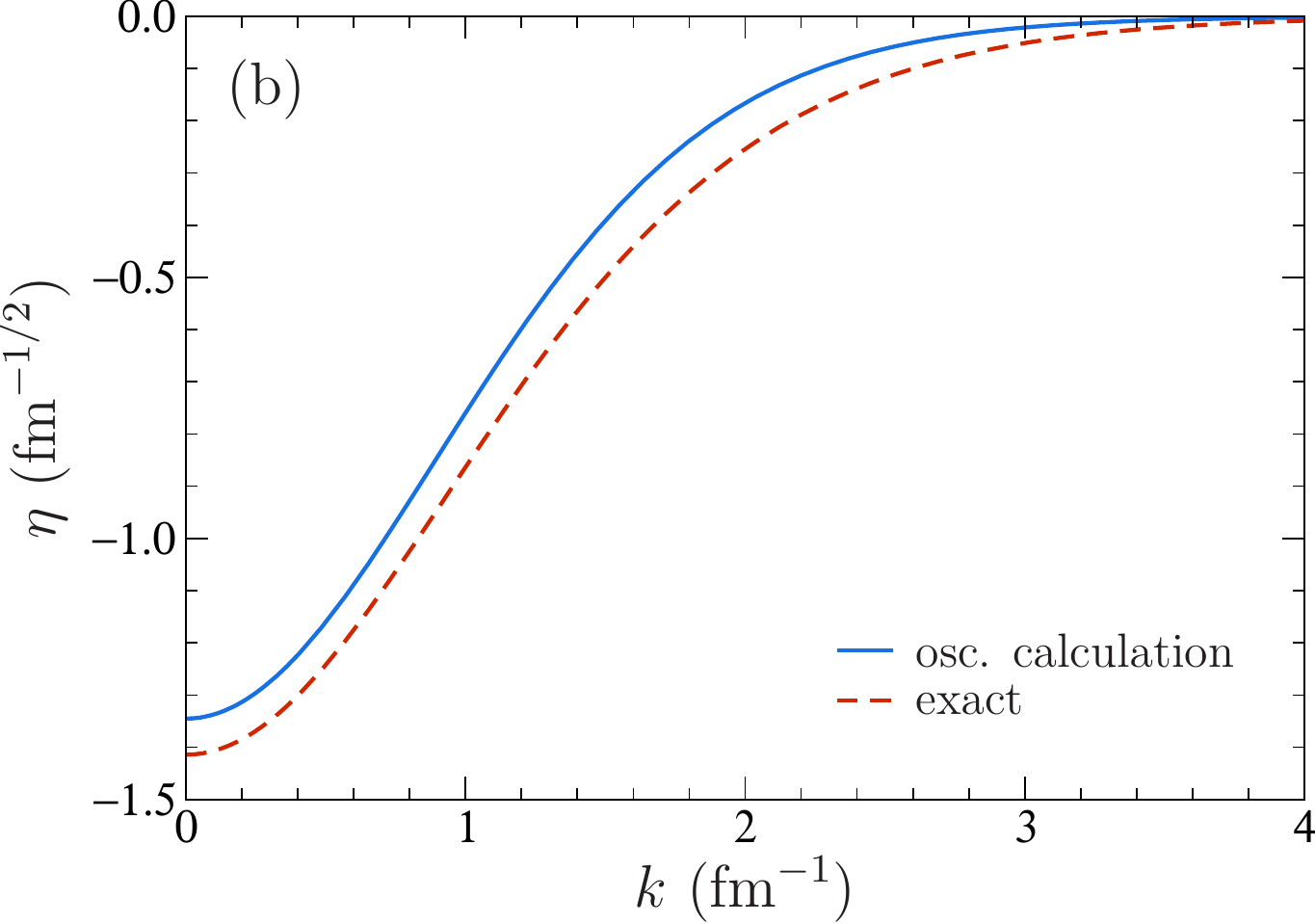}
 \caption{(Color online) (a) Wave functions and (b) corresponding separable 
   form factor for a Pöschl--Teller potential with $\alpha=2/3$ 
   and $\beta=3$.  Solid lines: results from oscillator calculation with 
   $b=4.0~\fm$ and $n=8$.  Dashed lines: exact (analytically known) results for 
   comparison.
 }
\label{fig:u-eta-PT3-23-n8}
\end{figure*}
%%%%%%%%%%%%%%%%%%%%%%%%%%%%%%%%%%%%%%%%%%%%%%%%%%%%%%%%%%%%%%%%%%%%%%%%%%%%%%%%

If we look at the definition of the form factors $\eta(k)$ and assume 
that the state $\ket{\psi}$ is an exact solution of the Schrödinger equation 
(without truncation artifacts), it is clear that we can rewrite
\begin{multline}
 \eta(k) = \mbraket{k}{V}{\psi}
 = \mbraket{k}{G_0(-\kappa_\infty^2)^{-1}}{\psi} \\
 = (-\kappa_\infty^2-k^2) \psi(k) \,.
\end{multline}
Using this, our extrapolation formulas can be rewritten in terms of $\psi(k)$ 
instead of $\eta(k)$, thus eliminating the explicit dependence on the 
potential.  In our numerical calculations, however, we only have approximate 
solutions to the Schrödinger equation.  While in principle one can carry out the 
above manipulations before making the \emph{approximation} of using the 
numerically-determined wavefunctions, it turns out that in practice it works 
much better to use the extrapolations based on $\eta(k)$ unless the calculation 
is pretty much converged already.

The reason for this is likely that while the momentum-space wavefunction that 
comes out of a non-UV-converged oscillator calculation exhibits some unphysical 
structure due to truncation artifacts, the form factors calculated from it are 
still very smooth; integrating $\psi$ with the potential $V$ essentially 
removes the truncation artifacts.  The effect is shown in 
Figs.~\ref{fig:u-eta-PT3-23-n4} and~\ref{fig:u-eta-PT3-23-n8} where we plot 
wavefunctions $u(k)$ and the corresponding separable form factors $\eta(k)$ as 
functions of $k$.  The results were obtained using a Pöschl--Teller potential 
with $\alpha=2/3$ and $\beta=3$ in truncated oscillator bases with $b = 
4.0~\fm$.  Clearly, even if the wavefunction is far from being converged, the 
corresponding $\eta(k)$ is smooth and close in shape to the known exact 
function (dashed curve).  Note also that the UV cutoffs $\Lambda_2 \approx 
1.2~\fmi$ ($n = 4$, Fig.~\ref{fig:u-eta-PT3-23-n4}) and $\Lambda_2 \approx 
1.6~\fmi$ ($n = 8$, Fig.~\ref{fig:u-eta-PT3-23-n8}) are clearly visible in 
the oscillator-based wavefunction.  Beyond the cutoff, they are essentially 
$0$, which means that they are not suitable for extrapolations to larger 
cutoffs.  This is different for the form factors, which still have 
high-momentum tails.

Finally, this analysis also shows that an extrapolation formula based on 
Eq.~\eqref{eq:quant-alt}, featuring the product $u(k)\eta(k)$ would not work 
well with wavefunctions obtained from the truncated oscillator calculation.

\section{Summary and outlook}
\label{sec:Outlook}

In this paper, we have developed a theoretical basis for UV errors
in truncated harmonic oscillator spaces.  We used the two-particle
system with model potentials and deuteron calculations with realistic potentials
as solvable theoretical laboratories to develop and test extrapolation schemes.
By studying the two-body system in great detail, we follow the
successful strategy of Refs.~\cite{More:2013rma} and~\cite{Furnstahl:2013vda}, 
which has recently led to successful extensions to the many-body 
sector~\cite{Furnstahl:2014aaa}.  First we established that the spectrum of the 
squared position operator in a finite oscillator basis is the same as that of a 
system with a hard cutoff in momentum.  This is the dual result to the IR, where 
the spectrum of the squared momentum operator in a truncated oscillator basis 
coincides with that of a spherical box with a hard wall at radius $L_2$.  
Matching the lowest eigenvalues establishes the cutoff $\Lambda_2$, which was 
determined in a $1/N$ expansion.  By duality, it is the same as $L_2$ (and 
beyond in $1/N$) when expressed in dimensionless units.  The appropriateness of 
$\Lambda_2$ was verified by model and deuteron calculations, which showed a 
smooth curve with little scatter compared to other choices.

Having transferred the problem from calculations in a truncated basis to 
calculations with an imposed sharp momentum cutoff, we turned to rank-one
separable potentials.  For these potentials, we could directly derive an 
analytic formula for the correction to a bound-state eigenvalue in terms of 
integrals over the potential that relied on the correction being small.
This formula was shown to be amenable to perturbation theory and asymptotic
expansions when $\Lambda_2$ is greater than the intrinsic UV scale of the 
potential.  This is useful for general tests and to establish that the UV 
correction depends on the high-momentum behavior of the potential.

But the true region of interest is when $\Lambda_2$ is comparable to or 
smaller than this scale.  In this case, the integral expressions can be used to 
parametrize extrapolation formulas to fit.  A new procedure was developed to 
generalize this extrapolation method to any potential by adapting the unitary 
pole approximation.  Tests for model potentials as well as for the deuteron with 
realistic potentials are very encouraging.  Finally, we showed how the simple 
Gaussian phenomenological extrapolation widely used in the past can be 
recovered from an expansion of the separable potential.

The IR and UV corrections exhibit a complementary mix of universal and
non-universal characteristics.  The IR corrections are dictated by asymptotic 
behavior and are consequently determined by observables, independent of the 
details of the interaction.  So unitarily equivalent potentials---such as those 
generated by renormalization-group running---will have the same corrections.  
In contrast, because they probe short-range features, UV corrections depend on 
the details of the interaction (and the state under consideration).  This was 
manifested here by the different corrections for the deuteron from SRG 
interactions at different resolutions as well as the explicit formulas with 
dependence on the high momentum behavior of the interaction.  

On the other hand, the IR corrections are non-universal with respect to the 
number of nucleons $A$, depending for example on the separation energy of the 
nucleus.  The dependence on $A$ for the UV corrections is not yet established 
theoretically, but fits of the Gaussian ansatz Eq.~\eqref{eq:gaussianform} 
to energies from the same SRG-evolved potential for different values of $A$ 
have been found to have roughly the same value of $b_1$ (approximately equal to 
$4/\lambda^2$, where $\lambda$ is the SRG flow 
parameter~\cite{Furnstahl:2012qg,Jurgenson:2013yya}).  Thus the $\Lambda_2$ 
dependence is the same with $\Delta E_\Lambda$ just scaled by an $A$-dependent 
overall constant.  For $A=2$, $\Delta E_\Lambda$ is determined by the 
short-distance or high-momentum behavior.  For $A>2$, the many-body wave 
function is expected to factorize into a two-body part and a remainder when 
those two particle coordinates are sufficiently close.  This can be understood 
from general considerations of short-range correlations~\cite{Kimball:1973aa} or 
more systematically using the operator product 
expansion~\cite{Bogner:2012zm,Hofmann:2013aa}.  If there is a common two-body 
part, it may determine the dominant $\Lambda_2$ dependence (using the 
separable-approximation approach or at the level of the Gaussian approximation) 
with the rest providing the $A$-dependent scale factor.  This behavior would be 
consistent with the observation of a universal shape for high-momentum tails in 
momentum distributions (or the corresponding short-distance 
behavior)~\cite{Feldmeier:2011qy}.  
This potential UV universality, as well as more direct approaches building on 
the discussion in Sec.~\ref{sec:Remarks}, is the subject of ongoing 
investigations.

\begin{acknowledgments} 
  We thank M.~Caprio, S.~Coon, H.-W.~Hammer, M.~Kruse, R.~Perry,
  R.~Roth, and K.~Wendt for useful discussions.  Ideas that led to this
  publication were exchanged at the conference ``Nuclear Theory in the
  Supercomputing Era (NTSE 2013)'' in Ames, Iowa.  This research was
  supported in part by the National Science Foundation under Grants
  No.~PHY-1068648 (Michigan State University) and No. PHY--1306250 (Ohio
  State University), and by the U.S. Department of Energy, Office of
  Science, Office of Nuclear Physics, under Award Numbers
  DE-FG02-96ER40963 (University of Tennessee), and No.
  DE-SC0008499/DE-SC0008511/DE-SC0008533 (SciDAC-3 NUCLEI
  Collaboration) and under Contract No. DE-AC05-00OR22725 (Oak Ridge
  National Laboratory).
\end{acknowledgments}

\appendix

\section{UV cutoff details}
\label{sec:Lambda-Details}

In this Appendix, we give a detailed derivation of the effective UV 
cutoff $\Lameff$ as a function of the HO parameters (basis 
size $N$ and frequency $\Omega$) that we report in Sec.~\ref{sec:Lambda}.

\subsection{Notation and conventions}

Consider the three-dimensional isotropic harmonic oscillator described
by the Hamiltonian (in natural units with $\hbar=c=1$)
\begin{equation}
 H_\mathrm{HO} = \frac{p^2}{2\mu} + \frac{\mu\Omega^2r^2}{2} \,,
\label{eq:H-HO}
\end{equation}
where $\mu$ is the reduced mass and $\Omega$ denotes the oscillator
frequency.  The eigenstates $\ket{n\ell m}$ of $H_\mathrm{HO}$ are
degenerate in the quantum number $m$,
\begin{equation}
 H_\mathrm{HO}\ket{n\ell m} = E_{n\ell m}\ket{n\ell m}
\label{eq:H-nlm}
\end{equation}
with
\begin{equation}
 E_{n\ell m} = \left(2n+\ell+\frac32\right)\hw \,.
\label{eq:E-nlm}
\end{equation}
We use a slightly modified version of the conventions and notation
from Ref.~\cite{Caprio:2012rv}.  The full three-dimensional wavefunction in 
configuration space is
\begin{equation}
 \psi_{n\ell m}(\vecr) = \braket{\vecr}{n\ell m}
 = \frac{u_{n\ell}(b;r)}{r}\YY_{\ell m}(\hat{\vecr})
\label{eq:psiR}
\end{equation}
with the reduced radial wavefunction
\begin{equation}
 u_{n\ell}(b;r) = N_{n\ell}(b)\times(r/b)^{\ell+1} \ee^{-(r/b)^2/2}
 L_n^{\ell+\nicefrac12}\big((r/b)^2\big) \,,
\label{eq:uR}
\end{equation}
where
\begin{equation}
 N_{n\ell}(b) = \sqrt{\frac{2n!}{b\,\Gamma(n+\ell+\nicefrac32)}} \,,
\label{eq:N}
\end{equation}
and
\begin{equation}
 b = (\mu\Omega)^{-1/2}
\label{eq:b}
\end{equation}
is the oscillator length.  The Fourier transform of Eq.~\eqref{eq:psiR} is
\begin{equation}
 \tilde{\psi}_{n\ell m}(\veck) = (2\pi)^{-\nicefrac32} \int\ddr\,
 \ee^{-\ii\veck\cdot\vecr} \psi_{n\ell m}(\vecr) \,.
\end{equation}
It can be written as
\begin{equation}
 \tilde{\psi}_{n\ell m}(\veck)
 = (-\ii)^\ell\frac{\wt{u}_{n\ell}(b;k)}{k}\YY_{\ell m}(\hat{\veck}) \,,
\label{eq:psiK}
\end{equation}
such that $\wt{u}_{n\ell}(b;k)$ is the Fourier--Bessel transform of 
$u_{n\ell}(b;r)$, \ie,
\begin{equation}
 \wt{u}_{n\ell}(b;k) = \sqrt{\frac{2}{\pi}}\int_0^\infty\dd r'\,
 kr' j_\ell(kr')\,u_{n\ell}(b;r') \,.
\label{eq:uK-FB}
\end{equation}
This gives
\begin{equation}
 \wt{u}_{n\ell}(b;k) = (-1)^n\wt{N}_{n\ell}(b)\times(kb)^{\ell+1}
 \ee^{-(kb)^2/2} L_n^{\ell+\nicefrac12}\big((kb)^2\big)
\label{eq:uK}
\end{equation}
with
\begin{equation}
 \wt{N}_{n\ell}(b) = \sqrt{\frac{2n!\,b}{\Gamma(n+\ell+\nicefrac32)}} \,.
\label{eq:N-tilde}
\end{equation}

\subsection{Smallest eigenvalue of $r^2$}

In the following derivation of $\Lameff$, we directly consider
subspaces with an arbitrary (but fixed) angular momentum $\ell$, but
quote S-wave ($\ell=0$) results explicitly for the sake of
illustration.  Denoting the square root of the smallest eigenvalue
of $r^2$ in the truncated oscillator subspace with angular momentum
$\ell$ by $\rho$,\footnote{Strictly, we should write $\rho_\ell$ here,
but we omit the additional subscript for notational simplicity.} the
localized momentum-space eigenfunction for a hard-wall (Dirichlet)
boundary condition in momentum space is
\begin{equation}
 \wt{\psi}_{\rho,\ell}(p) =
 \begin{cases}
  p\rho\,j_\ell(p\rho) \,, & 0 \leq p \leq x_\ell/\rho \,, \\
  0 \,, & p > x_\ell/\rho \,,
 \end{cases}
\label{eq:psi-rho-ell}
\end{equation}
where $x_\ell$ denotes the smallest positive zero of the spherical
Bessel function $j_\ell$.  For S-waves, one simply has
$\wt{\psi}_{\rho,\ell}(p)=\sin(p\rho)$ and $x_0=\pi$.
The eigenfunction can be expanded in terms of oscillator functions as
\begin{equation}
 \wt{\psi}_\rho(p)
 = \sum_{k=0}^\infty \wt{c}_k(\rho) \wt{u}_{k}(p) \,,
\label{eq:psi-rho-exp}
\end{equation}
without basis truncation so far.  We have used the short-hand 
notation
\begin{equation}
 \wt{u}_n(p) \equiv \wt{u}_{n\ell}(1;p) \,.
\label{eq:u-tilde}
\end{equation}
In particular, we set the oscillator length $b$ to unity for the time
being.  Exactly as in Ref.~\cite{More:2013rma}, the eigenvalue problem
\begin{equation}
 \left[r^2-\rho^2\right]\wt{\psi}_\rho(p) = 0
\label{eq:r2rho2-psi}
\end{equation}
becomes a set of coupled linear equations.  For S-waves, one can use
the fact that the three-dimensional oscillator wavefunctions are
directly related to the (odd) one-dimensional oscillator states and
write
\begin{equation}
 r^2 = a^\dagger a + \frac{1}{2}
 + \frac{1}{2}\left[a^2+(a^\dagger)^2\right] \,,
\label{eq:r2}
\end{equation}
where $a$ and $a^\dagger$ are ladder operators, to obtain (after
shifting some indices)
\begin{widetext}
\begin{multline}
 \left[r^2-\rho^2\right]\wt{\psi}_\rho(p) = 0
 \iff \sum_{k=0}^\infty\left[(2k+3/2-\rho^2)\wt{c}_k(\rho)
 - \frac12\sqrt{2k+1}\sqrt{2k+3}\,\wt{c}_{k+1}(\rho)\right. \\
 \left.-\,\frac12\sqrt{2k}\sqrt{2k+1}\,\wt{c}_{k-1}(\rho)\right]\wt{u}_k(p)
 = 0 \mathspace (\ell = 0) \,.
\label{eq:r2rho2-psi-exp}
\end{multline}
More generally, a direct evaluation yields (\cf the analogous results
for $p^2$ given in Ref.~\cite{Furnstahl:2013vda})
\begin{equation}
 \mbraket{k\ell m}{r^2}{j\ell m} = (2k+\ell+3/2)\delta_k^j 
 + \sqrt{k+1}\sqrt{k+\ell+3/2}\,\delta_k^{j+1} 
 + \sqrt{k}\sqrt{k+\ell+1/2}\,\delta_k^{j-1} \,,
\label{eq:r2-gen}
\end{equation}
and thus we get
\begin{multline}
 \left[r^2-\rho^2\right]\wt{\psi}_\rho(p) = 0
 \iff \sum_{k=0}^\infty\left[(2k+\ell+3/2-\rho^2)\wt{c}_k(\rho)
 - \sqrt{k+1}\sqrt{k+\ell+3/2}\,\wt{c}_{k+1}(\rho)\right. \\
 \left.-\sqrt{k}\sqrt{k+\ell+1/2}\,\wt{c}_{k-1}(\rho)\right]
 \wt{u}_k(p) = 0
\label{eq:r2rho2-psi-exp-ell}
\end{multline}
\end{widetext}
for arbitrary angular momentum $\ell$.

If the basis---and thus the sum in Eq.~\eqref{eq:r2rho2-psi-exp-ell}---is now
truncated at some maximum $k \equiv n$, the last 
equation of the coupled set reads
\begin{equation}
 (2n+\ell+3/2-\rho^2)\,\wt{c}_{n}(\rho)
 - \sqrt{n}\sqrt{n+\ell+1/2}\,\wt{c}_{n-1}(\rho) = 0 \,.
\label{eq:quant-cond-ell}
\end{equation}
Further following Ref.~\cite{Furnstahl:2013vda}, we introduce the 
Fourier--Bessel transform of $\wt{\psi}_\rho(p)$ as
\begin{equation}
 \wt{\psi}_\rho(p)
 = \sqrt{\frac{2}{\pi}}\int_0^\infty \dd r\,\psi_\rho(r)\,pr\,j_\ell(pr) \,,
\label{eq:psi-tilde-psi}
\end{equation}
and use
\begin{equation}
 pr\,j_\ell(pr) = \sqrt{\frac{\pi}{2}}\sum_{n=0}^\infty 
 \wt{u}_{n\ell}(b;p)u_{n\ell}(b;r) \mathtext{for arbitrary $b$}
\label{eq:j-uKuR}
\end{equation}
to infer
\begin{equation}
 \wt{c}_n(\rho) = \int_0^\infty\dd r\,\psi_\rho(r) u_n(r) 
\label{eq:c-tilde-1}
\end{equation}
from Eq.~\eqref{eq:psi-rho-exp}.  To proceed, we use the asymptotic 
approximation~\cite{Furnstahl:2013vda,Deano2013}
\begin{multline}
 u_{n\ell}(b;r) \approx
 \frac{2^{1-n}}{\pi^{1/4}} \sqrt{\frac{(2n+2\ell+1)!}{b\,(n+\ell)!n!}}
 (4n-2\ell+3)^{-\frac{\ell+1}{2}} \\
 \times\sqrt{4n+2\ell+3}\,(r/b)\,j_\ell\left(\sqrt{4n+2\ell+3}\,(r/b)\right) \,,
\label{eq:uR-asympt}
\end{multline}
valid for $n\gg 1$.  Defining
\begin{equation}
 \beta_\ell = \sqrt{4n+2\ell+3}
\label{eq:beta-ell}
\end{equation}
and still setting $b=1$ at this point, we get
\begin{equation}
\begin{split}
 \wt{c}_n(\rho) &\approx
 \frac{2^{1-n}}{\pi^{1/4}} \sqrt{\frac{(2n+2\ell+1)!}{(n+\ell)!n!}}
 \\ & \quad\null
 \times\beta_\ell^{-\ell-1}\int_0^\infty\dd r\,\psi_\rho(r)\,
 \beta_\ell r \, j_\ell(\beta_\ell r) \\
 &= \frac{2^{1-n}}{\pi^{1/4}} \sqrt{\frac{(2n+2\ell+1)!}{(n+\ell)!n!}}
 \times\beta_\ell^{-\ell-1} \sqrt{\frac\pi2}\,\wt{\psi}_\rho(\beta_\ell) \\
 &= \frac{\pi^{1/4}}{2^{n-1/2}}\sqrt{\frac{(2n+2\ell+1)!}{n!(n+\ell)!}}
 \times\beta_\ell^{-\ell} \rho\,j_\ell(\beta_\ell\rho)
 \;.
\end{split}
\label{eq:c-tilde-2}
\end{equation}
The intermediate and final steps here follow from
Eqs.~\eqref{eq:psi-tilde-psi} and~\eqref{eq:psi-rho-ell},
respectively, and we have the constraint $\rho<x_\ell/\beta_\ell$.
Inserting Eq.~\eqref{eq:c-tilde-2} into the quantization
condition~\eqref{eq:quant-cond-ell} gives an equation that is formally
exactly the same as given in Ref.~\cite{Furnstahl:2013vda} for the
IR case.\footnote{Some relative minus signs---compare, for example,
Eq.~\eqref{eq:quant-cond-ell} to Eq.~(31) in 
Ref.~\cite{Furnstahl:2013vda}---have dropped out along the way.  Note also
that Ref.~\cite{Furnstahl:2013vda} uses a slightly different convention for
the momentum-space oscillator wavefunctions that does not involve the phase
$(-1)^n$ in our Eq.~\eqref{eq:uK}.}

\subsection{Cutoff identification}

If we make the ansatz
\begin{equation}
 \rho = \frac{x_\ell}{\sqrt{4n+2\ell+3+2\Delta}} \,,
\label{eq:rho-ansatz}
\end{equation}
we get $\Delta=2$ in the limit $n \gg 1$ and $n\gg\ell$, independent
of $\ell$.  As we discuss in Appendix~\ref{sec:Subleading}, it is possible to 
derive subleading corrections to this result, which then depend on the
angular momentum $\ell$, but turn out to be numerically insignificant
for all present practical applications.

With $\Nmax=2n+\ell$, and restoring the oscillator length $b$ by 
dimensional analysis, our result can also be written as
\begin{equation}
 \rho = \frac{x_\ell b}{\sqrt2}\left(\Nmax+\frac32+2\right)^{\!-1/2} \,.
\label{eq:rho-Nmax}
\end{equation}
This implies that the UV cutoff $\Lameff$ corresponding to the basis
truncation at $\Nmax$ is not given by the naive estimate
\begin{equation}
 \Lambda_0 = \sqrt{2(\Nmax+3/2)}/b \,,
\label{eq:Lambda-0-app}
\end{equation}
which follows from $k = \sqrt{2\mu E}$ and Eq.~\eqref{eq:E-nlm}, but rather by
\begin{equation}
 \Lambda_2 = \frac{x_\ell}{\rho} = \sqrt{2(\Nmax+3/2+2)}/b \,,
\label{eq:Lambda-2-app}
\end{equation}
completely dual to the configuration-space box size $L_2$ given in 
Eq.~\eqref{eq:L2_def}.

\subsection{Subleading corrections to $L_2$ and $\Lambda_2$}
\label{sec:Subleading}

It is possible to derive subleading corrections to the result $\Delta=2$ that 
was derived in the previous subsection.  Because of the duality of 
configuration-space and momentum-space oscillator wavefunctions, the results 
derived in the following apply directly also to the effective box size $L_2$ 
used to calculate IR corrections.

\medskip
For the smallest eigenvalue $\rho^2$ of the operator $r^2$ in the (truncated) 
oscillator basis we now wish to make the general ansatz
\begin{equation}
 \rho = \frac{x_\ell}{\sqrt{4n+2\ell+3+2
 \left(\Delta_0+\dfrac{\Delta_1}{n}+\dfrac{\Delta_2}{n^2}+\cdots\right)}} \,.
\label{eq:rho-ansatz-gen}
\end{equation}
In principle, there is an infinite sum of terms with increasing inverse powers 
of $n$ in Eq.~\eqref{eq:rho-ansatz-gen}, but we only give explicit results here
up to $\OO(1/n^2)$.

In Sec.~\ref{sec:Lambda-2}, the result $\Delta=\Delta_0=2$ was found by 
inserting \eqref{eq:c-tilde-2} into the quantization 
condition~\eqref{eq:quant-cond-ell} and then considering the limits $n\gg1$ and 
$n\gg\ell$.  In practice, this is done by inserting the ansatz for 
$\rho=\rho(n)$ into $\tilde{c}_n(\rho)\sim \rho\,j_\ell(\beta_\ell\rho)$ and 
keeping only the leading term in an asymptotic expansion around $n=\infty$.

To obtain the desired subleading corrections, it is however not sufficient to 
simply keep higher-order terms in this asymptotic expansion.  Instead, one 
first has to go back a few steps and also keep higher-order corrections to the 
leading asymptotic approximation for the oscillator wavefunctions given in 
Eq.~\eqref{eq:uR-asympt}.  Note that this approximation follows from using 
Eq.~(15) of Ref.~\cite{Deano2013}, which states that the generalized 
Laguerre polynomials have the asymptotic expansion
%
%\begin{widetext}
\begin{multline}
 L_n^\alpha(z) = \frac{\Gamma(n+\alpha+1)}{n!}\ee^{z/2}
 \sum\limits_{m=0}^\infty \left(\frac{z}{2}\right)^m P_m(\alpha+1,z) \\
 \times(\kappa z)^{-\frac{m+\alpha}{2}} J_{m+\alpha}(2\sqrt{\kappa z})
\label{eq:Laguerre-expansion}
\end{multline}
%\end{widetext}
%
with
\begin{subequations}\begin{align}
 \kappa &= n + \frac{\alpha+1}{2} \\ &= n+\frac34 \mathtext{for} \alpha=1/2
\label{eq:kappa}
\end{align}\end{subequations}
and
\begin{equation}
 P_0(c,z) = 1 \mathtext{,} P_1(c,z) = z/6 \mathtext{,} \cdots \,.
\label{eq:P0-P1}
\end{equation}
Using this in Eq.~\eqref{eq:uR} and keeping only the first ($m=0$) term gives 
Eq.~\eqref{eq:uR-asympt}.  More generally, one finds that for large $n$ the 
oscillator wavefunctions $u_{n\ell}(r)$ can be expressed as a sum
\begin{equation}
 u_{n\ell}(r) = u_{n\ell}^{(0)}(r) + u_{n\ell}^{(1)}(r) + \cdots \,,
\label{eq:u-sum}
\end{equation}
where the individual terms involve (spherical) Bessel functions of 
increasing order.  Recalling Eq.~\eqref{eq:c-tilde-1}, it then follows that 
also
\begin{equation}
 \tilde{c}_n(\rho) = \tilde{c}_n^{(0)}(\rho)+\tilde{c}_n^{(1)}(\rho)+\cdots \,.
\label{eq:c-tilde-sum}
\end{equation}
We already know that
\begin{equation}
 \tilde{c}_n^{(0)}(\rho) = C(n)\,\beta_\ell^{-\ell}
 \times\rho\,j_\ell(\beta_\ell\rho)
\label{eq:c-tilde-j0}
\end{equation}
with
\begin{equation}
 C_\ell(n) = \frac{\pi^{1/4}}{2^{n-1/2}}\sqrt{\frac{(2n+2\ell+1)!}
 {n!(n+\ell)!}} \,.
\label{eq:C}
\end{equation}
The key step in deriving Eq.~\eqref{eq:c-tilde-j0} was to express 
$\tilde{c}_n^{(0)}(\rho)$ in terms of $\tilde{\psi}_\rho$ by using the 
Fourier--Bessel transform, which could be done since asymptotically 
$u_{n\ell}^{(0)}(r)$ is simply proportional to $j_\ell(\beta_\ell\rho)$.  More 
generally, for the individual terms in the expansion~\eqref{eq:u-sum} we have
\begin{multline}
 u_{n\ell}^{(k)}(r) = \frac{2^{1-n}}{\pi^{1/4}}
 \sqrt{\frac{(2n+2\ell+1)!}{(n+\ell)!n!}} \beta_\ell^{-(\ell+k)} \\
 \times P_k(\ell+3/2,r^2) r^{k+1} j_{\ell+k}(\beta_\ell r) \,.
\end{multline}
This means that to obtain a generalization of Eq.~\eqref{eq:c-tilde-2}, we have 
to calculate expressions of the form
\begin{multline}
 \tilde{c}_n^{(k)}(\rho) \sim \beta_\ell^{-(\ell+k)}
 \int_0^\infty\dd r\,\psi_\rho(r)\,  P_k(\ell+3/2,r^2) \\
 \times r^{k+1} j_{\ell+k}(\beta_\ell r) \,.
\label{eq:c-tilde-k-1}
\end{multline}
To evaluate these integrals, it is more convenient to work with 
Riccati--Bessel functions,
\begin{equation}
 \hat\jmath_\nu(z) = zj_\nu(z) \,,
\label{eq:j-hat-j}
\end{equation}
in terms of which we have
\begin{multline}
 \tilde{c}_n^{(k)}(\rho) \sim \beta_\ell^{-(\ell+k+1)}
 \int_0^\infty\dd r\,\psi_\rho(r)\,  P_k(\ell+3/2,r^2) \\
 \times r^k \hat\jmath_{\ell+k}(\beta_\ell r) \,.
\label{eq:c-tilde-k-1a}
\end{multline}
For the Riccati--Bessel functions one has the derivative relation~\cite{abramowitz1964}
\begin{equation}
 \frac{\partial \hat\jmath_\nu(z)}{\partial z}
 = \frac{\nu+1}{z} \hat\jmath_\nu(z) - \hat\jmath_{\nu+1}(z) \,,
\label{eq:j-hat-derivative}
\end{equation}
from which it follows straightforwardly that
\begin{equation}
 \hat\jmath_{\nu+1}(\beta r) = \frac{1}{r}
 \left[\frac{\nu+1}\beta-\frac\dd{\dd\beta}\right]\hat\jmath_\nu(\beta r) \,.
\label{eq:j-hat-beta}
\end{equation}

Using this relation $k$ times in Eq.~\eqref{eq:c-tilde-k-1}, we can eliminate 
the prefactor $r^k$ in favor of a differential operator with respect to a 
variable $\beta$,
\begin{multline}
 \tilde{c}_n^{(k)}(\rho) \sim \beta_\ell^{-(\ell+k+1)}
 \int_0^\infty\dd r\,\psi_\rho(r)\,  P_k(\ell+3/2,r^2)\, \\
 \times\left(\frac{\ell+k}{\beta}-\frac{\dd}{\dd\beta}\right)
 \left(\frac{\ell+k-1}{\beta}-\frac{\dd}{\dd\beta}\right)
 \cdots\left(\frac{\ell}{\beta}-\frac{\dd}{\dd\beta}\right) \\
 \null \times \hat\jmath_{\ell}(\beta r)\,\Big|_{\beta=\beta_\ell} \,.
\label{eq:c-tilde-k-1b}
\end{multline}
At this point, we have also conveniently reduced the order of 
the Riccati--Bessel functions so that we have the same function for each 
$\tilde{c}_n^{(k)}(\rho)$; all remaining additional $r$-dependence comes from 
the $P_k(\ell+3/2,r^2)$, which are polynomials in $r^2$.  This can also be 
eliminated by noting that
\begin{equation}
 r^2 \hat\jmath_\ell(\beta r) = \left(-\frac{\dd^2}{\dd\beta^2}
 +\frac{\ell(\ell+1)}{\beta^2}\right)\hat\jmath_\ell(\beta r) \,,
\label{eq:r2-j-hat}
\end{equation}
which follows immediately from the differential equations that defines 
the Riccati--Bessel functions and is formally just the free radial Schrödinger 
equation if one interchanges the variables $r$ and $\beta$.  Altogether, we 
have found that we can write
\begin{equation}
 \tilde{c}_n^{(k)}(\rho) \sim \beta_\ell^{-(\ell+k+1)}\,
 \int_0^\infty\dd r\,\psi_\rho(r)\,\mathcal{D}_{\beta,\ell}^{(k)}\,
 \hat\jmath_{\ell}(\beta r)\,\Big|_{\beta=\beta_\ell} \,,
\label{eq:c-tilde-k-1c}
\end{equation}
where $\mathcal{D}_{\beta,\ell}^{(k)}$ is some differential operator (with 
respect to $\beta$) which can be pulled out of the integral.  The precise form 
of this operator can be obtained from the equations above, but it is actually 
not important here.  At this point we can proceed exactly as in 
Eq.~\eqref{eq:c-tilde-2} and write, restoring the full prefactor,
\begin{equation}
\begin{split}
 \tilde{c}_n^{(k)}(\rho) &=
 \frac{2^{1-n}}{\pi^{1/4}} \sqrt{\frac{(2n+2\ell+1)!}{(n+\ell)!n!}}
 \,\beta_\ell^{-(\ell+k+1)}
 \nonumber \\ & \quad\null\times
 \mathcal{D}_{\beta,\ell}^{(k)} \int_0^\infty\dd r\,\psi_\rho(r)\,
 \hat\jmath_{\ell}(\beta r)\,\Big|_{\beta=\beta_\ell} \\
 &= \frac{2^{1-n}}{\pi^{1/4}} \sqrt{\frac{(2n+2\ell+1)!}{(n+\ell)!n!}}
 \,\beta_\ell^{-(\ell+k+1)} 
 \nonumber \\ & \quad\null\times
 \sqrt{\frac\pi2}\,
 \mathcal{D}_{\beta,\ell}^{(k)}\,\tilde{\psi}_\rho(\beta)\,
 \Big|_{\beta=\beta_\ell} \\
 &= C_\ell(n)\,\beta_\ell^{-(\ell+k+1)}
 \times \mathcal{D}_{\beta,\ell}^{(k)}\,\hat\jmath(\beta\rho)
 \Big|_{\beta=\beta_\ell} \\[0.8em]
 &= C_\ell(n)\,\beta_\ell^{-\ell-k} \times P_k(\ell+3/2,\rho^2)\,
 \rho^{k+1} j_{\ell+k}(\beta_\ell\rho) \,.
\end{split}
\label{eq:c-tilde-k-2}
\end{equation}
We have used here that $\tilde{\psi}_\rho(\beta) = \beta\rho\,j_\ell(\beta\rho)
= \hat\jmath(\beta\rho)$ for $\beta \leq x_\ell/\rho$, and that we can 
ultimately apply the operator $\mathcal{D}_{\beta,\ell}^{(k)}$ to get back the 
original expression as in Eq.~\eqref{eq:c-tilde-k-1}, only with $r$ replaced by 
$\rho$.  The coefficients $C_\ell(n)$ have been defined in Eq.~\eqref{eq:C}.

With these general expressions for all terms in the expansion of 
$\tilde{c}_n(\rho)$, we can now write the quantization 
condition~\eqref{eq:quant-cond-ell} as
\begin{align}
 (2n+\ell+3/2-\rho^2)\times\sum_{k=0}^{\kmax}&\tilde{c}_n^{(k)}(\rho)
 \nonumber \\
  \null - \sqrt{n}\sqrt{n+\ell+1/2}
 &\times\sum_{k=0}^{\kmax}\tilde{c}_{n-1}^{(k)}(\rho) = 0 \,.
\label{eq:quant-cond-ell-gen}
\end{align}
The appropriate truncation index $\kmax$ in this equation depends on both 
$\ell$ and the desired order for the subleading corrections.  To solve for 
these, we insert an ansatz of the form~\eqref{eq:rho-ansatz-gen} into 
Eq.~\eqref{eq:quant-cond-ell-gen} and solve for the coefficients $\Delta_0$, 
$\Delta_1$, etc. by performing an asymptotic expansion around $n=\infty$.  To 
do this consistently, it is important to keep all terms that can contribute to the 
maximum order we are interested in.  In general, there are cancellations 
between the polynomial prefactors $P_k(\ell+3/2,\rho^2)\times\rho^{k+1}$ and 
the spherical Bessel functions $j_{\ell+k}(\beta_\ell\rho)$ since the latter 
contribute inverse powers of $\beta_\ell\rho$, which become more prominent with 
increasing $\ell$.  At least for $\ell=0$ and $\ell=1$ we find that $\kmax=2$ 
is sufficient to get the corrections up to and including $\OO(1/n^2)$.  The 
results, obtained with computer algebra software (Wolfram Mathematica), are
\begin{align}
 \ell &= 0:  & \Delta_1 &= \frac{3-2\pi^2}{48} \;,
 & \Delta_2 &= \frac{-7(3-2\pi^2)}{192} \,, \\
 \ell &= 1:  & \Delta_1 &= \frac{1}{48}(3-2\pi^2) \;,
 & \Delta_2 &= \frac{3(5+2x_1^2)}{64} \,.
 \label{eq:lasteq}
\end{align}
One always has $\Delta_0=2$, independent of $\ell$.

%%%%%%%%%%%%%%%%%%%%%%%%%%%%%%%%%%%%%%%%%%%%%%%%%%%%%%%%%%%%%%%%%%%%%%%%%%%%%%%%
\begin{table}[htbp]
\centering
\begingroup
\renewcommand\arraystretch{1.25}
\caption{Comparison of the smallest distance scale $\rho$ at different orders 
in the $1/n$ expansion to the exact answer for several values of $n$.  S-wave 
results ($\ell = 0$).}
\label{tab:rho-ell-0}
\begin{tabular}{r|
 >{\centering\arraybackslash}p{5.1em}|
 >{\centering\arraybackslash}p{5.1em}|
 >{\centering\arraybackslash}p{5.1em}||
 >{\centering\arraybackslash}p{5.1em}
}
  \hline\hline
  $n$
  & $\rho$, $\OO(1/n^0)$ & $\rho$, $\OO(1/n^1)$ & $\rho$, $\OO(1/n^2)$
  & $\rho$, exact \\
  \hline
  1 & 0.94723 & 0.97876 & 0.92548 & 0.95857 \\
  2 & 0.81116 & 0.82075 & 0.81234 & 0.81629 \\
  3 & 0.72073 & 0.72518 & 0.72258 & 0.72355 \\
  4 & 0.65507 & 0.65756 & 0.65647 & 0.65681 \\
  5 & 0.60460 & 0.60617 & 0.60562 & 0.60576 \\
  6 & 0.56425 & 0.56531 & 0.56500 & 0.56507 \\
  7 & 0.53103 & 0.53178 & 0.53159 & 0.53163 \\
  8 & 0.50306 & 0.50362 & 0.50350 & 0.50352 \\
  9 & 0.47909 & 0.47952 & 0.47944 & 0.47945 \\
 10 & 0.45825 & 0.45859 & 0.45853 & 0.45854 \\
 11 & 0.43991 & 0.44018 & 0.44014 & 0.44015 \\
 12 & 0.42361 & 0.42384 & 0.42380 & 0.42381 \\
 \hline\hline
\end{tabular}
\endgroup
\end{table}
%%%%%%%%%%%%%%%%%%%%%%%%%%%%%%%%%%%%%%%%%%%%%%%%%%%%%%%%%%%%%%%%%%%%%%%%%%%%%%%%
\begin{table}[htbp]
\centering
\begingroup
\renewcommand\arraystretch{1.25}
\caption{Comparison of the smallest distance scale $\rho$ at different orders 
in the $1/n$ expansion to the exact answer for several values of $n$.  P-wave 
results ($\ell = 1$).}
\label{tab:rho-ell-1}
\begin{tabular}{r|
 >{\centering\arraybackslash}p{5.1em}|
 >{\centering\arraybackslash}p{5.1em}|
 >{\centering\arraybackslash}p{5.1em}||
 >{\centering\arraybackslash}p{5.1em}
}
  \hline\hline
  $n$
  & $\rho$, $\OO(1/n^0)$ & $\rho$, $\OO(1/n^1)$ & $\rho$, $\OO(1/n^2)$
  & $\rho$, exact \\
  \hline
  1 & 1.2462 & 1.3481 & 1.1464 & 1.2764 \\
  2 & 1.0898 & 1.1214 & 1.0860 & 1.1047 \\
  3 & 0.98054 & 0.99560 & 0.98424 & 0.98920 \\
  4 & 0.89868 & 0.90730 & 0.90242 & 0.90423 \\
  5 & 0.83441 & 0.83990 & 0.83741 & 0.83821 \\
  6 & 0.78220 & 0.78596 & 0.78455 & 0.78495 \\
  7 & 0.73871 & 0.74142 & 0.74055 & 0.74077 \\
  8 & 0.70175 & 0.70378 & 0.70321 & 0.70334 \\
  9 & 0.66984 & 0.67141 & 0.67101 & 0.67110 \\
  10 & 0.64192 & 0.64316 & 0.64288 & 0.64293 \\
  11 & 0.61722 & 0.61822 & 0.61802 & 0.61805 \\
  12 & 0.59517 & 0.59599 & 0.59584 & 0.59586 \\
 \hline\hline
\end{tabular}
\endgroup
\end{table}
%%%%%%%%%%%%%%%%%%%%%%%%%%%%%%%%%%%%%%%%%%%%%%%%%%%%%%%%%%%%%%%%%%%%%%%%%%%%%%%%

In Tables~\ref{tab:rho-ell-0} and~\ref{tab:rho-ell-1} we show (for $\ell=0$ and 
$\ell=1$, respectively) how subsequent inclusion of the correction terms makes 
the values for $\rho$ as defined in Eq.~\eqref{eq:rho-ansatz-gen} converge to 
the exact results, which have been calculated numerically.

\section{Pöschl--Teller states and form factors}
\label{sec:PT-Details}

In this Appendix we provide some details about the wavefunctions and separable 
form factors $\eta(k)$ for the Pöschl--Teller potential used in 
Sec.~\ref{sec:Sep-Approx}.  In the conventions of Flügge's
textbook~\cite{Fluegge:1999}, this potential can be written as\footnote{Compared 
to Ref.~\cite{Fluegge:1999} we have slightly changed the notation here by 
writing 
``$\beta$'' instead of ``$\lambda$'' (to avoid confusion with the scale 
$\lambda$ in Sec.~\ref{sec:RegContact}), and ``$\nu$'' instead of ``$m$'' (to 
label the states).}
\begin{equation}
 \VPT(r) = -\frac{\alpha^2 \beta(\beta-1)}{\cosh^2(\alpha r)}
\label{eq:V-PT-app}
\end{equation}
Labeling different states with an index $\nu$, we can write their wavefunctions 
as
\begin{eqnarray}
 \psi_{\beta\nu}(\alpha;r) &=& \frac{\sqrt{2}}{r}\cosh^\beta(\alpha r)
 \sinh(\alpha r)
 \nonumber \\ & & \null\times
 {_2F_1}\!\Bigl(\nu+\frac32,\beta-\nu-\frac12,\frac32;
 -\sinh^2(\alpha r)\Bigr) \,,
 \nonumber \\
\label{eq:psi-PT}
\end{eqnarray}
where $_2F_1$ is the hypergeometric function~\cite{Olver:2010:NHMF}.
The first odd bound state with nonzero energy occurs for $\beta=3,
\nu=0$, and has a binding momentum $\kappa=\alpha$.  For $\beta=4$,
there are two odd bound states, one of which has zero energy.  For
$\beta=5$ one finds two odd bound states at $\kappa=3\alpha$ ($\nu=0$)
and $\kappa=\alpha$ ($\nu=1$).  With Eq.~\eqref{eq:psi-PT} it is
straightforward to obtain the separable approximations for these
states.  We find
\begin{eqnarray}
 \eta_\PT(k) &=& -\frac{\sqrt{2}\,(k^2+\alpha^2)}
 {\alpha^2\cosh\!\left(\frac{\pi k}{2\alpha}\right)} \mathtext{,}
  \nonumber \\
 g_{\PT} &=& -\frac{5\pi}{16\alpha}
 \mathtext{for} \beta = 3 \mathtext{and} \nu = 0 \,,
\label{eq:eta-PT-3}
\end{eqnarray}
and
\begin{subequations}%
\begin{align}
 \eta_\PT(k) = -\frac{\sqrt{2}\,(k^2+\alpha^2)(k^2+9\alpha^2)}
 {6\alpha^4\cosh\!\left(\frac{\pi k}{2\alpha}\right)} &\mathtext{,}
 g_{\PT} = -\frac{63\pi}{256\alpha}
 \nonumber \\ 
 \mathtext{for} \beta = 5 \mathtext{and} \nu = 0 \,, \\
 \eta_\PT(k) = -\frac{\sqrt{2}\,(k^2+\alpha^2)(-7k^2+17\alpha^2)}
 {18\alpha^4\cosh\!\left(\frac{\pi k}{2\alpha}\right)} &\mathtext{,}
 g_{\PT} = -\frac{81\pi}{256\alpha}
 \nonumber \\
 \mathtext{for} \beta = 5 \mathtext{and} \nu = 1 \,.
\end{align}
\label{eq:eta-PT-5}%
\end{subequations}%
It is straightforward to obtain results also for higher values of $\beta$, but 
we restrict ourselves to these representative examples here.

\newpage

\end{document}